\documentclass[twocolumn]{aastex631}

\newcommand{\Mj}{$M_{\rm Jup}$}
\newcommand{\phip}{\phi_{p}}
\newcommand{\kms}{km\,s$^{-1}$}

\newcommand{\discminer}{\textsc{discminer}}

\newcommand{\bettermoments}{\textsc{bettermoments}}

\newcommand{\radmc}{\textsc{radmc3d}}
\newcommand{\exoalma}{exoALMA}
\newcommand{\filfinder}{\textsc{FilFinder}}
\newcommand{\ppdonet}{\textsc{ppdonet}}
\newcommand{\twCOfull}{$^{12}$CO\,$J=3\rightarrow2$}
\newcommand{\twCO}{$^{12}$CO}
\newcommand{\thCO}{$^{13}$CO}
\newcommand{\thCOfull}{$^{13}$CO\,$J=3\rightarrow2$}

\newcommand{\cqtau}{CQ\,Tau}

\newcommand{\hdone}{HD\,135344B}

\newcommand{\mwcsev}{MWC\,758}
\newcommand{\pdssix}{PDS\,66}

\hypersetup{
  colorlinks   = true, 
  urlcolor     = blue,
  linkcolor    = blue, 
  citecolor   = blue 
} 

\usepackage{booktabs} 
\usepackage{multirow} 
\usepackage{makecell} 
\usepackage{comment}
\usepackage{amsmath}
\usepackage{xcolor} 
\usepackage{ragged2e}
\usepackage[]{hyperref}

\graphicspath{{./}{figures/}}

\begin{document}

\title{exoALMA XX: Tomographic Detection of Embedded Planets in Protoplanetary Disks}

\author[0000-0001-8446-3026]{Andr\'es F. Izquierdo}
\altaffiliation{NASA Hubble Fellowship Program Sagan Fellow}
\affiliation{Department of Astronomy, University of Florida, Gainesville, FL 32611, USA; \textrm{\url{andres.izquierdo.c@gmail.com}}}
\affiliation{Leiden Observatory, Leiden University, P.O. Box 9513, NL-2300 RA Leiden, The Netherlands}

\author[0000-0001-7258-770X]{Jaehan Bae}
\affiliation{Department of Astronomy, University of Florida, Gainesville, FL 32611, USA; \textrm{\url{andres.izquierdo.c@gmail.com}}}

\author[0000-0003-4689-2684]{Stefano Facchini}
\affiliation{Dipartimento di Fisica, Universit\`a degli Studi di Milano, Via Celoria 16, 20133 Milano, Italy}

\author[0000-0001-7591-1907]{Ewine F. van Dishoeck}
\affiliation{Leiden Observatory, Leiden University, P.O. Box 9513, NL-2300 RA Leiden, The Netherlands}
\affiliation{Max-Planck Institut für Extraterrestrische Physik (MPE), Gießenbachstr. 1 , 85748 Garching bei München, Germany}

\author[0000-0001-6378-7873]{Marcelo Barraza-Alfaro}
\affiliation{Department of Earth, Atmospheric, and Planetary Sciences, Massachusetts Institute of Technology, Cambridge, MA 02139, USA}

\author[0000-0002-7695-7605]{Myriam Benisty}
\affiliation{Max-Planck Institute for Astronomy (MPIA), Königstuhl 17, 69117 Heidelberg, Germany}

\author[0000-0003-1534-5186]{Richard Teague}
\affiliation{Department of Earth, Atmospheric, and Planetary Sciences, Massachusetts Institute of Technology, Cambridge, MA 02139, USA}

\author[0000-0002-0491-143X]{Jochen Stadler} 
\affiliation{Universit\'e C\^ote d'Azur, Observatoire de la C\^ote d'Azur, CNRS, Laboratoire Lagrange, 06304 Nice, France}
\affiliation{European Southern Observatory, Karl-Schwarzschild-Str. 2, D-85748 Garching bei M\"unchen, Germany}


\author[0000-0003-2253-2270]{Sean M. Andrews}
\affiliation{Center for Astrophysics | Harvard \& Smithsonian, Cambridge, MA 02138, USA}

\author[0000-0002-2700-9676]{Gianni Cataldi} 
\affiliation{National Astronomical Observatory of Japan, 2-21-1 Osawa, Mitaka, Tokyo 181-8588, Japan}

\author[0000-0003-3713-8073]{Nicolás Cuello} 
\affiliation{Univ. Grenoble Alpes, CNRS, IPAG, 38000 Grenoble, France}

\author[0000-0003-2045-2154]{Pietro Curone} 
\affiliation{Departamento de Astronom\'ia, Universidad de Chile, Camino El Observatorio 1515, Las Condes, Santiago, Chile}

\author[0000-0002-1483-8811]{Ian Czekala} 
\affiliation{School of Physics \& Astronomy, University of St. Andrews, North Haugh, St. Andrews KY16 9SS, UK}

\author[0000-0003-4679-4072]{Daniele Fasano} 
\affiliation{Universit\'e C\^ote d'Azur, Observatoire de la C\^ote d'Azur, CNRS, Laboratoire Lagrange, 06304 Nice, France}

\author[0000-0002-9298-3029]{Mario Flock} 
\affiliation{Max-Planck Institute for Astronomy (MPIA), Königstuhl 17, 69117 Heidelberg, Germany}

\author[0000-0003-1117-9213]{Misato Fukagawa} 
\affiliation{National Astronomical Observatory of Japan, Osawa 2-21-1, Mitaka, Tokyo 181-8588, Japan}

\author[0000-0002-5503-5476]{Maria Galloway-Sprietsma}
\affiliation{Department of Astronomy, University of Florida, Gainesville, FL 32611, USA; \textrm{\url{andres.izquierdo.c@gmail.com}}}

\author[0000-0002-8138-0425]{Cassandra Hall} 
\affiliation{Department of Physics and Astronomy, The University of Georgia, Athens, GA 30602, USA}
\affiliation{Center for Simulational Physics, The University of Georgia, Athens, GA 30602, USA}
\affiliation{Institute for Artificial Intelligence, The University of Georgia, Athens, GA, 30602, USA}

\author[0000-0001-6947-6072]{Jane Huang} 
\affiliation{Department of Astronomy, Columbia University, 538 W. 120th Street, Pupin Hall, New York, NY, USA}

\author[0000-0003-1008-1142]{John~D.~Ilee} 
\affiliation{School of Physics and Astronomy, University of Leeds, Leeds, UK, LS2 9JT}

\author[0000-0001-8061-2207]{Andrea Isella}
\affiliation{Department of Physics and Astronomy, Rice University, 6100 Main St, Houston, TX 77005, USA}
\affiliation{Rice Space Institute, Rice University, 6100 Main St, Houston, TX 77005, USA}

\author[0009-0007-5371-3548]{Jensen Lawrence}
\affiliation{Department of Earth, Atmospheric, and Planetary Sciences, Massachusetts Institute of Technology, Cambridge, MA 02139, USA}

\author[0000-0002-8896-9435]{Geoffroy Lesur} 
\affiliation{Univ. Grenoble Alpes, CNRS, IPAG, 38000 Grenoble, France}

\author[0000-0002-2357-7692]{Giuseppe Lodato} 
\affiliation{Dipartimento di Fisica, Universit\`a degli Studi di Milano, Via Celoria 16, 20133 Milano, Italy}

\author[0000-0003-4663-0318]{Cristiano Longarini} 
\affiliation{Institute of Astronomy, University of Cambridge, Madingley Road, CB3 0HA, Cambridge, UK}

\author[0000-0002-8932-1219]{Ryan A. Loomis}
\affiliation{National Radio Astronomy Observatory, 520 Edgemont Rd., Charlottesville, VA 22903, USA}

\author[0000-0002-1637-7393]{François Ménard}
\affiliation{Univ. Grenoble Alpes, CNRS, IPAG, 38000 Grenoble, France}

\author[0000-0001-5907-5179]{Christophe Pinte}
\affiliation{Univ. Grenoble Alpes, CNRS, IPAG, 38000 Grenoble, France}
\affiliation{School of Physics and Astronomy, Monash University, Clayton VIC 3800, Australia}

\author[0000-0002-4716-4235]{Daniel J. Price} 
\affiliation{School of Physics and Astronomy, Monash University, Clayton VIC 3800, Australia}

\author[0000-0003-4853-5736]{Giovanni Rosotti} 
\affiliation{Dipartimento di Fisica, Universit\`a degli Studi di Milano, Via Celoria 16, 20133 Milano, Italy}

\author[0000-0003-1859-3070]{Leonardo Testi}
\affiliation{Dipartimento di Fisica e Astronomia, Università di Bologna, I-40129 Bologna, Italy}

\author[0000-0003-1526-7587	]{David J. Wilner} 
\affiliation{Center for Astrophysics | Harvard \& Smithsonian, Cambridge, MA 02138, USA}

\author[0000-0002-7501-9801]{Andrew J. Winter}
\affiliation{Astronomy Unit, School of Physics and Astronomy, Queen Mary University of London, London E1 4NS, UK}

\author[0000-0002-7212-2416]{Lisa W\"olfer} 
\affiliation{Department of Earth, Atmospheric, and Planetary Sciences, Massachusetts Institute of Technology, Cambridge, MA 02139, USA}

\author[0000-0001-9319-1296]{Brianna Zawadzki} 
\affiliation{Department of Astronomy, Van Vleck Observatory, Wesleyan University, 96 Foss Hill Dr., Middletown, CT, 06459, USA}

\begin{abstract}

The \exoalma{} Large Program has revealed a wealth of substructures in the dust and molecular line emission of several protoplanetary discs, suggesting that planet formation may unfold within highly dynamic environments. Using synthetic observations of planet-disc interactions and disc instabilities, we demonstrate how the origin of these substructures can be investigated through a tomographic study of molecular lines, extending the scope of the analysis beyond line-centroid kinematics alone. Our results indicate that with only a few hours of ALMA integration at moderate angular resolution ($0\farcs{15}-0\farcs{30}$), it is possible to identify the key signatures driven by planets more massive than 0.1\% of the stellar mass. These signatures manifest not only as deviations from Keplerian motion but also as localized line broadening, enabling accurate constraints on the orbital radius and azimuthal location of the planets. We further show that a diagnostic based on line skewness in spectrally resolved observations can help distinguish between planetary and instability-driven signatures, owing to the distinct degrees of velocity coherence associated with each mechanism. Finally, we apply this tomographic analysis to \exoalma{} CO line data for the discs of \hdone{} and \mwcsev{}. In \hdone{}, we identify strongly localized velocity and line-width perturbations, suggesting the possibility of three massive planets embedded in the disc: one at $R=95$\,au, exterior to the continuum substructures, and two within dust gaps at $R=41$\,au and $R=73$\,au. For \mwcsev{}, the dominance of vertical-velocity spirals over localized signatures is consistent with predictions from models of moderate disc eccentricities or warps, potentially induced by a substellar companion in the inner regions of the system.

\end{abstract}


\section{Introduction}
\label{sec:intro}

Sensitive and high-angular resolution ALMA observations of protoplanetary discs are revolutionizing our understanding of the physicochemical mechanisms of planet formation. These data have revealed different families of dust and gas substructures in numerous discs \citep{andrews+2018, oberg+2021, teague+2025, fukagawa+2026}, potentially arising from the interplay between thermochemical processes \citep[e.g.][]{miotello+2023, oberg+2023} and physical mechanisms, including planet-- and star-disc interactions \citep[see][for a review]{pinte+2023}, as well as gravitational and (magneto)hydrodynamical disc instabilities \citep[e.g.][]{lesur+2023, bae+2023}.

In particular, the dynamical interaction between discs and planets is known to induce distinct signatures in molecular-line data, which can be broadly categorized into two groups according to their observed symmetry. The first group comprises axisymmetric substructures dominated by azimuthal velocity flows associated with pressure modulations around planet-carved gaps \citep[e.g.][]{teague+2018a, rosotti+2020, izquierdo+2023, stadler+2025}. The second group includes non-axisymmetric patterns extending across the disc, typically in the form of spirals \citep[e.g.][]{teague+2021, calcino+2022, woelfer+2023}, as well as localized kinematic perturbations \citep[e.g.][]{pinte+2018_kink, casassus+2019, izquierdo+2022, stadler+2023} and chemical anomalies detected in the vicinity of embedded planets \citep{law+2023, izquierdo+2026}. A comparatively less explored feature in this group involves localized meridional flows {induced by a planet, which lead to enhanced velocity dispersion. These flows} arise from the interplay between planet-induced gravitational torques and viscous torques that act to refill the planet-driven gap from disc layers above the midplane \citep{dong+2019, izquierdo+2023}. 

Similar observational features may, however, also arise from fluid instabilities that develop under specific disc conditions. For instance, nearly axisymmetric flows are a common outcome of the vertical shear instability (VSI; e.g. \citealt{barraza+2021}), which operates in thermally stratified discs with short cooling times that sustain differential rotation above the midplane. Spiral-like structures, by contrast, are more characteristic of gravitational instability (GI; e.g. \citealt{hall+2020, speedie+2024}), which occurs in sufficiently massive discs where self-gravity overcomes thermal and rotational support. Large-scale spiral morphologies, together with localized eddies, can also arise from the magnetorotational instability (MRI; e.g. \citealt{barraza+2025}), which develops in regions where the gas is sufficiently ionized for magnetic fields to couple to the flow and drive turbulence.

In this work, we quantify the detectability of planet-driven perturbations using exoALMA-like observations of three-dimensional planet-disc interaction simulations. We adopt a tomographic approach in which line broadening and line asymmetries are used as diagnostics of coherent velocity flows in the vicinity of embedded planets, and compare these signatures with those produced in VSI, MRI, and GI simulations. We then apply this analysis to two distinct \exoalma{} targets: the disc of \hdone{}, where multiple localized perturbations are identified, and \mwcsev{}, where large-scale substructures predominate.

This Letter is organized as follows. Section \ref{sec:data_models} describes the observational data and channel-map models of the \exoalma{} targets analyzed in this study. Section \ref{sec:hydro} presents the extraction and characterization of observable features from models of planet-disc interactions and fluid instabilities. Section \ref{sec:residuals} illustrates the application of this methodology to the discs of \hdone{} and \mwcsev{}. Section \ref{sec:discussion} discusses the interpretation of the detected signatures, and Section \ref{sec:conclusions} highlights the main results of the work. 

\vspace{0.5cm}
\section{Data and Models}
\label{sec:data_models}

   \begin{figure*}
   \centering
    \includegraphics[width=1.0\textwidth]{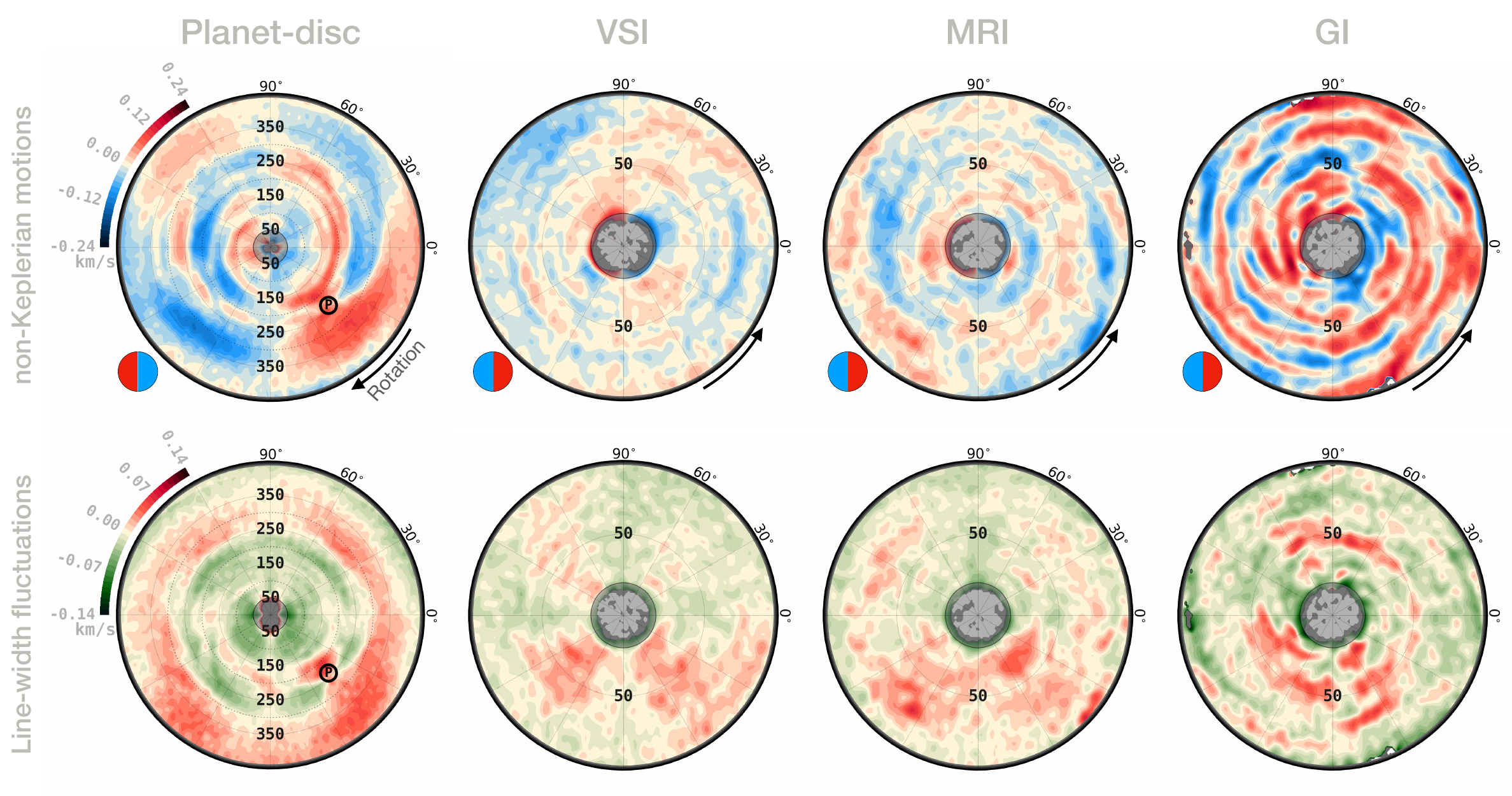}
      \caption{Velocity and line-width residuals derived from \discminer{} models applied to synthetic observations of planet-disc interaction ($q=2\times10^{-3}$, $\phi_p=-45^\circ$) and disc instabilities. Details of the simulation setups are provided in Sect. \ref{sec:planet_disc}. All discs are inclined by $-30^\circ$, and the residuals from all mechanisms are shown on the same color scale indicated on the left. Semi red-blue circles in the bottom left corners of the top-row panels indicate the locations of the redshifted and blueshifted sides of the simulated disc, which are useful to understand the velocity structure of the non-Keplerian flows (see Fig. \ref{fig:velocity_patterns}). 
              }
         \label{fig:mechanisms}
   \end{figure*}

   \begin{figure*}
   \centering
    \includegraphics[width=1.0\textwidth]{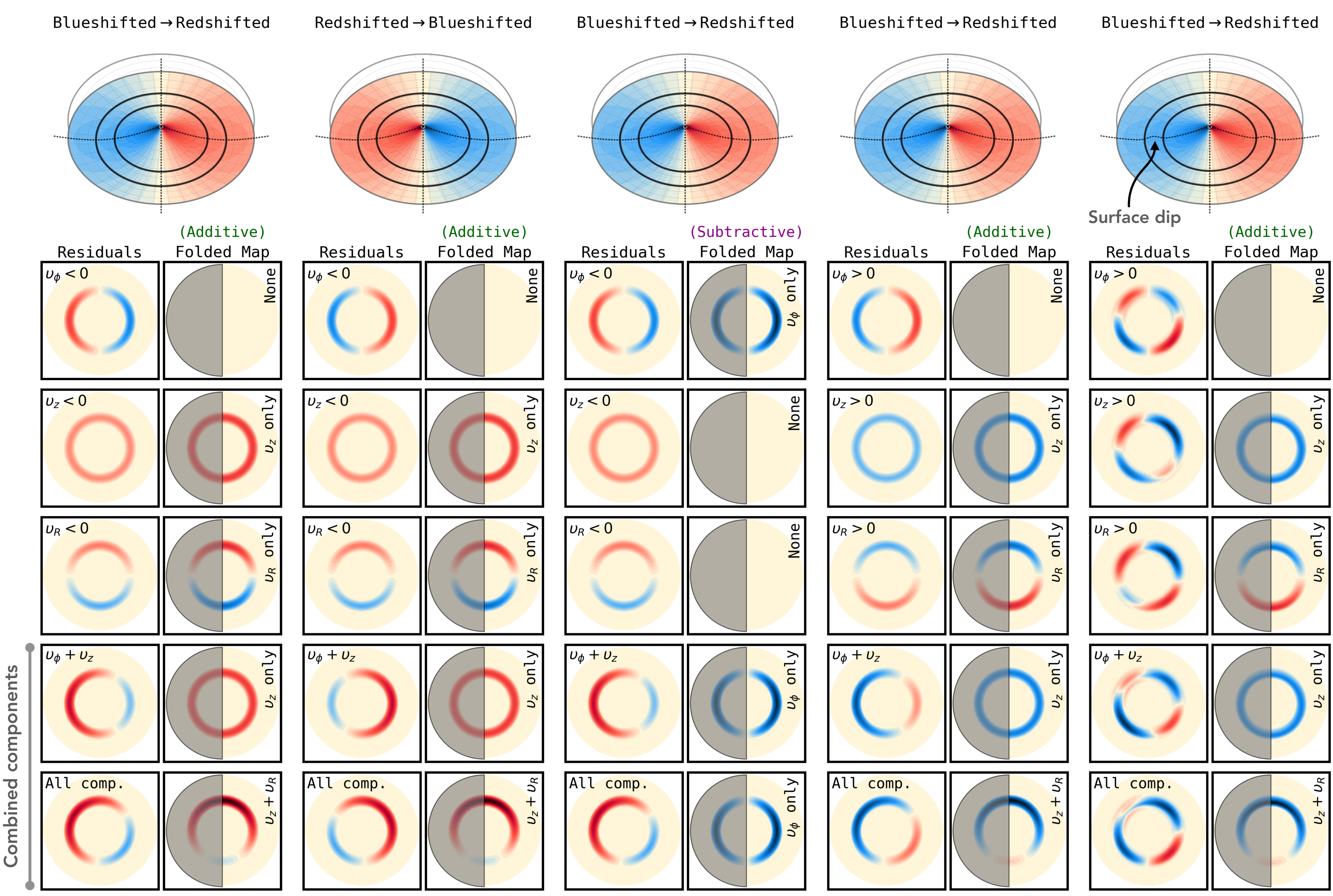}
      \caption{Illustrating axisymmetric velocity residuals for different combinations of velocity components, projected disc rotation, and the resulting folded maps obtained using either the additive or subtractive folding method. The first three columns show residuals for negative velocity flows after subtraction of a Keplerian model, while the following two focus on positive ones. In this context, \textit{positive} denotes super-Keplerian motion for $v_\phi$, outward motion in cylindrical radius for $v_R$, and upward motion relative to the disc midplane for $v_z$. The rightmost column shows a scenario where the underlying disc emission surface includes a dip (30\,au wide and 20\,au deep) that is not considered by the subtracted model. This substructure introduces an anti-symmetric contribution to the residuals around the disc minor axis, which is consequently filtered out after additive folding. 
              }
         \label{fig:velocity_patterns}
   \end{figure*}

The ALMA data used in this Letter consist of \exoalma{} continuum-subtracted Fiducial Images of \twCO{} and \thCOfull{}
for the discs of \hdone{} and \mwcsev{}, as defined in \citet{teague+2025}. These data cubes have a circular synthesized beam of $0\farcs{15}$ and a channel spacing of 0.1\,\kms{}. Details of the calibration and imaging procedures are provided in \citet{loomis+2025}.
We also use \exoalma{} Fiducial Continuum Images centered at 0.9\,mm from \citet{curone+2025}, and VLT/SPHERE scattered-light images presented in \citet{stolker+2016} and \citet{benisty+2015}.
Channel-map models of the targets produced with \discminer{}, as reported by \citet{izquierdo+2025}, are used as references to compute residual maps. These residuals enable the analysis of gas substructures that manifest as non-Keplerian motions and, more generally, as perturbations in the shape of the molecular line profiles. The Keplerian stellar masses obtained from the channel-map models are 1.61 and 1.40\,$M_\odot$ for \hdone{} and \mwcsev{}, respectively, assuming Gaia-derived distances of 135 and 156\,pc \citep{gaia+2023}. 
The low disc inclinations of $-16^\circ$ and $19^\circ$, as inferred with \discminer{}, render the backside emission contribution negligible for both targets, resulting in line profiles that are generally symmetric about their centroids. We therefore characterize the line peaks, centroid velocities, and line widths of both the observed and modeled spectra by fitting single-Gaussian profiles along the spectral axis at each image pixel.

\section{Analysis of Synthetic observations}
\label{sec:hydro}

In this section, we present our analysis methodology for identifying gas substructures in molecular line images produced by planet-disc interactions and fluid instabilities including VSI, MRI, and GI. The simulations are postprocessed to match the observational constraints of \exoalma{}, allowing us to assess the detectability of the associated signatures under realistic conditions. Residual maps of the analyzed mechanisms, obtained after subtracting smooth Keplerian \discminer{} models, are shown in Figure \ref{fig:mechanisms}, with selected \twCO{} intensity channels presented in Figure \ref{fig:channels_mechanisms} of the Appendix.

\subsection{Simulation setup and radiative transfer} \label{sec:planet_disc}

The planet-disc interaction simulations explored in this work were performed using the \textsc{fargo3d} code \citep{benitez+2016}. This choice is not critical, however, as other hydrodynamic packages produce consistent observable kinematic signatures when analyzed through the same pipeline applied here \citep{bae+2025}.  
The simulations are set up following \citet{teague+2019nat}, which aimed to reproduce the gas kinematics in the disc of HD\,163296. In brief, the disc is initialized with a density distribution that satisfies the hydrostatic equilibrium and adopts a uniform alpha viscosity of 10$^{-3}$. The disc midplane and atmosphere follow independent power-law temperature distributions with radius, connected in the vertical direction by a squared-cosine function \citep{dartois+2003}.
Instead of inserting multiple planets as in \citet{teague+2019nat}, we run two single-planet simulations with planet masses of $M_p = 2$\,\Mj{} and $4$\,\Mj{} orbiting a $2$\,$M_\odot$ star, corresponding to planet-to-star mass ratios of $q\!\sim\!1\times10^{-3}$ and $\sim\!2\times10^{-3}$, respectively. The orbital radius of the planet is fixed at $R_p=240$\,au, while the azimuth $\phi_p$ is varied in eight increments of $45^{\circ}$ to study the effects of projection and radiation transport on the retrieved perturbations along the entire disc circumference. 

Radiative transfer calculations of CO line intensities were performed with \radmc{} \citep{dullemond+2012}, assuming a distance to the source of 101\,pc and three disc inclinations: $i = -15^\circ$, $-30^\circ$, and $-45^\circ$. The central star is assumed to have mass and luminosity of $2~M_\odot$ and $17~L_\odot$, respectively, similar to those of HD~163296 \citep{fairlamb2015}. We assume that CO is fully photodissociated in surface layers of the disc where the vertically integrated gas density falls below $10^{21}~{\rm cm}^{-2}$ \citep{visser2009}, and that CO is frozen onto dust grains in regions where the temperature drops below 21~K \citep{schwarz2016}. For the remainder of the gas disc we adopt a CO/H$_2$ abundance ratio of $10^{-5}$. 

Synthetic observations replicating the main properties of the Fiducial Images from \exoalma{} were generated using the \textsc{syndisk} code\footnote{\url{https://github.com/richteague/syndisk}}, adopting a channel spacing of 0.1\,\kms{}, and circular beams of $0\farcs{15}$ and $0\farcs{3}$. Two rms noise levels of 1.2 and 2.4\,K were adopted for the fiducial $0\farcs15$ beam, corresponding to signal-to-noise ratios (SNR) of $\sim\!15$ and $\sim\!30$ relative to the azimuthally averaged peak intensity of 35\,K measured at the planet’s orbital radius in the $q\!\sim\!2\times10^{-3}$, $i=15^\circ$ simulation. Under the same observational constraints, the larger $0\farcs3$ beam also used here yields an $8\times$ increase in SNR.

The simulations and synthetic observations of the disc instabilities studied in this work are described in detail by \citet{barraza+2025}. They were performed using the magnetohydrodynamics code PLUTO \citep{mignone+2007} and were originally presented by \citet{flock+2020} for the vertical shear instability (VSI), \citet{flock+2015} for the magnetorotational instability (MRI), and \citet{bethune+2022} for gravitational instability (GI). In all cases, the adopted stellar mass was $M_\star = 0.5\,M_\odot$, with a self-gravitating disc mass of $0.1\,M_\odot$ in the GI simulation, and a disc inclination of $i=-30^\circ$. For the postprocessing stage, we generate synthetic \twCOfull{} emission for each mechanism using \radmc{}, assuming a source distance of 140\,pc, a channel spacing of 0.1\,\kms{}, and a \twCO{} abundance of $10^{-4}$, reduced by a factor of $10^{5}$ in regions of the disc where the temperature falls below 21\,K (to emulate CO freeze-out) or where the vertically integrated gas column density drops below $10^{21}$\,cm$^{-2}$ (to account for CO photodissociation). Since the physical domains of the simulations are typically smaller ($\sim\!100$\,au) than those of most \exoalma{} discs, the angular resolution of the synthetic observations was set to $0\farcs035$ in order to sample the disc radial extent with $\sim\!20$ resolution elements, consistent with the average of our targets. The rms noise level was set to 1.3\,K, yielding signal-to-noise ratios between 15 and 30 relative to the measured peak intensities across the disc.

   \begin{figure*}
   \centering
    \includegraphics[width=1.0\textwidth]{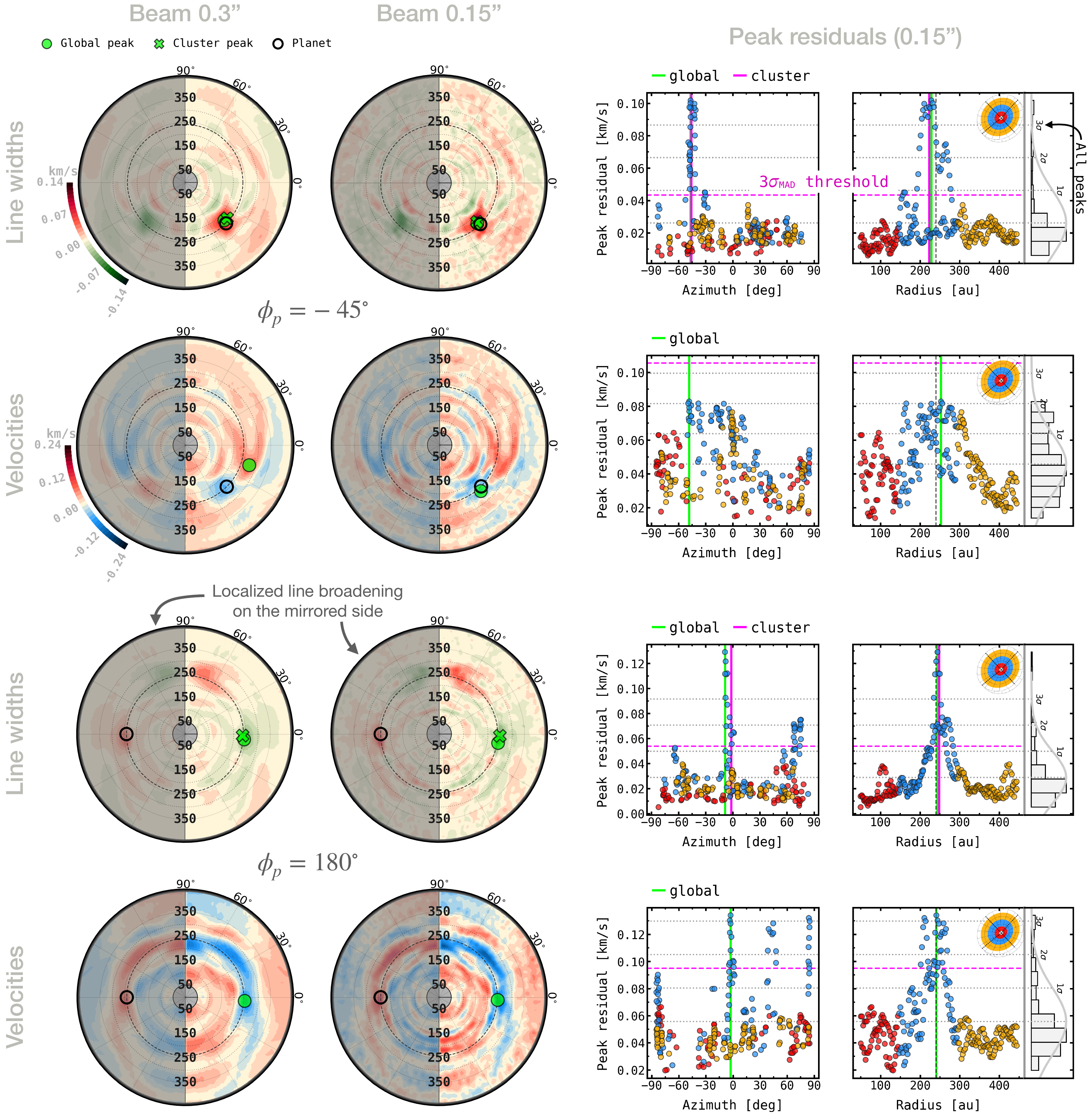}
      \caption{Illustration of the localized signatures identified in folded line-width and velocity residuals traced by \twCO{} emission from a planet-disc interaction simulation with a planet-to-star mass ratio of $q\!\sim\!2\times10^{-3}$, assuming a disc inclination of $i=-30^\circ$. Results are shown for two planet azimuths, $\phip=-45^\circ$ and $180^\circ$, and for two beam sizes, $0\farcs{3}$ and $0\farcs{15}$. Green circles and crosses mark the locations of peak and clustered residuals, while the unfilled circle denotes the true planet position. A signal is considered detected when its global peak exceeds a significance of 3$\sigma$, as indicated by the horizontal dotted and dashed lines, corresponding to thresholds defined using all peak residuals and the MAD estimator, respectively. Detections based on line widths have a higher success rate due to their stronger spatial localization (clustered residuals), compared to their velocity counterparts, which in this example remain undetected for the $\phi_p = -45^\circ$ planet and only marginally detected for $\phi_p = 180^\circ$. Since embedded planets are expected to produce localized line-width enhancements rather than reductions, the sign of the line-width signal also helps resolve the ambiguity between opposite sides of the disc introduced by the folding process.
              }
         \label{fig:folded_residuals_radmc}
   \end{figure*}

\subsection{Folded maps and Localized signatures} \label{sec:localized_signatures}

Embedded planets are known to induce localized perturbations in the dynamical and physical structure of their host discs {\citep[e.g.][]{perez+2018, dong+2019}.} To identify such signatures in our simulated data, we search for high-amplitude, spatially clustered residuals in folded velocity and line-width maps following the approach of \citet{izquierdo+2021}. By applying this folding procedure, which compares two-dimensional maps of these line-profile properties on the redshifted and blueshifted sides of the disc, we can selectively isolate non-axisymmetric perturbations driven by planets and other mechanisms. 

\paragraph{Folded maps} For velocities, folded maps can be constructed either by adding or subtracting one side of the disc from the other, depending on which velocity components are to be suppressed. As illustrated in Figure \ref{fig:velocity_patterns}, additive folding minimizes the contribution of axisymmetric azimuthal velocity flows \citep[e.g. those prominent around gas gaps,][]{izquierdo+2023, benisty+2026}, whereas subtractive folding is more effective at filtering out axisymmetric vertical or radial velocity components \citep[e.g.][]{stadler+2023}. Since azimuthal velocity flows typically dominate around surface density gaps \citep{stadler+2025} and on global scales primarily due to the disc Keplerian rotation \citep{longarini+2025}, we adopt additive folding as the default method for reducing velocity maps unless otherwise specified. Conversely, because line width is a scalar quantity, we search for localized enhancements in this property using subtractive folding. Peak intensities are excluded from this analysis, as for the optically thick line adopted they respond more strongly to large-scale temperature fluctuations than to the local influence of the planet. However, they may still be valuable for optically thinner tracers that probe closer to the midplane, where they can capture both temperature and density anomalies in the planet’s vicinity.   

\paragraph{Localized signatures} {We} take the locations of global peak residuals as proxies for embedded planets. These are derived from a two-dimensional distribution of peak velocity and line-width values extracted per annulus from the folded maps, with annuli spaced at intervals of one-fifth of the beam size. A global peak is considered significant when it exceeds $3\sigma$ relative to the mean value of all identified peaks, where $\sigma$ denotes the standard deviation of the peak distribution. As a less conservative alternative, and to avoid bias from the candidate itself inflating the noise value, we additionally evaluate a scatter estimator of the peak distribution ($p$) based on the median absolute deviation (MAD), defined as
\begin{equation} \label{eq:mad}
    \sigma_{\rm MAD} = 1.4826\,\times\,\mathrm{median}(|p-\mathrm{median}(p)|) 
\end{equation}
which provides a Gaussian-equivalent estimate of the standard deviation for normally distributed data while remaining insensitive to a small number of outliers \citep[e.g.][]{leys+2013}, such as planet-induced peaks. Hence, $\sigma_{\rm MAD}$ offers a more representative measure of the background disc perturbations away from the planet vicinity.

We also apply a K-means algorithm to identify localized clusters of residuals across both the azimuthal and radial extents of the disc as additional indicators of planet locations. The clustering analysis uses the same distribution of peak velocity and line-width residuals extracted from the folded maps. A cluster is classified as significantly localized and coherent when its variance exceeds the mean variance of all other clusters identified in the disc by more than 3$\sigma$, where $\sigma$ denotes the standard deviation of the cluster-to-cluster variance distribution.

\subsubsection{Planet-driven signatures} \label{sec:hydro_linewidths}


Figure \ref{fig:folded_residuals_radmc} illustrates a subset of the folded maps obtained from our planet-disc interaction models, highlighting the localized line-width and velocity signatures induced by the embedded planet. These maps correspond to models with a planet-to-star mass ratio of $q\!\sim\!2\times10^{-3}$ and are extracted from images convolved to beam sizes of $0\farcs15$ (with SNR$\!\sim\!30$) and $0\farcs3$, assuming a disc inclination of $i=-30^\circ$. 

\paragraph{Line width vs. velocity signals} For the $0\farcs15$ beam, detections based on line-width residuals are successful for all eight planet azimuths considered, with median offsets of 6.2\,au in radius (i.e. less than half a beam width) and 6.0$^\circ$ in azimuth relative to the true planet position. 
The corresponding median detection significances across planet azimuths are 4.3$\sigma$ with respect to all peak residuals and 8.3$\sigma$ when referenced to the scatter estimate $\sigma_{\rm MAD}$ (Eq. \ref{eq:mad}), which is less sensitive to the strong outliers associated with the planet signal. Under the same setup, these metrics are considerably improved for an inclination of $i = -15^\circ$, yielding median offsets of 0.9\,au in radius and 1.5$^\circ$ in azimuth, and median detection significances of 5.3$\sigma$ relative to all peaks and 15.5$\sigma$ with the MAD estimator. For $i = -30^\circ$, the line-width signals are detected as highly clustered in six of the eight planet azimuths; planets at $\phi_p = -135^\circ$ and $\phi_p = -90^\circ$ are identified only through global peak values. For $i = -15^\circ$, planets at all azimuths are detected using both global peak and clustered residuals.

   \begin{figure*}
   \centering
    \includegraphics[width=1.0\textwidth]{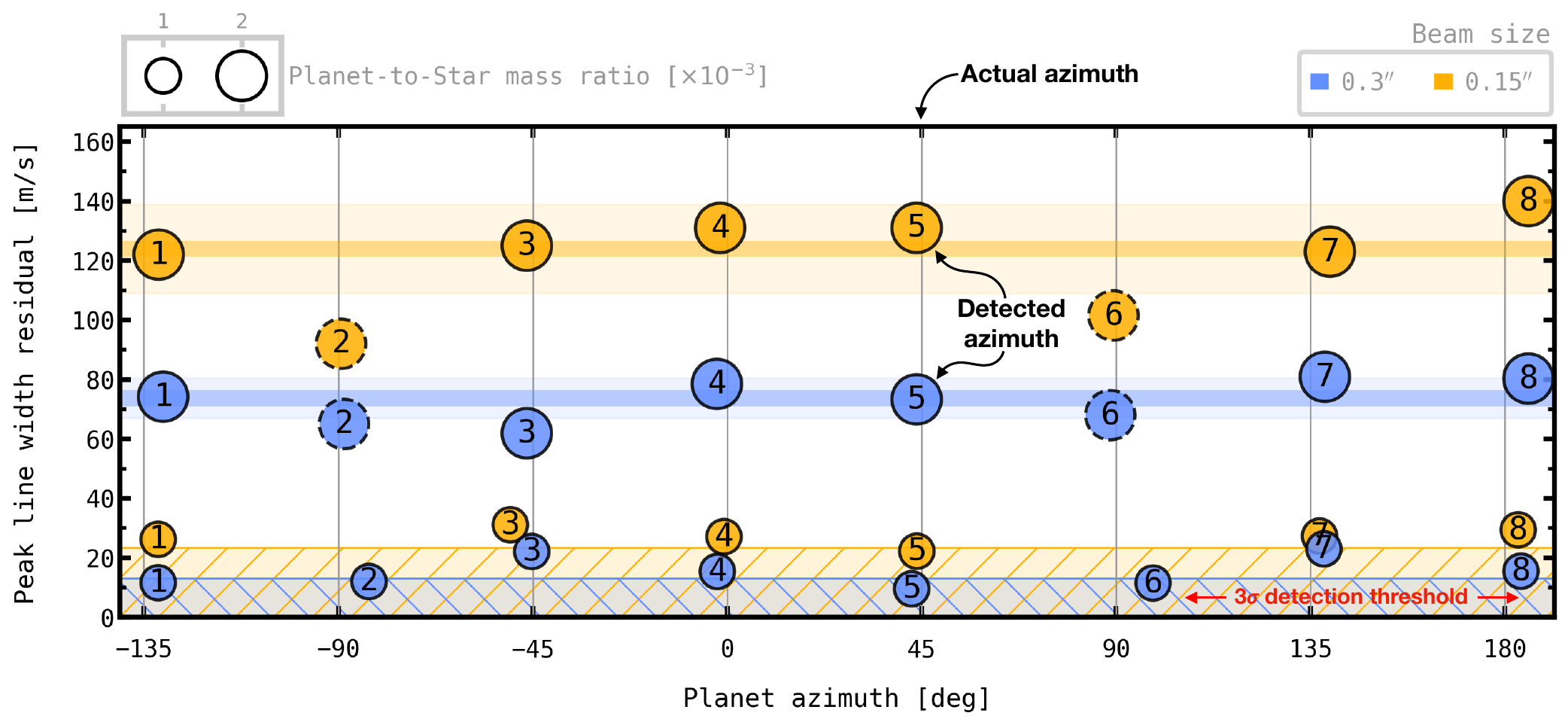}
      \caption{Detected amplitude of folded line-width perturbations as a function of the inferred planet azimuth in synthetic observations of the planet-disc interaction simulations introduced in Sect.~\ref{sec:planet_disc}, for different planet masses and beam sizes, and a disc inclination of $i = -15^\circ$. Although increasing the beam size from $0\farcs{15}$ to $0\farcs{3}$ weakens the observed signal around the planet, its detectability is not compromised, suggesting that relatively inexpensive ALMA observations are sufficient to robustly detect planets with 0.2\% the mass of the star and marginally detect those with 0.1\%. Line-width amplitudes shown as filled circles were measured using subtractive folding of the disc, whereas dotted circles, corresponding to planets located along the disc minor axis, were obtained using additive folding (followed by division by two) to avoid self-subtraction of the signal in the former method. Vertical lines indicate the true planet azimuths used in each of the eight simulation snapshots. Horizontal lines and shaded regions denote the median and standard deviation of the peak values, respectively, while hatched areas mark the median $3\sigma$ detection threshold for the low-mass planet simulations.
              }
         \label{fig:peak_linewidth_vs_azimuth}
   \end{figure*}

   \begin{figure*}
   \centering
    \includegraphics[width=1.0\textwidth]{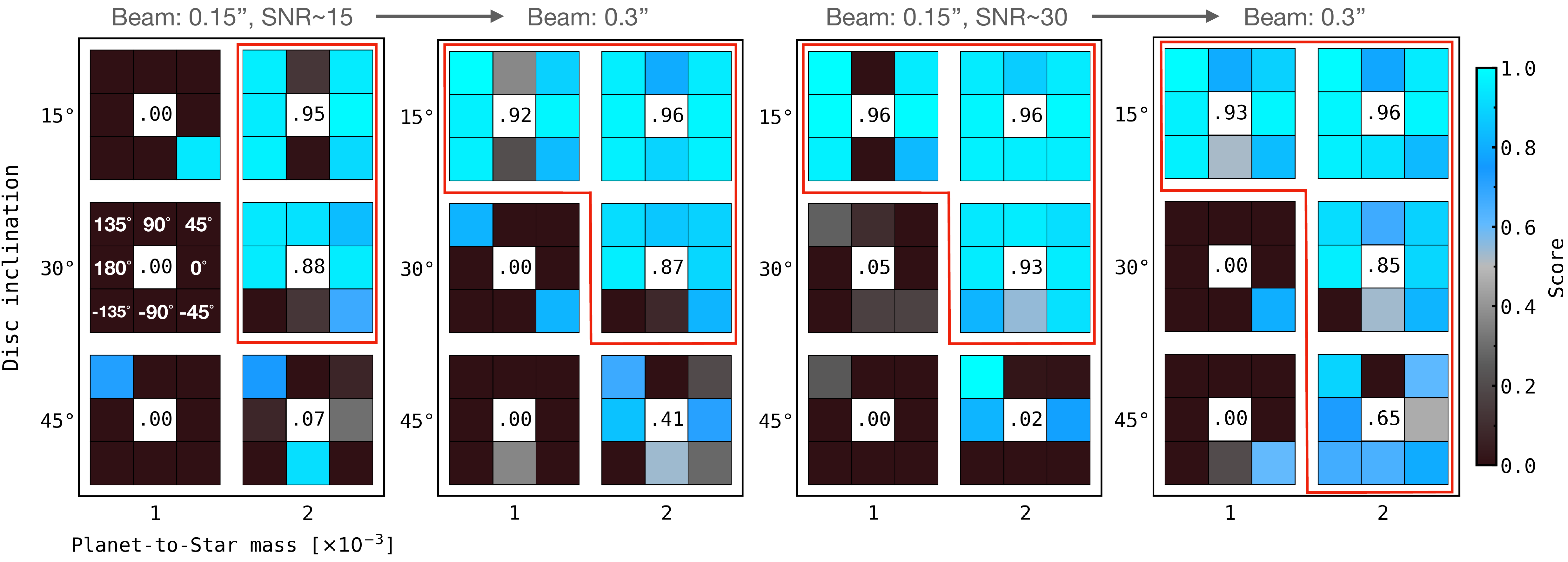}
      \caption{Goodness of planet detections based on localized line widths across all disc inclinations, planet masses and azimuths, beam sizes, and signal-to-noise ratios (SNR) considered. Sub-squares represent the eight adopted planet azimuths (from $-135^\circ$ to $180^\circ$ in steps of $45^\circ$). A score of 1 corresponds to a planet detected at its true  location with a significance of at least $3\sigma$. The score is defined as the product of three Gaussian functions evaluating the proximity of the detection in radial and azimuthal coordinates, as well as its significance. Planets with scores above 0.6 (blue sub-squares) are detected within their radial pressure bump and azimuthal sector, with a significance of at least $3\sigma$. Central sub-squares indicate the median score across all planet azimuths, with those exceeding 0.6 outlined in red to highlight the inclination-mass configurations where planet-driven line broadening is detectable for each adopted beam size and SNR.
              }
         \label{fig:scores_linewidth}
   \end{figure*}

   \begin{figure}
   \centering
    \includegraphics[width=0.99\columnwidth]{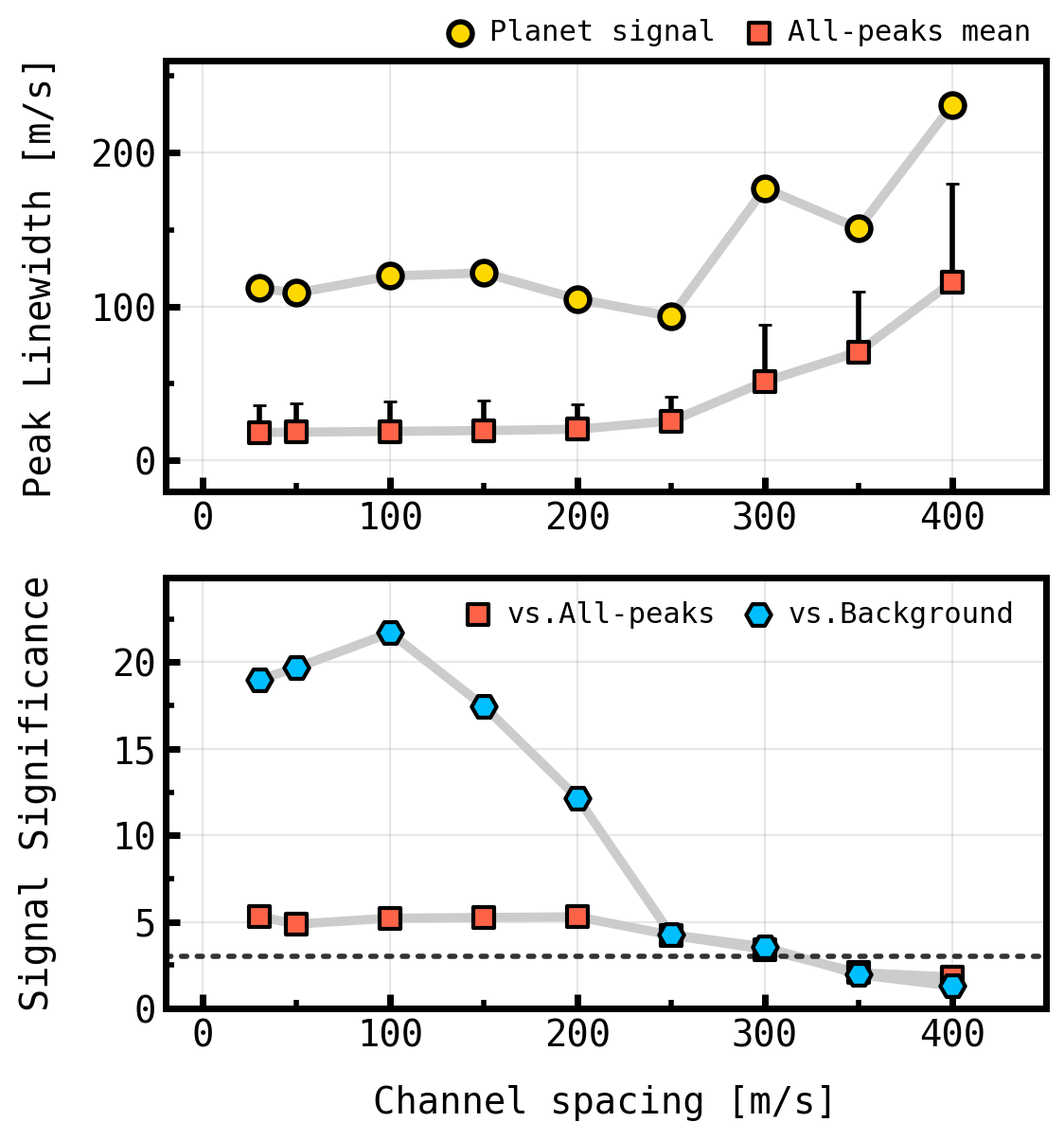}
      \caption{Peak line-width residual as a function of channel spacing for the planet-disc interaction model with a planet-to-star mass ratio of $q = 2 \times 10^{-3}$, an inclination of $i = -15^\circ$, and a planet azimuth of $\phi_p = -45^\circ$, imaged with a beam size of $0\farcs15$ and an SNR of $\sim\!30$ (at $\Delta\upsilon=0.1$\,\kms{}). Also shown is the detection significance of the signal relative to the full distribution of peak residuals and to the median absolute deviation (MAD) of the background, which is considerably less affected by planet-driven outliers.
              }
         \label{fig:channelsp_linewidth_significance}
   \end{figure}

   \begin{figure}
   \centering
    \includegraphics[width=0.99\columnwidth]{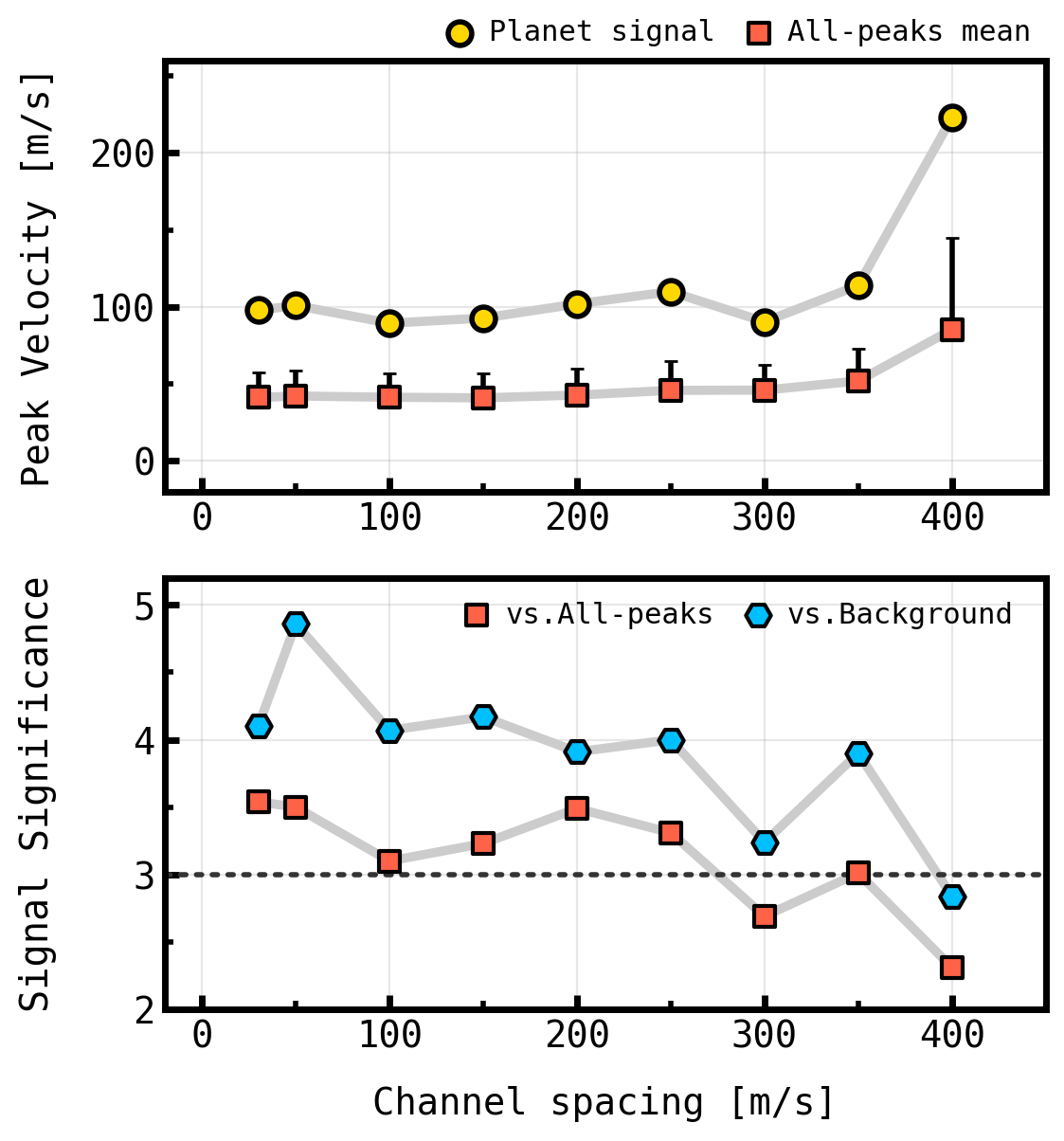}
      \caption{As Fig. \ref{fig:channelsp_linewidth_significance} but for peak velocity residuals.
              }
         \label{fig:channelsp_velocity_significance}
   \end{figure}

   \begin{figure*}
   \centering
    \includegraphics[width=1.0\textwidth]{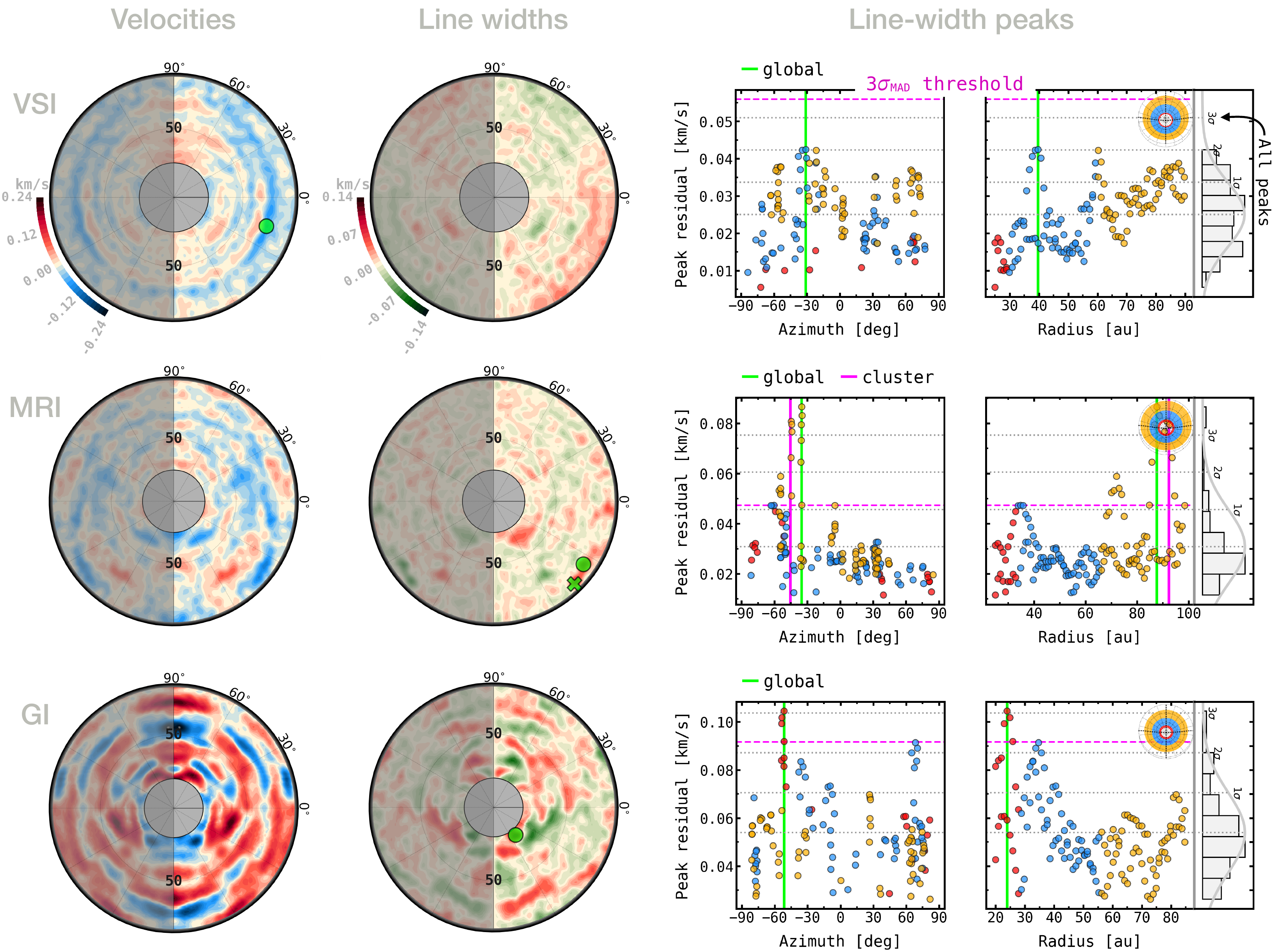}
      \caption{Folded velocity and line-width maps for the disc-instability simulations, showing the detected global peaks (green circles) and clusters (crosses) above the $3\sigma$ threshold. The right-hand panels display the distribution of peak line-width residuals as a function of azimuth and radius, which are used for signal extraction and are color-coded according to the radial region of the disc from which they were derived. In Sect. \ref{sec:line_asymmetries}, we examine the spectral origin of these signals in comparison with the planet-disc interaction scenario.
              }
         \label{fig:folded_instabilities}
   \end{figure*}

Detections based on velocity residuals are significantly less robust, with success rates of 2/8 and 4/8 for inclinations of $i = -30^\circ$ and $i = -15^\circ$, respectively. They also yield considerably lower significances than those based on line widths. For the detected signals at $i = -30^\circ$, we obtain a median significance of 3.1$\sigma$ relative to all peaks and 6.1$\sigma$ using the MAD estimator, while at $i = -15^\circ$ the corresponding values are 3.1$\sigma$ and 4.3$\sigma$. 
The similarity between the all-peak and MAD significances indicates that the peak velocity residuals are more nearly normally distributed than the line-width residuals, implying a weaker dominance of the planet-driven signal with respect to background disc fluctuations. 

This behavior arises because strong velocity perturbations develop not only in the immediate vicinity of the planet but also at larger radial and azimuthal separations, particularly for high-mass planets or in the presence of fluid instabilities. A similar result was reported by \citet{izquierdo+2021}, who showed that the signals induced by their most massive planet ($3$\,\Mj{}, or $q \sim 3 \times 10^{-3}$) were less localized in velocity than those produced by lower-mass planets. Crucially, we note that this effect is likely further exacerbated by the large-scale vertical motions driven by the planet, which were not included in the two-dimensional simulations of \citet{izquierdo+2021}.
Additionally, as illustrated in Figure \ref{fig:vlos_profiles_planet} in the Appendix, this effect can be amplified by projection: depending on the planet’s azimuthal location, multiple velocity components may add coherently (or destructively) along the line of sight at regions far from (or near) the planet, thereby competing with the localized signal associated with the embedded object. 

Despite these challenges, velocity residuals remain a valuable diagnostic of planet presence. As shown in Figure \ref{fig:hydro_signatures_12co} of the Appendix, although the velocity perturbations around the modeled planet are not strongly localized relative to background fluctuations, they still respond to the large-scale pressure modulation associated with the carved gap, and produce pronounced azimuthal gradients linked to the planet-induced Doppler flip, defined as a local reversal in the sign of the velocity signal across the planet location \citep[e.g.][]{perez+2018, rabago+2021, izquierdo+2022}. 

\paragraph{Line-width signal vs. planet azimuth and beam size} By contrast with velocity residuals, the amplitude of the detected line-width signals shows a minimal dependence on the planet’s azimuth, as demonstrated in Figure \ref{fig:peak_linewidth_vs_azimuth}, making it a more direct proxy for planet mass. An additional advantage of this diagnostic is its robustness to beam size; although degrading the angular resolution reduces the observed amplitude of the perturbations, their detectability remains largely unaffected. {For a planet-to-star mass ratio of $q\!\sim\!2 \times 10^{-3}$, we estimate a decrease in the median peak line width across planet azimuths from 124\,m\,s$^{-1}$ at a beam size of $0\farcs{15}$ to 74\,m\,s$^{-1}$ at $0\farcs{3}$.} However, in all cases, the coarser beam does not compromise the accuracy of the recovered planet location, highlighting the feasibility of systematically searching for planets even in lower-resolution data or in images deliberately degraded to improve the SNR. 

\paragraph{Planet mass detection limit} The line-width peaks from the $q\!\sim\!1 \times 10^{-3}$ hydro models are typically identified near the true planet azimuths but lie close to the 3$\sigma$ threshold relative to both the all-peak and background (MAD) fluctuations. This suggests that a planet with a mass comparable to Jupiter orbiting a Solar-mass star provides a {reasonable lower bound for the detection of} planet-driven line broadening {at the adopted orbital radius and} under the observational constraints of \exoalma{}. This {estimate} is consistent with the lowest-mass planet independently {inferred} by \citet{pinte+2025} to reproduce kink-like features in channel maps of the \exoalma{} targets.

\paragraph{Detection robustness across parameter space} We extend this analysis by summarizing the robustness of line-width peak detections across all adopted disc inclinations, signal-to-noise ratios, planet masses, azimuths, and beam resolutions in Figure \ref{fig:scores_linewidth}. This is quantified using a scoring system ranging from 0 to 1, where a score of 1 corresponds to a detection located exactly at the planet’s true position with a significance of at least $3\sigma$. The score is defined as the product of three unit-amplitude Gaussian functions that evaluate the detection's proximity to the planet in radial and azimuthal coordinates, as well as its significance. The two spatial Gaussians adopt standard deviations of $\sigma_r=35$\,au, corresponding to half the radial width of the pressure modulation induced by the planet (or $1-2$ beams), and $\sigma_\phi=20^\circ$, equivalent to an azimuthal slice of $1/18$th of the disc circumference. The significance Gaussian is capped at $3\sigma$ and decreases normally for lower values, making the metric only weakly dependent on the choice of scatter estimator ($\sigma$ or $\sigma_{\rm MAD}$) introduced in Sect. \ref{sec:localized_signatures}. Under these definitions, any planet with a score above 0.6 (blue squares) satisfies the condition of being detected within its radial pressure bump and azimuthal sector, with a significance of at least $3\sigma$. The same analysis is performed using peak velocity residuals, as illustrated in Figure \ref{fig:scores_velocity} of the Appendix.

\paragraph{Effect of SNR, beam size, and channel spacing} Regardless of the signal-to-noise ratio {(SNR)} or angular resolution employed, line-width features are robustly detected for disc inclinations of $-15^\circ$ and $-30^\circ$ for the highest planet mass considered ($q\!\sim\!2 \times 10^{-3}$). Importantly, doubling the SNR at fixed beam size substantially enhances the detectability of these signatures for the lower-mass planet ($q \sim 1 \times 10^{-3}$) and for the higher inclination of $-45^\circ$.
This analysis further indicates that lower angular resolution does not necessarily hinder detectability; in some cases, it reveals features that remain undetected in higher-resolution (but lower-SNR) images and can even enhance the significance of the planet signal at larger disc inclinations. 

Similarly, adopting coarser channel spacings by factors of 2--3 does not compromise the detectability of either line-width or velocity signals, as illustrated in Figures \ref{fig:channelsp_linewidth_significance} and \ref{fig:channelsp_velocity_significance}. However, in contrast to degrading angular resolution, the amplitudes of the planet signals remain largely unaffected until channelization effects begin to dominate. This suggests that relaxing the spectral resolution may be a viable observational alternative for boosting SNR without sacrificing planet detectability or angular resolution, particularly when probing the innermost regions of the disc where beam smearing can significantly suppress perturbations in the planet's vicinity.


   \begin{figure*}
   \centering
    \includegraphics[width=1.0\textwidth]{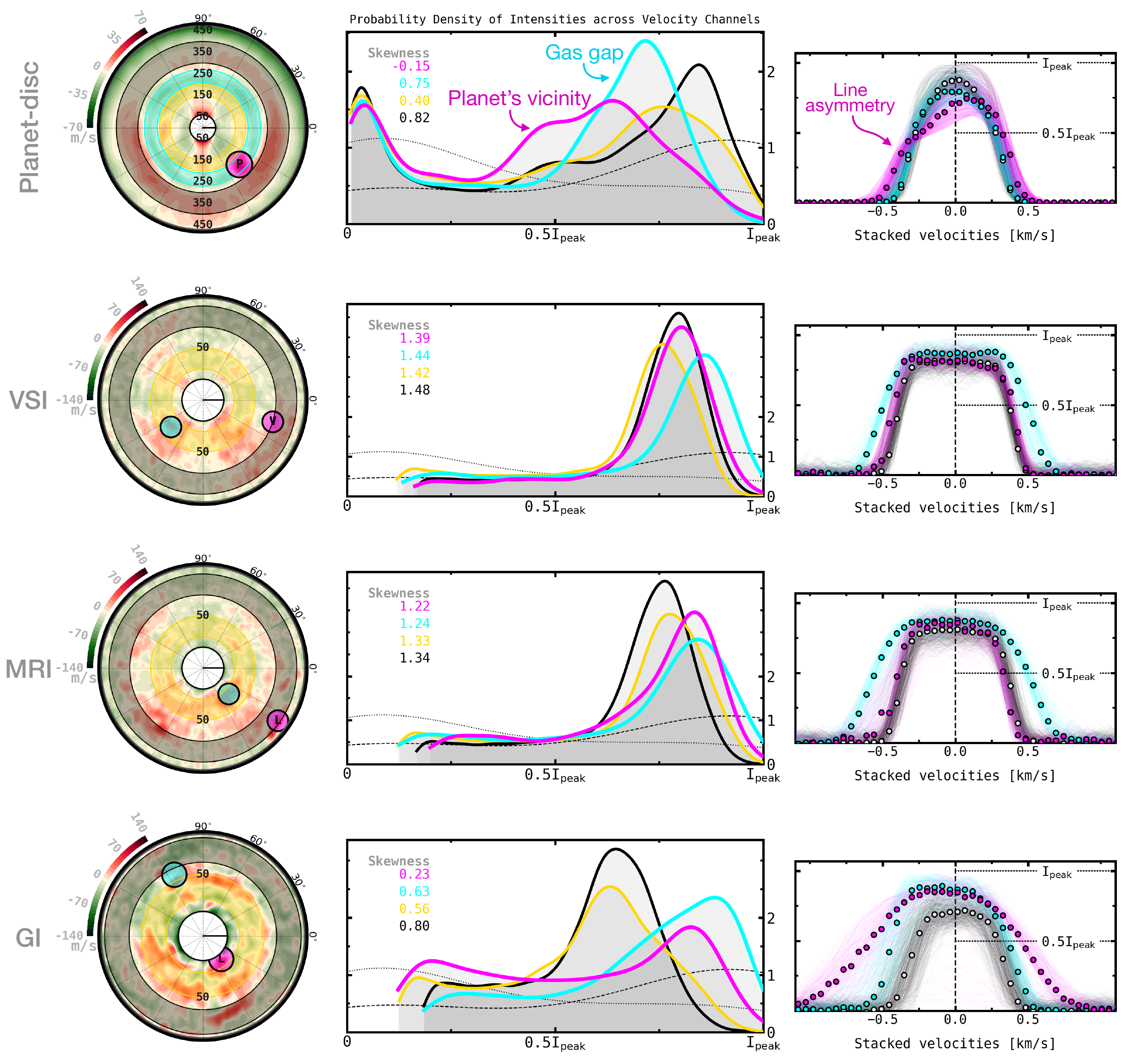}
      \caption{Comparison of \twCO{} intensity distributions and line-profile morphologies for the mechanisms explored in this work. The \textit{left column} shows line-width residuals overlaid with the regions analyzed. The \textit{middle column} presents the probability density functions of intensities from the regions highlighted on the left with the same color code, computed across all velocity channels. For reference, the dotted and dashed lines show the probability densities obtained for a Gaussian and a thick-Gaussian profile, respectively, assuming the same rms level and channel  spacing of the synthetic data. In the planet-disc scenario, the magenta curve corresponds to the intensity distribution in the planet’s vicinity (``P'' on the left), while the blue curve traces the gap region. For the disc instabilities, the magenta curves indicate regions of enhanced line widths or velocities (``L'' or ``V'' on the left) detected above the $3\sigma$ threshold (in Fig. \ref{fig:folded_instabilities}), while the blue curves represent other selected areas with increased line width for comparison. In all cases, the yellow and black curves represent the innermost and outermost parts of the disc. The reported skewness values are computed directly from the probability densities, with low and high values corresponding to flatter and more peaked distributions, respectively. The \textit{right column} shows 500 randomly drawn spectra from each region, stacked in velocity after deprojection assuming Keplerian rotation. These comparisons highlight that the line broadening associated with disc instabilities is relatively isotropic, whereas planet-disc interactions produce extended tails toward low intensities and velocities, responding to the presence of high velocity gradients and coherent flows around the planet.
              }
         \label{fig:intensdistrib}
   \end{figure*}

\subsubsection{Localized signatures of disc instabilities} \label{sec:line_instabilities}

Although the kinematic structures produced by disc instabilities are predominantly characterized by large-scale residuals, consistent with previous studies (e.g. \citealt{hall+2020} for GI; \citealt{barraza+2021} for VSI), localized perturbations can also arise in observations of these mechanisms. This is illustrated in Figure \ref{fig:folded_instabilities}, which presents folded velocity and line-width residuals together with the locations of global peaks and clusters above the $3\sigma$ detection threshold. Among the localized signals identified, the highest significance (in the MRI line-width maps) reaches only $3.5\sigma$, lower than the typical values ($>4\sigma$) obtained for the planet-disc simulation with $q \sim 2 \times 10^{-3}$, though comparable when the planet mass is reduced by half. Nevertheless, we do not identify complementary signatures that could plausibly mimic a planetary origin, as summarized in Fig. \ref{fig:hydro_signatures_12co} of the Appendix. For instance, variations in the rotation velocity associated with pressure modulations, or local minima in azimuthally averaged line widths indicative of a gap, are absent at the locations of the detected features.

Crucially, we find that the morphology of the line broadening associated with the signals detected in the disc-instability models is substantially more symmetric about the line centroid than in the case of an embedded planet, indicating an intrinsic difference in the projected motions responsible for the observed line-width enhancements. As discussed in the next section, this behavior provides a robust diagnostic for distinguishing localized planet-driven signatures from substructures induced by disc instabilities.

\subsection{Line asymmetries and intensity distributions: planets vs. disc instabilities} \label{sec:line_asymmetries}

To investigate the origin of the localized features in the planet-disc and disc-instability simulations, we examine the observed intensity distributions and stacked line morphologies around the detected signals in Figure \ref{fig:intensdistrib}. The most prominent difference is that the planet-driven line broadening arises from intensity distributions that are significantly more skewed toward low intensities and blueshifted velocities than those found in quiescent regions of the disc. This pattern is consistently observed around the line-width features detected across all other planet-disc simulations, regardless of the planet azimuth, beam size, or signal-to-noise ratio. In contrast, regions of enhanced line width in disc instabilities exhibit intensity distributions centered at higher intensity values relative to other areas of the disc and show no significant asymmetries in velocity. 

The pronounced line asymmetries induced by the planet in its vicinity point to intricate yet structured motions probed by the observations. They indicate that coherent velocity flows capable of producing multi-component spectra dominate in the planetary case, whereas more isotropic motions prevail in the disc instabilities. For reference, in planet-disc synthetic observations with a beam size of $0\farcs3$ and a disc inclination of $i = -15^\circ$, we measure a blue-to-redshifted line asymmetry\footnote{Defined as the difference between the velocities at the 16th (blueshifted) and 84th (redshifted) percentiles of the line profile, computed with \bettermoments{}, after subtraction of the same quantity measured from the reference \discminer{} model.} of $103 \pm 3$\,m\,s$^{-1}$ for a planet with 0.2\% the mass of the star, and $50 \pm 3$\,m\,s$^{-1}$ for a planet of 0.1\%. 

We attribute these asymmetries to meridional flows converging toward the planet, consistent with the velocity structure predicted by the simulations (see Fig. \ref{fig:edgeon_velocities_tau} in the Appendix). These flows are traced in the warm upper layers on the front side of the disc, where they predominantly recede relative to the observer and therefore produce a bright redshifted component in the \twCO{} line profile. In contrast, the mirrored velocity counterpart on the opposite side of the midplane (i.e. the back side) is largely suppressed by the high optical depth of the line, resulting in a deficit of blueshifted intensity. If the disc is thermally stratified, as observations systematically suggest \citep{galloway+2025}, this asymmetry is expected to be even more pronounced, since only gas from the colder lower layers on the back side (at velocities sampling the optically thinner line wings) can contribute to the observed blueshifted emission.

To demonstrate this behavior, we ran simulations using the optically thinner tracer \thCOfull{} and extracted line profiles at the same locations for all mechanisms (see Fig. \ref{fig:intensdistrib_13co} in the Appendix). In the planet–disc interaction model, localized line broadening is still detected at the planet location, but the line-intensity asymmetries vanish, confirming that, in this case, converging motions from above and below the midplane contribute more equally to the observed emission. For the disc instabilities, the \thCO{} lines remain similarly symmetric about their centroids as in the \twCO{} models, indicating that the velocity fluctuations responsible for the line-width enhancements in these mechanisms are also vertically isotropic.

\begin{figure*}
   \centering
    \includegraphics[width=1.0\textwidth]{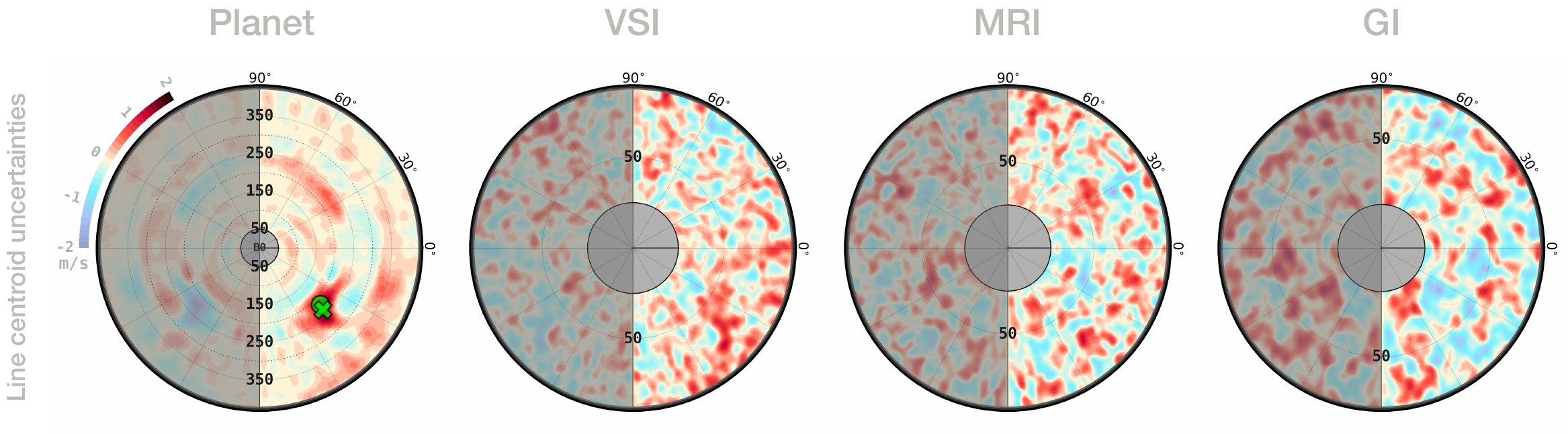}
      \caption{Folded maps of line-centroid uncertainties extracted from the Gaussian least-squares fits to the line profiles of the planet-disc and disc-instability simulations. This quantity is computed as the square root of the absolute value of the diagonal elements of the covariance matrix returned by the fit. A localized and coherent signal is detected only around the planet’s location in this observable, resulting from the strong line asymmetries induced by the embedded planet in its vicinity, as demonstrated in Fig. \ref{fig:intensdistrib} (top right). No strong correlation is found between these residuals and the velocity or line-width substructures observed in disc instabilities.
              }
         \label{fig:folded_dv}
   \end{figure*}

Another consequence of the velocity asymmetries induced by planets in optically thick observations is an increase in the uncertainty of the line-centroid measurements in the planet's vicinity. As illustrated in Figure \ref{fig:folded_dv}, this effect also produces a highly localized and clustered signal at the planet’s location, whereas for the disc-instability models, the corresponding uncertainties are distributed more broadly across the disc and show no significant correlation with the velocity or line-width substructures identified in the residual maps.


\section{Application to the discs of \hdone{} and \mwcsev{}} \label{sec:residuals}

   \begin{figure*}
   \centering
    \includegraphics[width=1.0\textwidth]{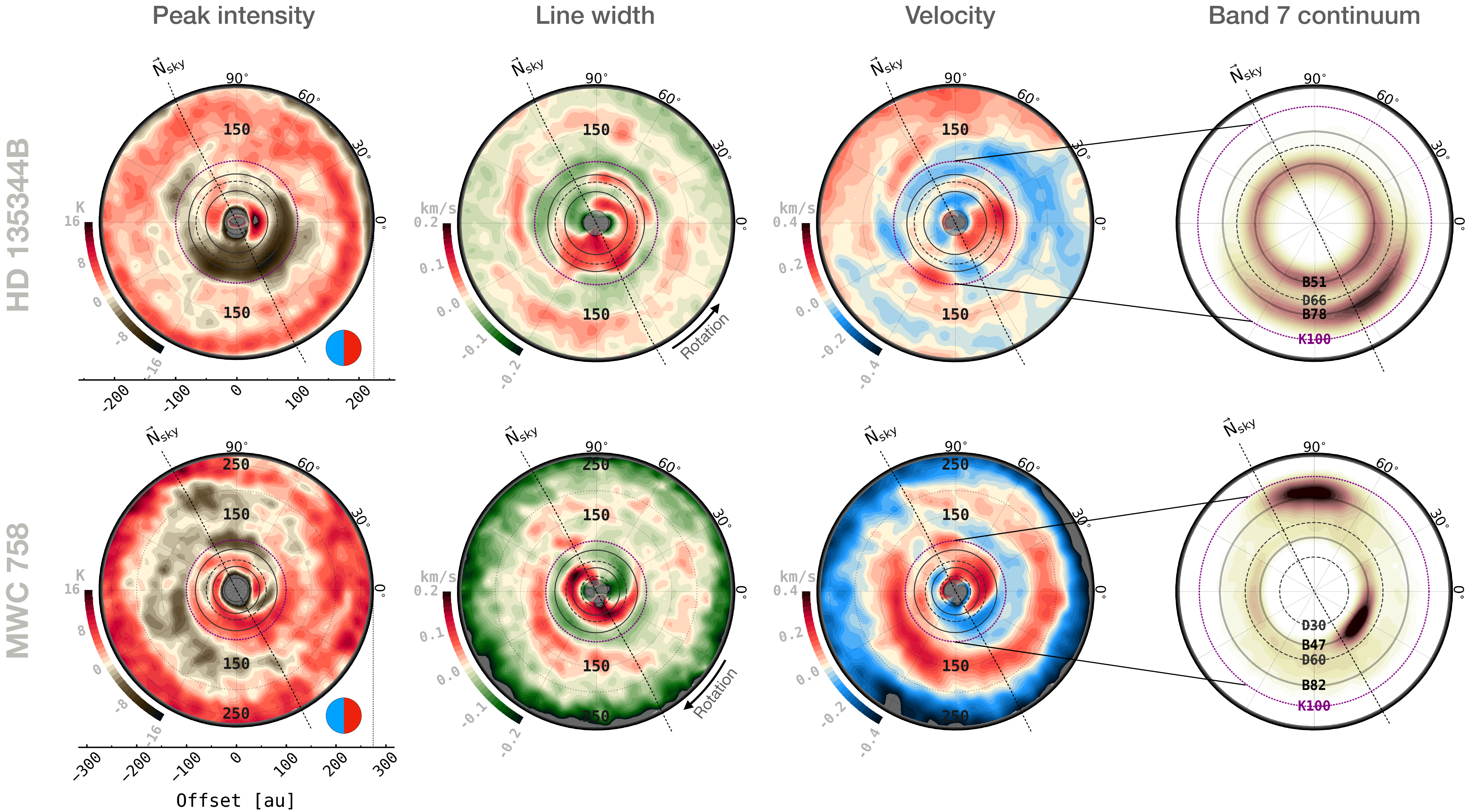}
      \caption{Peak intensity (left column), line width (middle left) and velocity (middle right) residuals traced by \twCOfull{}, along with Band 7 continuum emission (right), from the discs of \hdone{} (top row) and \mwcsev{} (bottom), deprojected onto each disc's reference frame. In this frame, the vertical axis is aligned with the projected minor axis of the disc on the sky. Also shown is the north axis (where $\rm{PA}=0^\circ$) of the sky for reference. The blue-red circle in the bottom right corner of the leftmost panels marks the blueshifted and redshifted sides of each disc. Solid and dashed lines indicate the radial location of millimeter dust rings and gaps, respectively, and the purple dotted lines highlight the orbits of the outermost planet candidates proposed in the literature to explain the observed dust features. These targets exhibit a high degree of substructure in all three types of residuals, suggesting the presence of strong pressure fluctuations and deviations from Keplerian motion.
              }
         \label{fig:residuals_12co}
   \end{figure*}

   \begin{figure}
   \centering
    \includegraphics[width=0.94\columnwidth]{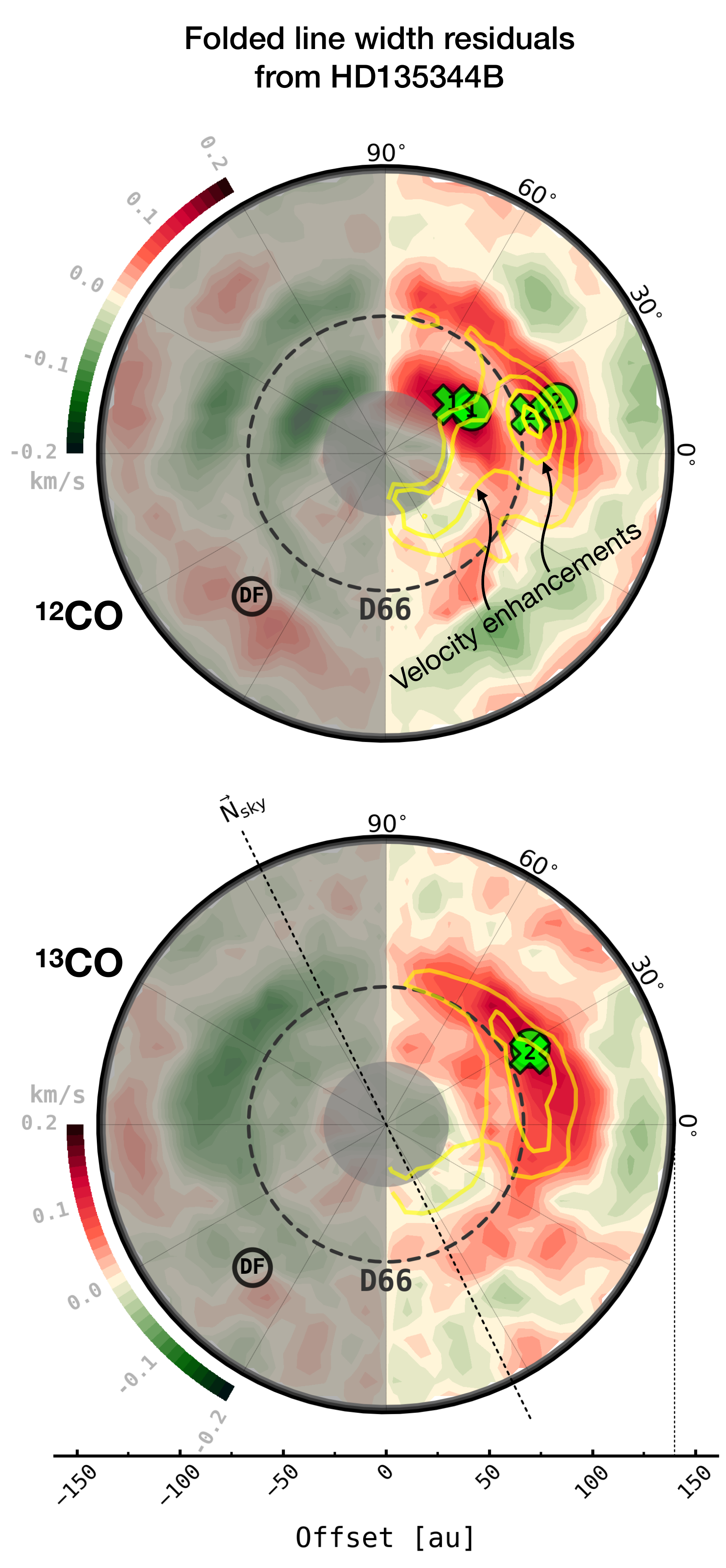}
      \caption{Folded line width residuals from \twCO{} and \thCOfull{} emission for the disc of \hdone{}. The region between $R=30-90$\,au in the first quadrant displays significant line-width enhancements in both tracers, indicating perturbations spanning multiple altitudes above the disc midplane, potentially driven by a massive planet. Green circles mark the locations of global peaks, while green crosses represent detected clusters associated with highly localized signals. Yellow contours highlight co-spatial peak velocity residuals in the range of $+50$ to $+300$\,m\,s$^{-1}$. The outermost \twCO{} cluster coincides with both the global velocity peak and the detected cluster in \thCO{}. 
              }
         \label{fig:folded_residuals_hd135344}
   \end{figure}

The discs of \hdone{} and \mwcsev{} are characterized by the presence of gaps, rings, arcs, and spiral patterns in continuum emission, potentially induced by massive yet unseen embedded planets \citep[e.g.][]{garufi+2013, dong+2015}. Here, we apply our tomographic analysis of molecular-line data to investigate the possible planetary origin of the gas and dust structures identified in these systems.

Figure \ref{fig:residuals_12co} presents a gallery of the residual maps derived for these sources, obtained by subtracting the best-fit \discminer{} models of the \twCOfull{} line emission following \citet{izquierdo+2025}. Both discs exhibit a high degree of substructure across all residual diagnostics: peak intensity, which closely traces temperature variations due to the high optical depth of the tracer; line width, which responds to surface-density and temperature fluctuations as well as unresolved velocity flows; and velocity centroids, which probe deviations from Keplerian rotation. 

Localized features are particularly prominent in the velocity map of \hdone{}, manifested as sharp sign reversals (or Doppler flips) that closely overlap with regions of enhanced line broadening. In contrast, \mwcsev{} is dominated by extended signatures, manifesting as spirals, annular structures, and arcs throughout the different line diagnostics. In the remainder of this section, we focus on the retrieval and interpretation of these features. For applications to the full \exoalma{} sample, we refer the reader to \citet{fukagawa+2026} and \citet{benisty+2026}, which present a systematic identification and analysis of residuals across all targets.

\subsection{Localized features in \hdone{}} \label{sec:localized_hd135344}

We follow the approach described in Section \ref{sec:localized_signatures} to search for planet candidates associated with localized signatures in the line-width and velocity maps of \hdone{}. In this source, we identify strongly localized line-width peaks and clusters, shown in Figure \ref{fig:folded_residuals_hd135344} as circles and crosses, respectively. Two such features, labeled ``1'' and ``2'', are detected in \twCO{} at ($R=41$\,au, $\phi=36^{\circ}$) and ($R=73$\,au, $\phi=15^{\circ}$), and in \thCO{} at ($R=78$\,au, $\phi=26^{\circ}$), where $R$ denotes the cylindrical radius and $\phi$ is the azimuthal angle measured from the redshifted major axis in the deprojected disc frame. {Although these features dominate the peak extraction, line-width enhancements in \twCO{} are also observed near the velocity Doppler flip labeled ``DF'', reported in Sect.~\ref{sec:dopper_flip} at ($R=95$\,au, $\phi=-133^{\circ}$).} Table \ref{tab:candidate_clusters} summarizes the locations of these planet candidates, together with their sky-projected coordinates relative to the disc center from the best-fit \discminer{} model of \citet{izquierdo+2025}.

The amplitudes of the detected line-width signals in \twCO{} are $0.18$ and $0.13$\,\kms{} for candidates ``1'' and ``2'', respectively, and $0.16$\,\kms{} for the \thCO{} feature associated with candidate ``2''. Prominent velocity residuals are also found in the same region of the disc (shown as yellow contours), reaching amplitudes of up to $0.26$\,\kms{} in \twCO{}, and $0.14$\,\kms{} in \thCO{}. However, these residuals do not form clusters that are significantly more localized than those observed elsewhere in the disc, indicating that velocity perturbations are spatially more extended than the corresponding line-width enhancements. This behavior is fully consistent with the predictions from our massive planet-disc interaction simulations analysed in Sect. \ref{sec:hydro} (see e.g. Fig. \ref{fig:folded_residuals_radmc}).

\setlength{\tabcolsep}{4.5pt} 
\begin{table}[h]
    \centering
    \caption{Planet candidates identified in the disc of \hdone{}, including their locations in polar and sky-projected coordinates.}
    \label{tab:candidate_clusters}
    \begin{tabular}{crrrrc}
        \hline
        Candidate & $R$\,[au] & $\phi$\,[$^\circ$] & $R$\,[$''$] & PA\,[$^\circ$] & Tracer \\
        \hline
        C1 & 41 & 36 & 0.30 & 278 & \twCO{} \\
        C2 & 73 & 15 & 0.54 & 258 & \twCO{} \& \thCO{} \\
        DF & 95 & $-133$ & 0.69 & 109 & \twCO{} \& \thCO{} \\
        
        \hline
    \end{tabular}
\end{table}

{\citet{izquierdo+2023} reported similar localized line broadening traced by CO emission in the disc of HD\,163296, spatially coincident with a planet candidate associated with a velocity Doppler flip and chemical anomalies attributed to its circumplanetary region \citep{izquierdo+2022, izquierdo+2026}; in the disc of MWC\,480, at the radial location of candidate planet-driven buoyancy spirals \citep{teague+2021}; and in AS\,209, in the region where a circumplanetary disc was proposed based on localized \thCO{} intensity signals \citep{bae+2022}.}

\subsubsection{A prominent Doppler flip at R=95\,au}
\label{sec:dopper_flip}

Doppler-flip signatures also serve as indicators of the presence of embedded planets in discs. These features manifest as sharp transitions between negative and positive velocity residuals, arising from the gravitational interaction between the planet and the surrounding gas in its vicinity \citep[e.g.][]{perez+2018}. They also provide valuable diagnostics of the planet mass and orbital parameters and, to a lesser extent, of the local disc viscosity and temperature structure \citep{rabago+2021}.

   \begin{figure*}
   \centering
    \includegraphics[width=1\textwidth]{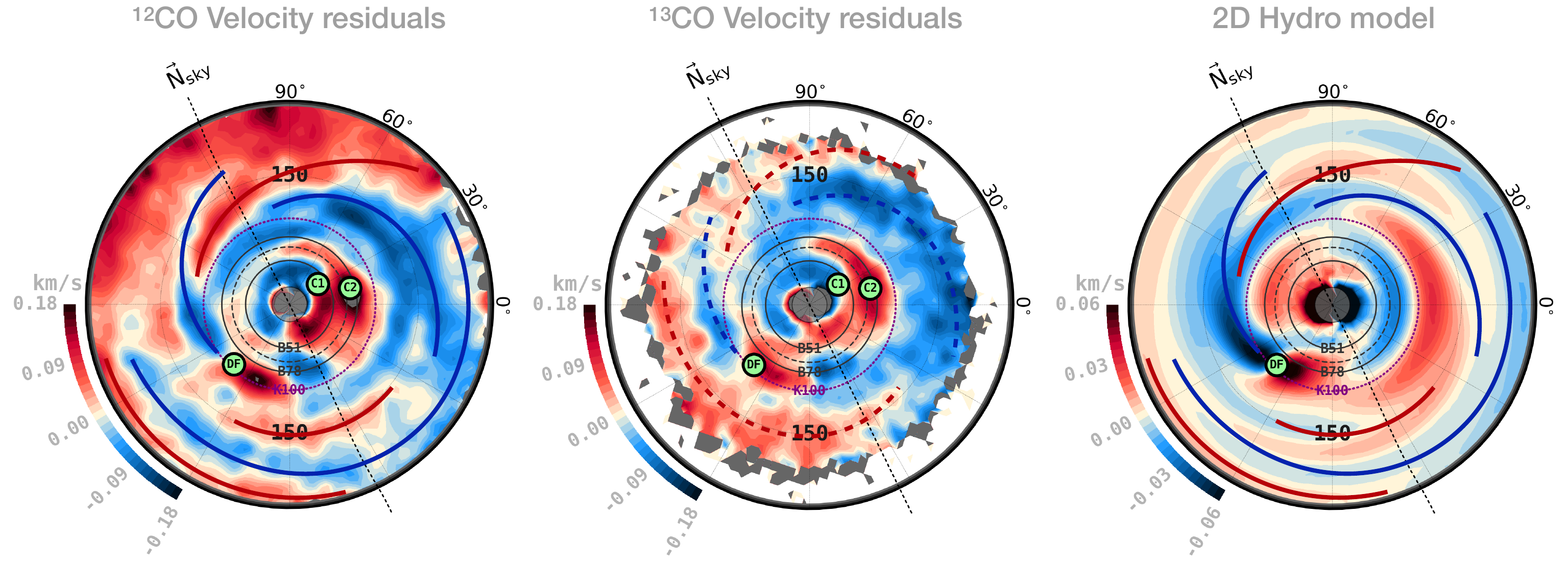}
      \caption{Centroid velocity residuals observed in the \twCO{} and \thCO{} data of the \hdone{} disc, compared with a two-dimensional hydrodynamical model tailored to this system. The Doppler flip identified in Sect. \ref{sec:dopper_flip}, centered at the DF marker, spatially overlaps in both tracers and is accompanied by large-scale velocity substructures spanning a broad radial and azimuthal extent of the disc. These features are qualitatively reproduced by the hydrodynamical model, generated with the \ppdonet{} neural network introduced by \citet{mao+2023} and postprocessed with \discminer{}, assuming a 6\,\Mj{} planet at the center of the Doppler flip, an aspect ratio of 0.075, and an $\alpha$ viscosity of $10^{-2}$. The red and blue solid lines overlaid on the model residuals trace the large-scale \twCO{} signatures, emphasizing the similarity in both the velocity sign flips and the spatial extent of the structures near and far from the planet, despite the simplicity of the adopted model. 
              }
         \label{fig:hydro_hd135344}
   \end{figure*}

   \begin{figure*}
   \centering
    \includegraphics[width=1.0\textwidth] {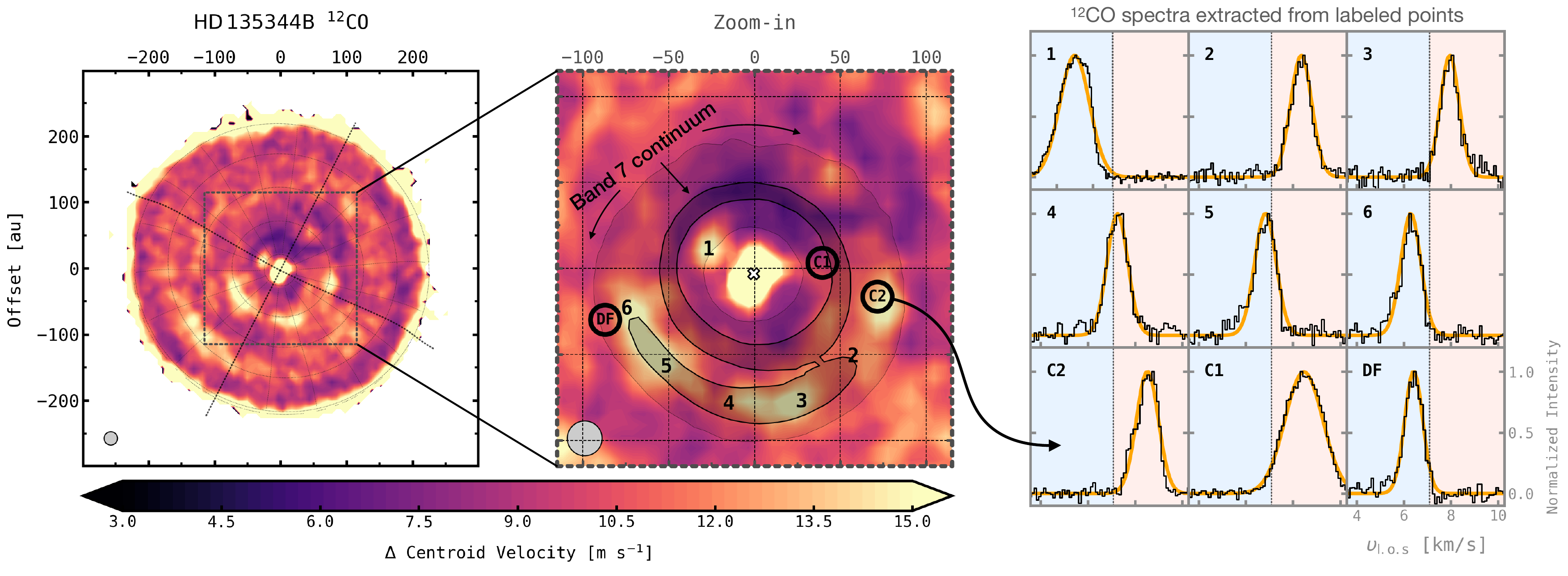}
      \caption{Statistical uncertainties in centroid velocities derived from Gaussian fits to the \twCOfull{} line profiles for the disc of \hdone{}. The zoom-in panel highlights localized signals near the mm continuum crescent (outlined by black contours to the south), around the proposed planet associated with line-width clusters and velocity peaks at $R=73$\,au (denoted by the C2 circle), and near the planet candidate linked to the velocity Doppler-flip signature at $R=95$\,au (marked by the DF circle). In the rightmost panels, spectra extracted from the numbered points in the middle panel illustrate the asymmetric line profiles responsible for the increased uncertainties. These morphologies suggest the presence of spatially unresolved velocity flows and are reminiscent of the line shapes predicted by synthetic observations of planet-disc interaction simulations (see Sect. \ref{sec:line_asymmetries}).  
              }
         \label{fig:velocity_errors_hd135344}
   \end{figure*}

In cases with high substructure complexity, where Doppler-flip perturbations are not the strongest but remain visually detectable as in the disc of \hdone{}, we apply a peak-finding algorithm on gradient residual maps to pinpoint the central location of these signatures. 
This is illustrated in Figure \ref{fig:peak_gradients_hd135344} of the Appendix, where we present azimuthal gradient maps of velocity residuals in \twCO{} and \thCO{}. Pluses and minuses overlaid on the reference residual map denote positive and negative peak gradients, respectively, to mark the locations of Doppler flips in the velocity field. Azimuthal gradient peaks with values below $50$\,m\,s$^{-1}$\,beam$^{-1}$ were excluded from the analysis, retaining only the steepest gradients corresponding to velocity shifts exceeding half the channel spacing of the data per beam. The radial section of the disc where the Doppler-flip signatures overlap in \twCO{} and \thCO{} is highlighted. Among the peaks that are co-spatial within a beam size ($\sim\!20$\,au at the source distance) for these tracers, three correspond to sequential velocity sign flips centered at $(R, \phi) = (95\,{\rm au}, -133^{\circ})$, $(89\,{\rm au}, -93^{\circ})$, and $(80\,{\rm au}, -60^{\circ})$ in the disc reference frame, and one at $(72\,{\rm au}, 24^{\circ})$ co-spatial with the line-width clusters and velocity perturbations reported earlier in Fig. \ref{fig:folded_residuals_hd135344}.

We propose the presence of a massive embedded planet at the location of the most prominent Doppler flip, centered at $R = 95$\,au and $\phi = -133^{\circ}$. The peak negative-to-positive amplitude of this signal is $0.32 \pm 0.02$,\kms{} in \twCO{} and $0.24 \pm 0.04$\,\kms{} in \thCO{}, implying a modest decrease toward the disc midplane. 
Additional evidence supporting this candidate includes a spiral-like structure seen in positive line-width residuals {(Figs. \ref{fig:residuals_12co} and \ref{fig:folded_residuals_hd135344}, top row)} emerging from the location of the Doppler flip, as well as a local minimum in the azimuthally averaged line widths at this orbital separation \citep[Fig.~10 in][]{izquierdo+2025}, which may indicate a surface-density gap traced by reduced line saturation owing to the lower optical depth \citep{hacar+2016, izquierdo+2021}.

Variations in the amplitude of this velocity feature across lines may provide additional clues about the vertical location and physical mechanism driving the signal. Given that the disc of \hdone{} is close to face-on ($i=-16^{\circ}$), we speculate that a substantial fraction of the observed velocity signals, and their modulation with altitude, are likely shaped by vertical motions. \citet{pinte+2019} and \citet{rabago+2021} found in their simulations that vertical downward velocities are highest near an embedded planet peak at approximately one pressure scale height above the midplane. In our proposed scenario, this would imply that \twCO{} traces gas closer to this elevation than \thCO{}, suggesting a relatively flat disc compared to the average $z/r$ of the \exoalma{} sample. Interestingly, this interpretation is consistent with the low $z/r$ values reported for discs with similar radial extents, such as \cqtau{} or \pdssix{} \citep[see e.g. Fig. 6 of][]{galloway+2025}. The assumption of a shallow disc to explain the reduction in velocity amplitude toward the midplane could be relaxed, however, if the planet also excites buoyancy resonances. In that case, vertical motions would begin to weaken from even higher elevations \citep{bae+2021}.

To further investigate the influence of this planet on the disc structure, we generated a velocity map from hydrodynamical models of planet-disc interaction tailored to this system using the \ppdonet{} neural network \citep{mao+2023}. We then subtracted the best-fit Keplerian rotation derived with \discminer{} for this target in \citet{izquierdo+2025} to obtain velocity residuals. The adopted setup assumes a gas-disc alpha viscosity of $10^{-2}$, an aspect ratio of 0.075, and a 6\,\Mj{} planet located at the center of the Doppler flip\footnote{{These parameters were not derived from a formal best-fit procedure but were manually adjusted to qualitatively reproduce the localized and large-scale features observed in the data. Moreover, because the neural network does not account for vertical motions or vertically stratified temperature structures, the adopted parameters may not correspond to the optimal physical properties required to reproduce the observed disc kinematics.}}, consistent with previous models used to reproduce the spiral arms observed in scattered light (see Sect. \ref{sec:discussion_hd135344b} for discussion). 

Figure \ref{fig:hydro_hd135344} compares the residuals from the \twCO{} and \thCO{} data with those predicted by the hydrodynamical model. In addition to the Doppler flip signature, the model exhibits large-scale substructures including arcs and spiral-like patterns, that closely resemble those observed in \twCO{} (traced by the solid blue and red lines) and in \thCO{}. The consistency of the velocity patterns across different molecular lines suggests that the observed substructures share a common origin and that the associated perturbations propagate over multiple scale heights within the disc’s vertical structure.

We note, however, that although the overall dynamical structure of the disc is well reproduced by invoking this planet, the amplitude of the model residuals is lower by a factor of $\sim\!2-3$ compared to the observations. This discrepancy may be mitigated by incorporating vertical motions and temperature gradients into the models, which are expected to enhance the planet-driven velocity features \citep[e.g.][]{bae+2021} without requiring a prohibitively high planet mass that would otherwise conflict with direct-imaging constraints ruling out planets more massive than $3$\,\Mj{} at the orbital radius of the Doppler Flip \citep{cugno+2024}. Full three-dimensional simulations are therefore essential to reliably constrain the masses of embedded planets from kinematic signatures, particularly in low-inclination discs.

\subsubsection{Line asymmetries around line-width signals}

   \begin{figure*}
   \centering
    \includegraphics[width=1.0\textwidth] {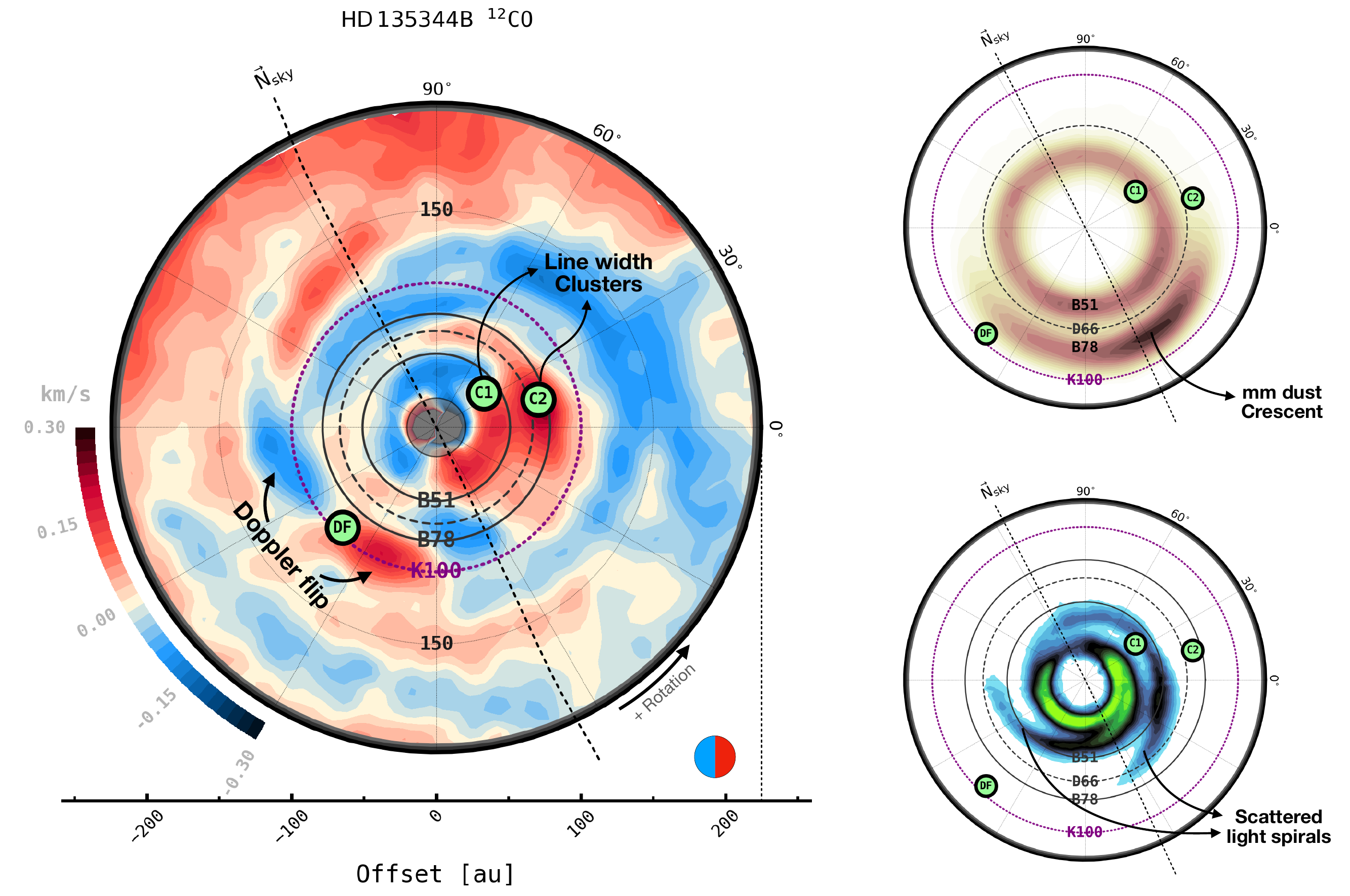}
      \caption{Deprojected view of velocity and dust substructures in the disc of \hdone{}. \textit{Left}: \twCOfull{} velocity residuals illustrating deviations from Keplerian rotation in the gas disc. \textit{Bottom right}: VLT/SPHERE near-infrared observations \citep{stolker+2016} probing scattered light from micron-sized dust grains. \textit{Top right}: ALMA Band 7 continuum observations \citep{teague+2025, curone+2025} tracing thermal emission from millimeter-sized dust grains. Solid and dashed lines mark the radial locations of peaks and troughs in the millimeter continuum, while the purple dotted K100 line indicates the orbital radius of the planet candidate proposed by \citet{dong+2015} and \citet{bae+2016} to reproduce the near-infrared spirals and the millimeter dust crescent. The green circles indicate the locations of the Doppler flip and line-width clusters identified in \twCO{} and \thCO{} in this work, potentially associated with embedded planets.
              }
         \label{fig:summary_hd135344}
   \end{figure*}

We also find evidence for asymmetric line profiles at the locations of the outermost line width clusters detected in \twCO{} and \thCO{}, consistent with the predictions from our planet-disc interaction simulations discussed in Sect. \ref{sec:line_asymmetries}. Figure \ref{fig:velocity_errors_hd135344} presents a map of 1$\sigma$ statistical uncertainties in centroid velocities derived from Gaussian fits to the \twCO{} line emission of this source. Typical uncertainties are $\sim\!10$\,m\,s$^{-1}$, with enhanced values partially tracing the millimeter continuum crescent and localized increases around the line-width cluster at $R=73$\,au and near the Doppler Flip identified at $R=95$\,au. An additional localized feature is found interior to the millimeter ring B51, at {$R=30$\,au}. 

To illustrate the origin of these elevated uncertainties, we extracted spectra at the positions of the outer line-width cluster and at pixels co-spatial with the millimeter crescent. These profiles exhibit clear asymmetries relative to their line centroids and, in the most prominent cases, appear double-peaked, indicating the presence of multiple overlapping velocity components. In all cases, however, the asymmetries favor downward flows, consistent with predictions from our planet-disc simulations (see Figs. \ref{fig:intensdistrib} and \ref{fig:edgeon_velocities_tau}). Near the outer edge of the disc, at $R\!\sim\!170$\,au, elevated uncertainties also align with the spiral structure identified in line-width residuals (Fig. \ref{fig:residuals_12co}), suggesting spatially unresolved and asymmetric velocity flows along this feature as well.

Figure \ref{fig:summary_hd135344} summarizes the localized signatures identified in \hdone{}. In Section \ref{sec:discussion}, we examine the possible connection between these candidate planet-driven perturbations and the dust substructures revealed by millimeter and near-infrared continuum observations of this source.

\subsection{Large-scale signatures in \mwcsev{}} \label{large_scale_signatures}

   \begin{figure*}
   \centering
    \includegraphics[width=0.96\textwidth]{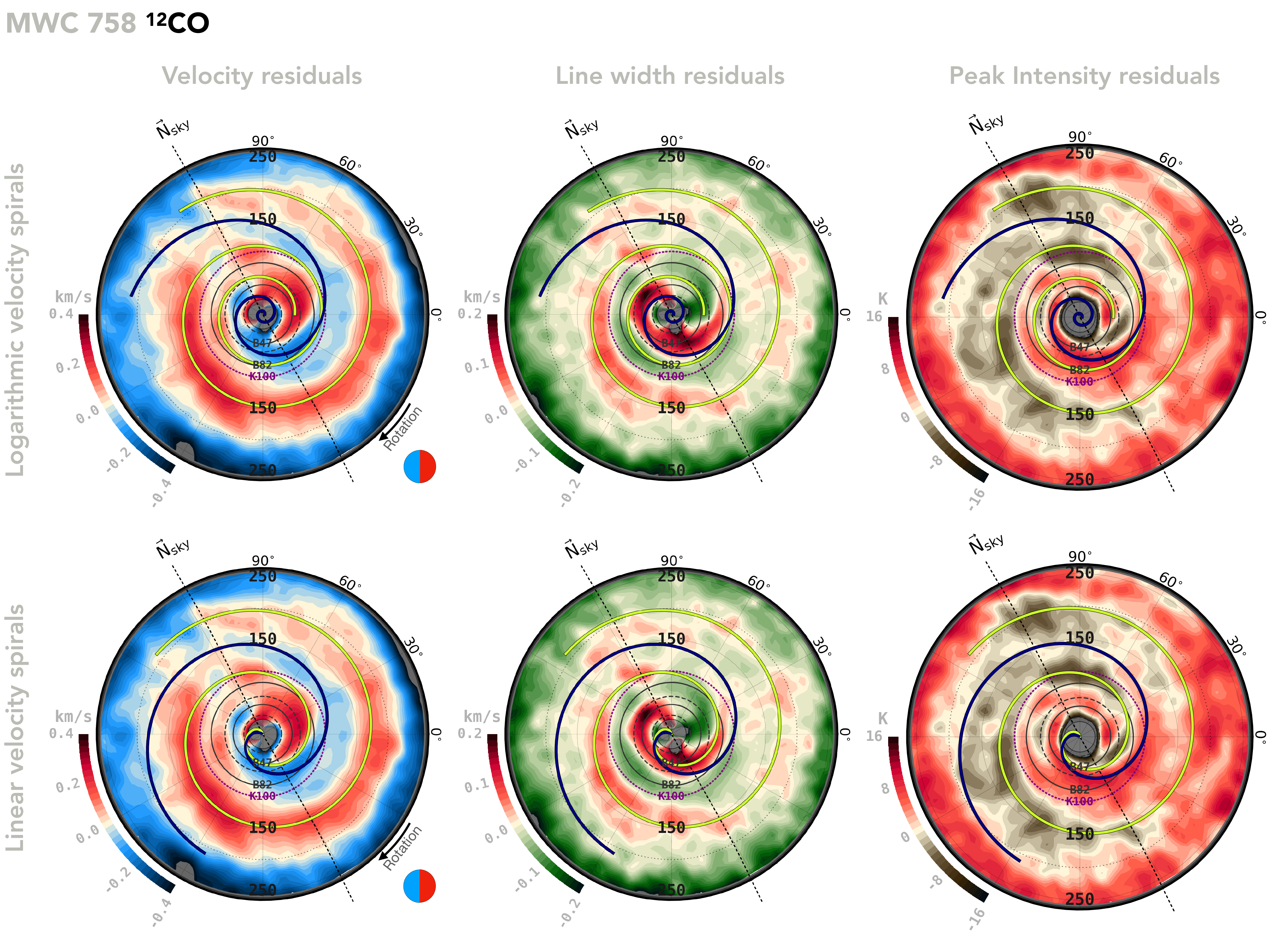} \caption{Logarithmic ($r = a \text{e}^{b\phi}$, top) and linear ($r = a + b\phi$, bottom) spiral models fitted to the spirals identified in \twCO{} velocity residuals for the disc of \mwcsev{}, overlaid on the velocity (left), line-width (middle), and peak-intensity (right) residual maps. Table \ref{tab:mwc758_spirals} summarizes the best-fit parameters for the spirals. While both prescriptions reproduce the overall morphology of the outer signatures in velocity and line width, the logarithmic model additionally captures the inner features in velocity and peak intensity near the radial locations of the millimeter dust crescents B47 and B82.
              }
         \label{fig:spirals_mwc758_12co}
   \end{figure*}

The disc of \mwcsev{} exhibits prominent spiral-like signatures across all three residual diagnostics. To identify and quantify these perturbations, we employ the \filfinder{} code \citep{koch+2015}, which applies an adaptive thresholding algorithm to isolate pixels belonging to coherent substructures (referred to as filaments) over a wide dynamic range in the observable of interest. Each detected substructure is assigned a two-dimensional mask, which is subsequently reduced to a one-dimensional skeleton via a Medial Axis Transform \citep[first introduced by][]{blum+1967} to serve as a shape descriptor. The resulting medial axes are then used to characterize the morphology and spatial configuration of the filaments \citep[see e.g.][for an application to the disc kinematics of HD\,163296]{izquierdo+2022}. For this purpose, we consider several geometric descriptors provided by \filfinder{}, including filament length, width, and orientation. 

\setlength{\tabcolsep}{3.5pt} 
\begin{table}[h]
    \centering
    \caption{Best-fit parameters for the logarithmic and linear spiral prescriptions fitted to the velocity structures traced by \twCOfull{} emission in the disc of \mwcsev{}. The parameter $b$ in the linear fit is expressed in units of au\,rad$^{-1}$.} \label{tab:mwc758_spirals}
    \begin{tabular}{lccc}
        \hline
        \textbf{Type} & \textbf{Component} & \boldmath{$a$} \textbf{(au)} & \boldmath{$b$} \\
        \hline
        Logarithmic & Positive & $170.0 \pm 1.1$ & $5.5 \pm 0.3^\circ$ \\
         ($r = a \text{e}^{b\phi}$) & Negative & $95.8 \pm 2.3$ & $15.1 \pm 1.4^\circ$ \\
        \hline
        Linear & Positive & $171.7 \pm 1.1$ & $15.8 \pm 0.6$\\
          ($r = a + b\phi$) & Negative & $99.4 \pm 1.9$ & $26.2 \pm 1.7$\\
        \hline
    \end{tabular}
\end{table}

The results of this analysis are presented in Figure \ref{fig:filaments_mwc758_12co} in the Appendix, which shows the filaments identified in the \twCO{} velocity residuals, together with their measured pitch angles and projected widths. We find that the longest velocity feature, highlighted in cyan across all panels, exhibits a nearly uniform pitch angle of $\sim\!0.3^\circ$ at all radii, forming an annular pattern that coherently traces blueshifted velocities consistent with outflowing material in the outer disc regions \citep[see][for details]{benisty+2026}. In contrast, the larger and radially varying pitch angles retrieved for the inner filaments suggest that these correspond to non-axisymmetric substructures forming spiral patterns, possibly driven by embedded companions (see Sect. \ref{sec:discussion}). We also find that the two outermost filaments are spatially resolved, with an average full width at half maximum of $\sim\!1.5$ beams measured perpendicular to their medial axes. The outermost blueshifted filament, however, may extend beyond this width in regions of lower signal-to-noise that were excluded from the analysis.

The velocity spirals manifest as coherent redshifted and blueshifted patterns extending over more than $\sim\!270^\circ$ in azimuth, without sign reversals across the disc's minor and major axes. As illustrated in Fig. \ref{fig:velocity_patterns}, this behavior suggests that the signatures are likely dominated by downward and upward vertical motions, respectively.

To further characterize the morphology of these substructures, we fitted logarithmic ($r = a \text{e}^{b\phi}$) and linear ($r = a + b\phi$) spiral models to their medial axes using \textsc{scipy}'s least-squares minimization algorithm. Figure \ref{fig:spirals_mwc758_12co} shows the resulting analytic spirals overlaid on the velocity, line-width, and peak-intensity residual maps, with the best-fit parameters listed in Table \ref{tab:mwc758_spirals}. Both logarithmic and linear prescriptions reproduce the overall geometry of the velocity spirals reasonably well. However, the extrapolated logarithmic spiral provides a better match to velocity and intensity features in the inner disc, particularly near the millimeter dust crescents B47 and B82, which the linear model fails to capture.

This analysis also demonstrates that the redshifted velocity spiral spatially coincides with elongated regions of enhanced line width, regardless of the adopted fitting prescription. The absence of corresponding temperature increases along this spiral, as traced by the \twCO{} peak intensity residuals, indicates that the observed line broadening most likely arises from unresolved velocity flows, high optical depths, or a combination of both, rather than from shock or radiative heating of the gas. 

\section{Discussion} \label{sec:discussion}

\subsection{Planets in the disc of \hdone{}} 
\label{sec:discussion_hd135344b}
As illustrated in Figure \ref{fig:summary_hd135344}, all kinematic and line-width signatures identified in the disc of \hdone{} are located near the dust ring (B51), gap (D66), and azimuthal dust trap (B78) observed in the millimeter continuum \citep{vandermarel+2016, cazzoletti+2018, curone+2025}, suggesting a strong connection between these substructures and pressure modulations potentially induced by embedded companions. Notably, the same radial region of the disc hosts a nearly symmetric two-armed spiral structure observed in the near infrared \citep{muto+2012, garufi+2013, stolker+2016, ren+2023, maio+2025}, consistent with expectations from tidal interactions between discs and planets \citep[e.g.][]{dong+2015, bae+2016, baruteau+2019}. From a kinematic perspective, and based on the small and large-scale signatures identified in Sect. \ref{sec:localized_hd135344}, we propose three plausible locations for embedded planets in this disc.

First, a planet candidate (C2) at a radial separation of $R=73$\,au, near the inner edge of the millimeter dust asymmetry B78 and azimuthally within the first half of the first quadrant of the disc, centered at $\phi=15^{\circ}$ as inferred by our clustering analysis (or $R=0\farcs{54}$, $\rm{{PA}}=258^\circ$ in projected sky coordinates). This object would produce line broadening and blueshifted asymmetries, as well as strong velocity perturbations at different heights above the midplane traced by \twCO{} and \thCO{} (Figs. \ref{fig:folded_residuals_hd135344}, \ref{fig:hydro_hd135344}, and \ref{fig:velocity_errors_hd135344}). Its radial location is close to that of the planet proposed by \citet{xie+2024} at an orbital radius of $R=66$\,au to explain the pattern speed of the near-infrared spirals\footnote{The reference Keplerian velocity adopted in the spiral-pattern analysis is consistent with that used in this work, as \citet{xie+2024} assumed a nearly identical stellar mass (1.6\,M$_\odot$) and disc inclination ($16.7^\circ$), matching the values independently inferred from the disc kinematics in \citet{izquierdo+2025}.} observed in this source (Fig. \ref{fig:summary_hd135344}). 

The second candidate (C1) corresponds to a closer-i planet located at $R=41$\,au and $\phi=36^{\circ}$ (or $R=0\farcs{30}$, $\rm{{PA}}=278^\circ$), interior to the inner dust ring B47, and possibly responsible for the localized line broadening in the disc upper layers traced by \twCO{} (Fig. \ref{fig:folded_residuals_hd135344}). However, the absence of \twCO{} line asymmetries (Fig. \ref{fig:velocity_errors_hd135344}) and the only moderate broadening detected in \thCO{} at this location suggest that, if such a planet exists, it is likely less massive than $1.6$\,\Mj{} (i.e. $<0.1$\% of the stellar mass), as predicted by our planet-disc interaction models. Alternatively, these signatures may reflect outlier perturbations from an even more interior planet that cannot be robustly constrained due to beam smearing at small radii, given the finite angular resolution of our data ($\sim\!20$\,au at the source distance). This scenario would be broadly consistent with the findings of \citet{maio+2025}, who propose a $2$\,\Mj{} planet to explain a compact infrared intensity signal at an orbital radius of 29\,au and an azimuth of $\phi = 31^\circ$, which overlaps in azimuth with the \twCO{} line-width feature reported here, though at a slightly smaller radial separation.

This also aligns with \citet{garufi+2013} and \citet{vandermarel+2016}, who suggested that a massive planet at $\sim\!30$\,au could generate the pressure maximum responsible for the millimeter dust ring B51. \citet{vandermarel+2016} further proposed that this inner pressure bump could induce a secondary pressure maximum at larger radii, triggering a vortex via the Rossby wave instability and leading to the formation of the millimeter dust crescent B78 as an asymmetric dust trap. In their interpretation, one of the spiral arms observed in scattered light would originate from the vortex overdensity, while the other would be driven by the inner planet candidate. Interestingly, \citet{cazzoletti+2018} argued that the vortex alone could potentially excite the entire spiral structure, having accumulated sufficient mass (up to a few Jupiter masses) to perturb the disc kinematics. Such a scenario could account for the enhanced line widths (Fig. \ref{fig:residuals_12co}) and spectral asymmetries (Fig. \ref{fig:velocity_errors_hd135344}) observed around the dust crescent. This is also partly consistent with the findings of \citet{woelfer+2025}, who show that the kinematic pattern near the millimeter crescent resembles that expected from a vortex, although it cannot, on its own, reproduce the more extended features.

Finally, we identify a third planet candidate (DF) associated with a Doppler-flip velocity signature in the third quadrant of the disc at $R=95$\,au and $\phi=-133^\circ$ (or $R=0\farcs{69}$, $\rm{{PA}}=109^\circ$), detected in both \twCO{} and \thCO{} velocity maps. This planet may also be responsible for a surface-density dip traced by reduced azimuthally averaged line widths at the same orbital radius \citep[see Fig. 10 of][and Fig. \ref{fig:hydro_signatures_12co} of this work]{izquierdo+2025}. Moreover, its inferred location is in excellent agreement with the planet proposed by \citet{dong+2015} and \citet{bae+2016} to explain both the two-armed spiral observed in scattered light and the millimeter dust crescent B78. 

Notably, this Doppler flip is intersected by the innermost tip of a large-scale line-width spiral extending between $\sim$100-200\,au (see Fig. \ref{fig:residuals_12co}, top row), and may therefore be associated with the spiral density waves driven by the planet. The absence of elevated temperatures along this spiral further suggests a non-thermal origin, likely associated with enhanced turbulence, or more generally, unresolved velocity flows, increased optical depth, or a combination of both. This complements the findings of \citet{casassus+2021}, who reported turbulent spiral features traced by \twCO{}\,$J=2\rightarrow1$ emission in this source, albeit located closer to the inner disc and spatially coincident with the near-infrared spirals.

\subsection{Origin of the spirals structures in \mwcsev{}}

   \begin{figure*}
   \centering
    \includegraphics[width=1.0\textwidth]{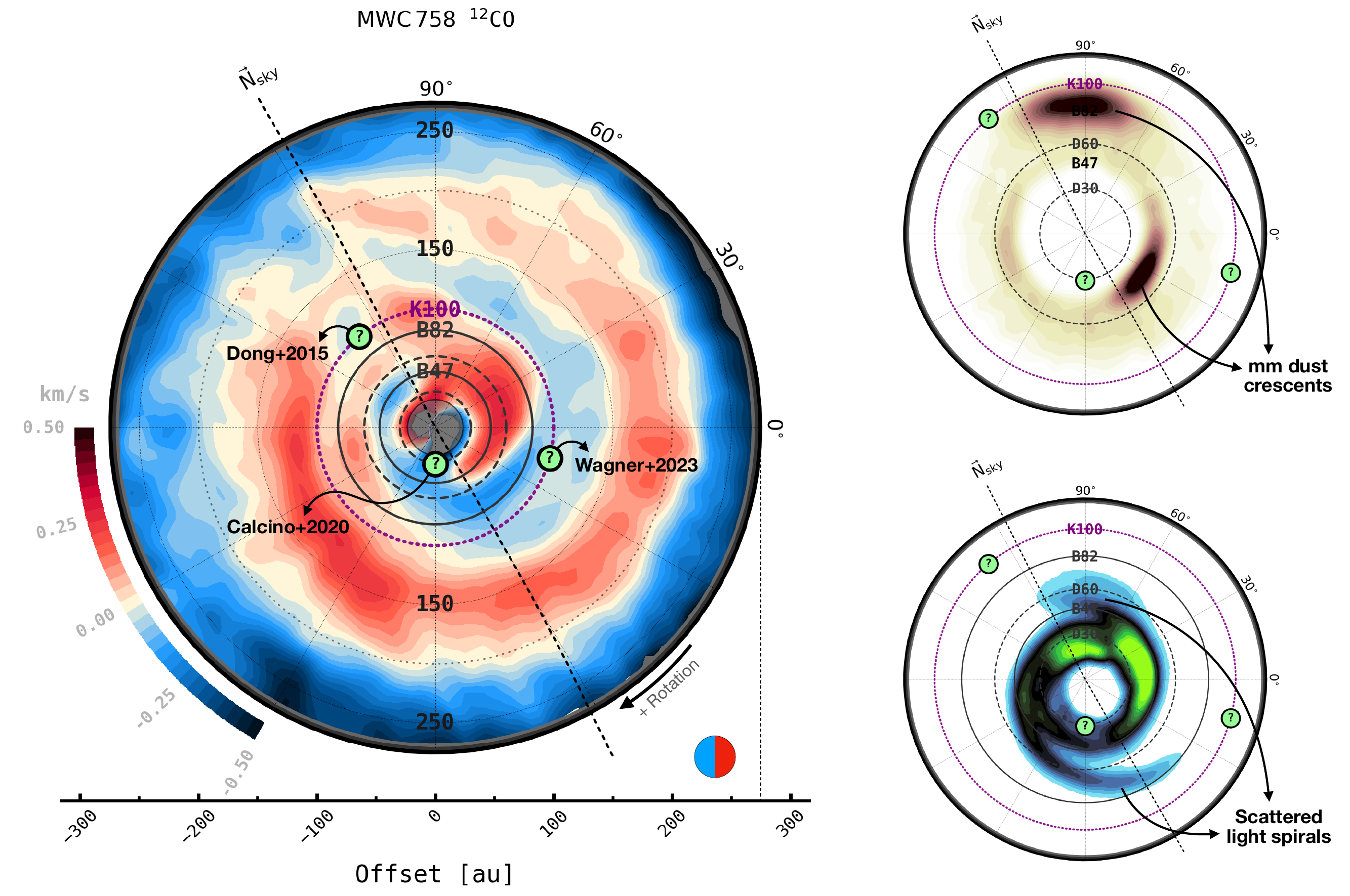}
      \caption{Deprojected view of velocity and dust substructures in the disc of \mwcsev{}. \textit{Left}: \twCOfull{} velocity residuals illustrating deviations from Keplerian rotation in the gas disc. Bottom right: VLT/SPHERE near-infrared observations \citep{benisty+2015} probing scattered light from micron-sized dust grains. Top right: ALMA Band 7 continuum observations \citep{teague+2025, curone+2025} tracing thermal emission from millimeter-sized dust grains. Solid and dashed lines mark the radial locations of peaks and troughs in the millimeter continuum, while the purple dotted K100 line indicates the orbital radius of the giant planet proposed by \citet{wagner+2023} based on a compact signal identified at infrared wavelengths. The green circles indicate the locations of selected planet candidates reported in the literature to explain the observed dust substructures.
              }
         \label{fig:summary_mwc758}
   \end{figure*}

Previous studies of the disc around \mwcsev{} have revealed spiral structures traced by scattered light \citep{grady+2013, benisty+2015} and molecular line emission \citep{boehler+2018}, as well as pronounced asymmetries in the millimeter continuum \citep{dong+2018} within the inner $\sim\!100$\,au. From a kinematic perspective, this source stands out within our sample due to its prominent velocity and line width spirals, which extend from the inner disc regions out to $\sim\!200$\,au in both \twCO{} and \thCO{} lines. Figure \ref{fig:summary_mwc758} illustrates the spatial configuration of these velocity substructures together with the near-infrared spirals and the locations of the millimeter continuum crescents. 

Two main scenarios involving embedded companions have been proposed to explain the origin of the continuum substructures in this source. The first invoke the presence of at least one massive planet or stellar companion inside the millimeter dust cavity ($R_p < 49$\,au), with the near-infrared two-armed spirals located exterior to the companion's orbit. The second scenario proposes a massive planet outside the cavity, capable of exciting the spirals interior to its orbit.

Within the first group, \citet{calcino+2020} showed that a massive planet on an eccentric orbit ($e=0.4$) with a semi-major axis of 33.5\,au---close to the point-like source reported by \citet{reggiani+2018} at $\sim\!20$\,au---can independently reproduce the two-armed spiral structure. 
In the second group, \citet{dong+2015} demonstrated that a single massive planet ($M_p/M_{\star} = 6\times10^{-3}$) located at $R_p\!\sim\!100$\,au could similarly excite the observed spirals. 
More recently, \citet{wagner+2023} reported a localized infrared intensity signal at a similar orbital radius, though on the opposite side of the disc, spatially connected to the southern near-infrared spiral arm. While such a planet would be broadly consistent with the line-width minimum identified in our data after azimuthal averaging at $R\!\sim\!90$\,au, the absence of a localized velocity or line-width perturbation at this orbital radius disfavors the presence of a high-mass companion.

By measuring the pattern speed of the near-infrared spirals, \citet{ren+2020} also proposed the presence of an outer planet, though at a larger orbital radius of 172\,au. Since that study adopted higher values for both the stellar mass and disc inclination ($M_\star=1.9$\,M$_\odot$ and $i=21^\circ$) than those inferred from our kinematic analysis ($M_\star=1.4$\,M$_\odot$ and $i=19.4^\circ$; \citealt{izquierdo+2025}), the corresponding planet location would shift to smaller radii when recalculated using our updated parameters.
Similarly, the absence of localized perturbations in these outer regions argues against the presence of a high-mass planet at such separations.

Turning to the kinematical features revealed in our data, vertical motions appear to play a dominant role in shaping the tightly wound velocity spirals observed in this source (Fig. \ref{fig:summary_mwc758}). This effect is likely enhanced by the nearly face-on orientation of the disc, in which radial and azimuthal velocity components are strongly suppressed by projection effects. If these spirals are of planetary origin and intrinsically dominated by vertical motions, their morphology and velocity structure would be broadly consistent with spiral waves launched at buoyancy resonances by an embedded companion \citep{bae+2021}. Spirals driven at Lindblad resonances by an inner massive planet could, in principle, account for the large pitch angles traced by the scattered light spirals in the inner disc, while also reproducing the gradual decrease in pitch angle of the outer velocity spirals with radius ($<10^\circ$, see Fig. \ref{fig:filaments_mwc758_12co}). Alternatively, a hybrid scenario involving at least one inner massive planet ($R_p<40$\,au) responsible for the large-pitch spirals in the inner disc, and an outer planet ($R_p\!\sim\!100$\,au) driving the tightly wound spirals in the outer regions could explain the observations. A similar two-planet configuration was proposed by \citet{baruteau+2019}, who invoked a Jupiter-mass planet at 35\,au, and a companion five times more massive at 140\,au to simultaneously explain the scattered-light spirals and the two compact crescents in the millimeter continuum.  

Another scenario consistent with the observed vertical spiral motions involves an eccentric circumstellar disc induced by a substellar companion in the innermost regions of the system. \citet{ragusa+2024} showed that a companion with a mass of about 10\% that of the central star can excite moderate eccentricities in the outer disc ($e\lesssim0.2$), which in turn generate $m=1$ blueshifted and redshifted spiral patterns in the vertical velocity component, closely resembling the spiral structures observed in \mwcsev{}. This interpretation is further supported by the measurable eccentricity of the dust cavity in this source \citep[$e\!\sim\!0.1$][] {dong+2018}, as well as by the comparable amplitudes of the vertical motions predicted in eccentric-disc models ($\sim\!0.4$\,\kms{}) and those inferred from our velocity spirals ($\sim\!0.3$\,\kms{}). Such large vertical velocities are difficult to reproduce via the planet-buoyancy spirals in the outer disc without invoking a very massive planet \citep{bae+2021}, making the eccentric-disc scenario one of the most compelling explanations for the observed features. We refer the reader to \citet{benisty+2026} for a broader discussion of vertical motions across the \exoalma{} sample.

An alternative explanation that does not necessarily require embedded companions was recently proposed by \citet{winter+2025}, who demonstrated that the observed $m=1$ spiral structures in \mwcsev{} and other \exoalma{} targets can arise from moderately warped disc geometries. In these models, small radial variations of only a few degrees in inclination and position angle relative to the mean disc orientation can produce azimuthal asymmetries in the projected Keplerian rotation that closely mimic the spiral kinematic patterns observed in this system.

\section{Conclusions} \label{sec:conclusions}

In this work, we present a tomographic analysis of molecular-line properties designed to extract and characterize gas substructures in protoplanetary discs observed with ALMA. We first apply this methodology to synthetic observations of planet-disc interaction and disc-instability simulations to evaluate the detectability of localized substructures and line-profile perturbations driven by these mechanisms. We then apply the same framework on \exoalma{} CO observations of the discs around \hdone{} and \mwcsev{}, where a variety of dust substructures have long been associated with planet-disc interactions. Our main findings are summarized as follows:

\begin{enumerate}
   
    \item Through synthetic observations of planet-disc interaction simulations, we identify a correlation between the location of embedded planets and localized line broadening in the disc (Fig. \ref{fig:folded_residuals_radmc}). Using line-width enhancements as tracers of planets offers significant advantages over kinematic analyses based solely on line-centroid perturbations, as this observable is largely insensitive to the planet's azimuthal location and provides a higher degree of spatial clustering (Figs. \ref{fig:peak_linewidth_vs_azimuth}, \ref{fig:channelsp_linewidth_significance}, and \ref{fig:channelsp_velocity_significance}). 

    \item Planets with masses greater than 0.2\% of the stellar mass (e.g. 2\,\Mj{} for a 1\,M$_\odot$ star) can be robustly detected through line-broadening signatures in discs of moderate inclination ($i<45^\circ$) under fiducial \exoalma{} observational setups. Importantly, achieving a high signal-to-noise ratio (SNR) at the planet’s location is more critical for detection than high angular or spectral resolution (Figs. \ref{fig:scores_linewidth}, \ref{fig:channelsp_linewidth_significance}, and \ref{fig:channelsp_velocity_significance}). Increasing the SNR by a factor of $\sim\!8$ (e.g. by degrading the beam size from $0\farcs{15}$ to $0\farcs{3}$) enables the detection of planets with masses as low as 0.1\% of the stellar mass, which would otherwise remain undetected even in discs at low inclinations ($i<30^\circ$). 

    \item Although disc instabilities can also produce localized velocity and line-width perturbations (Fig. \ref{fig:folded_instabilities}), they can be distinguished from planet-driven signals by examining the morphology of the corresponding line profiles. Broadened optically thick lines in the vicinity of an embedded planet are asymmetric around the line centroid and exhibit an intensity deficit toward blueshifted velocities, consistent with partially resolved meridional flows preferentially traced from the upper disc surface. In contrast, line broadening associated with disc instabilities is comparatively more isotropic, with intensity distributions concentrated closer to the line centroid (Figs. \ref{fig:intensdistrib} and \ref{fig:folded_dv}).
    
    \item We identify evidence for the presence of unseen planets in the disc of \hdone{}. Based on the analysis of peak and clustered perturbations in velocity and line-width residuals, we propose three candidate locations for these companions. The first corresponds to a planet at $R=73$\,au and $\phi=15^\circ$, responsible for the observed line broadening across multiple vertical layers of the disc and exhibiting line asymmetries consistent with our planet-disc synthetic observations. The second is a more interior planet at $R=41$\,au and $\phi=36^{\circ}$, triggering increased velocity dispersions in the disc's upper layers (Fig. \ref{fig:folded_residuals_hd135344}). The third candidate lies exterior to the continuum substructures and induces a Doppler-flip signature centered at $R=95$\,au and $\phi=-133^\circ$, accompanied by a spiral-like line-width perturbation and large-scale velocity structures qualitatively reproduced by a \ppdonet{} hydrodynamical model (Fig. \ref{fig:hydro_hd135344}). While these companions have been previously invoked to explain the substructures observed in the dust continuum emission, this work provides the first gas-based evidence supporting their presence (see Fig. \ref{fig:summary_hd135344} for a summary).
    
    \item The disc of \mwcsev{} is dominated by large-scale perturbations in the form of spiral structures across all line diagnostics. The velocity spirals are tightly wound in the outer disc and spatially coincide with regions of enhanced line broadening, indicative of unresolved velocity flows or increased optical depths (Fig. \ref{fig:spirals_mwc758_12co}). The most prominent spiral exhibits velocity amplitudes of $\sim\!0.2-0.3$\,\kms{}, predominantly shaped by downward vertical motions, with a weaker interior blueshifted counterpart consistent with upward flows. Although a massive planet located exterior to the dust continuum substructures could, in principle, generate a similar velocity spiral, we do not identify any significantly localized perturbation that could be attributed to such a companion. Instead, scenarios involving a warped outer disc or a substellar object in the inner regions provide a more natural explanation for both the observed spiral velocities and the absence of localized signatures (see Fig. \ref{fig:summary_mwc758} for a summary). 
    
\end{enumerate}

\section*{Acknowledgments}

We thank the anonymous referee for the constructive comments and suggestions, which greatly improved the quality of this work. 
This Letter makes use of the following ALMA data: ADS/JAO.ALMA\#2021.1.01123.L. ALMA is a partnership of ESO (representing its member states), NSF (USA) and NINS (Japan), together with NRC (Canada), MOST and ASIAA (Taiwan), and KASI (Republic of Korea), in cooperation with the Republic of Chile. The Joint ALMA Observatory is operated by ESO, AUI/NRAO and NAOJ. The National Radio Astronomy Observatory is a facility of the National Science Foundation operated under cooperative agreement by Associated Universities, Inc. We thank the North American ALMA Science Center (NAASC) for their generous support including providing computing facilities and financial support for student attendance at workshops and publications. Support for AFI was provided by NASA through the NASA Hubble Fellowship grant No. HST-HF2-51532.001-A awarded by the Space Telescope Science Institute, which is operated by the Association of Universities for Research in Astronomy, Inc., for NASA, under contract NAS5-26555. JB acknowledges support from NASA XRP grant No. 80NSSC23K1312. SF is funded by the European Union (ERC, UNVEIL, 101076613), and acknowledges financial contribution from PRIN-MUR 2022YP5ACE. EFvD acknowledges support from the ERC grant 101019751 MOLDISK. MB, JS, DF have received funding from the European Research Council (ERC) under the European Union’s Horizon 2020 research and innovation programme (PROTOPLANETS, grant agreement No. 101002188). PC and LT acknowledge support by the Italian Ministero dell'Istruzione, Universit\`a e Ricerca through the grant Progetti Premiali 2012 – iALMA (CUP C52I13000140001) and by the ANID BASAL project FB210003. NC has received funding from the European Research Council (ERC) under the European Union Horizon Europe research and innovation program (grant agreement No. 101042275, project Stellar-MADE). M. Flock has received funding from the European Research Council (ERC) under the European Unions Horizon 2020 research and innovation program (grant agreement No. 757957). M. Fukagawa is supported by a Grant-in-Aid from the Japan Society for the Promotion of Science (KAKENHI: No. JP22H01274). CH acknowledges support from NSF AAG grant No. 2407679. JDI acknowledges support from an STFC Ernest Rutherford Fellowship (ST/W004119/1) and a University Academic Fellowship from the University of Leeds. AI acknowledges support from the National Aeronautics and Space Administration under grant No. 80NSSC18K0828.
G. Lodato has received funding from the European Union's Horizon 2020 research and innovation program under the Marie Sklodowska-Curie grant agreement No. 823823 (DUSTBUSTERS), from PRIN-MUR 20228JPA3A and the European Union Next Generation EU, CUP: G53D23000870006. CL has received funding from the European Union's Horizon 2020 research and innovation program under the Marie Sklodowska-Curie grant agreement No. 823823 (DUSTBUSTERS) and by the UK Science and Technology research Council (STFC) via the consolidated grant ST/W000997/1. FMe has received funding from the European Research Council (ERC) under the European Union's Horizon Europe research and innovation program (grant agreement No. 101053020, project Dust2Planets). CP acknowledges Australian Research Council funding via FT170100040, DP18010423, DP220103767, and DP240103290. DP acknowledges Australian Research Council funding via DP18010423, DP220103767, and DP240103290. GR acknowledges funding from the Fondazione Cariplo, grant no. 2022-1217, and the European Research Council (ERC) under the European Union’s Horizon Europe Research \& Innovation Programme under grant agreement no. 101039651 (DiscEvol). AJW has received funding from the European Union's Horizon 2020 research and innovation programme under the Marie Skłodowska-Curie grant agreement No 101104656. Support for BZ was provided by The Brinson Foundation. This work was partly supported by the Deutsche Forschungsgemein- schaft (DFG, German Research Foundation) - Ref no. 325594231 FOR 2634/2 TE 1024/2-1, and by the DFG Cluster of Excellence Origins (www.origins-cluster.de). This project has received funding from the European Research Council (ERC) via the ERC Synergy Grant ECOGAL (grant 855130). Views and opinions expressed by ERC-funded scientists are however those of the author(s) only and do not necessarily reflect those of the European Union or the European Research Council. Neither the European Union nor the granting authority can be held responsible for them.  

\software{\textsc{astropy} \citep{Astropy_2022}, \bettermoments{} \citep{teague+2018_bettermoments}, \textsc{casa} \citep{casa}, \textsc{cmasher} \citep{cmasher+2020}, \discminer{} \citep{izquierdo+2021}, \textsc{emcee} \citep{foreman+2013}, \textsc{matplotlib} \citep{Hunter_mpl}, \textsc{numpy} \citep{harris_np}, \textsc{scikit-image} \citep{scikit-image}, \textsc{scipy} \citep{Virtanen_scipy}, \textsc{radio-beam} \citep{radio-beam}, \textsc{spectral-cube} \citep{spectral-cube}}

\bibliography{main}{}

@ARTICLE{benisty+2026,
   author = {{Benisty}, M. and {Izquierdo}, A. and {Stadler}, J.},
         title = "{exoALMA XX. Vertical Motions}",
         journal = {\apjl},
         year = 2026,
         volume = {TBD},
         number = {TBD},
}

@ARTICLE{fukagawa+2026,
   author = {{Fukagawa}, M. and {Izquierdo}, A. and {Facchni}, S.},
         title = "{exoALMA XX. Gas substructures}",
         journal = {\apjl},
         year = 2026,
         volume = {TBD},
         number = {TBD},
}

@ARTICLE{fairlamb2015,
       author = {{Fairlamb}, J.~R. and {Oudmaijer}, R.~D. and {Mendigut{\'\i}a}, I. and {Ilee}, J.~D. and {van den Ancker}, M.~E.},
        title = "{A spectroscopic survey of Herbig Ae/Be stars with X-shooter - I. Stellar parameters and accretion rates}",
      journal = {\mnras},
         year = 2015,
        month = oct,
       volume = {453},
       number = {1},
        pages = {976-1001},
          doi = {10.1093/mnras/stv1576},
archivePrefix = {arXiv},
       eprint = {1507.05967},
 primaryClass = {astro-ph.SR},
       adsurl = {https://ui.adsabs.harvard.edu/abs/2015MNRAS.453..976F}
}

@ARTICLE{schwarz2016,
       author = {{Schwarz}, Kamber R. and {Bergin}, Edwin A. and {Cleeves}, L. Ilsedore and {Blake}, Geoffrey A. and {Zhang}, Ke and {{\"O}berg}, Karin I. and {van Dishoeck}, Ewine F. and {Qi}, Chunhua},
        title = "{The Radial Distribution of H$_{2}$ and CO in TW Hya as Revealed by Resolved ALMA Observations of CO Isotopologues}",
      journal = {\apj},
         year = 2016,
        month = jun,
       volume = {823},
       number = {2},
          eid = {91},
        pages = {91},
          doi = {10.3847/0004-637X/823/2/91},
archivePrefix = {arXiv},
       eprint = {1603.08520},
 primaryClass = {astro-ph.SR},
       adsurl = {https://ui.adsabs.harvard.edu/abs/2016ApJ...823...91S}
}

@ARTICLE{visser2009,
       author = {{Visser}, R. and {van Dishoeck}, E.~F. and {Black}, J.~H.},
        title = "{The photodissociation and chemistry of CO isotopologues: applications to interstellar clouds and circumstellar disks}",
      journal = {\aap},
         year = 2009,
        month = aug,
       volume = {503},
       number = {2},
        pages = {323-343},
          doi = {10.1051/0004-6361/200912129},
archivePrefix = {arXiv},
       eprint = {0906.3699},
 primaryClass = {astro-ph.GA},
       adsurl = {https://ui.adsabs.harvard.edu/abs/2009A&A...503..323V}
}

@ARTICLE{andrews+2018,
       author = {{Andrews}, Sean M. and {Huang}, Jane and {P{\'e}rez}, Laura M. and {Isella}, Andrea and {Dullemond}, Cornelis P. and {Kurtovic}, Nicol{\'a}s T. and {Guzm{\'a}n}, Viviana V. and {Carpenter}, John M. and {Wilner}, David J. and {Zhang}, Shangjia and {Zhu}, Zhaohuan and {Birnstiel}, Tilman and {Bai}, Xue-Ning and {Benisty}, Myriam and {Hughes}, A. Meredith and {{\"O}berg}, Karin I. and {Ricci}, Luca},
        title = "{The Disk Substructures at High Angular Resolution Project (DSHARP). I. Motivation, Sample, Calibration, and Overview}",
      journal = {\apjl},
         year = 2018,
        month = dec,
       volume = {869},
       number = {2},
          eid = {L41},
        pages = {L41},
          doi = {10.3847/2041-8213/aaf741},
archivePrefix = {arXiv},
       eprint = {1812.04040},
 primaryClass = {astro-ph.SR},
       adsurl = {https://ui.adsabs.harvard.edu/abs/2018ApJ...869L..41A}
}

@ARTICLE{bae+2016,
       author = {{Bae}, Jaehan and {Zhu}, Zhaohuan and {Hartmann}, Lee},
        title = "{Planetary Signatures in the SAO 206462 (HD 135344B) Disk: A Spiral Arm Passing through Vortex?}",
      journal = {\apj},
         year = 2016,
        month = mar,
       volume = {819},
       number = {2},
          eid = {134},
        pages = {134},
          doi = {10.3847/0004-637X/819/2/134},
archivePrefix = {arXiv},
       eprint = {1601.04976},
 primaryClass = {astro-ph.EP},
       adsurl = {https://ui.adsabs.harvard.edu/abs/2016ApJ...819..134B}
}

@ARTICLE{bae+2021,
       author = {{Bae}, Jaehan and {Teague}, Richard and {Zhu}, Zhaohuan},
        title = "{Observational Signature of Tightly Wound Spirals Driven by Buoyancy Resonances in Protoplanetary Disks}",
      journal = {\apj},
         year = 2021,
        month = may,
       volume = {912},
       number = {1},
          eid = {56},
        pages = {56},
          doi = {10.3847/1538-4357/abe45e},
archivePrefix = {arXiv},
       eprint = {2102.03899},
 primaryClass = {astro-ph.EP},
       adsurl = {https://ui.adsabs.harvard.edu/abs/2021ApJ...912...56B}
}

@ARTICLE{bae+2022,
       author = {{Bae}, Jaehan and {Teague}, Richard and {Andrews}, Sean M. and {Benisty}, Myriam and {Facchini}, Stefano and {Galloway-Sprietsma}, Maria and {Loomis}, Ryan A. and {Aikawa}, Yuri and {Alarc{\'o}n}, Felipe and {Bergin}, Edwin and {Bergner}, Jennifer B. and {Booth}, Alice S. and {Cataldi}, Gianni and {Cleeves}, L. Ilsedore and {Czekala}, Ian and {Guzm{\'a}n}, Viviana V. and {Huang}, Jane and {Ilee}, John D. and {Kurtovic}, Nicolas T. and {Law}, Charles J. and {Gal}, Romane Le and {Liu}, Yao and {Long}, Feng and {M{\'e}nard}, Fran{\c{c}}ois and {{\"O}berg}, Karin I. and {P{\'e}rez}, Laura M. and {Qi}, Chunhua and {Schwarz}, Kamber R. and {Sierra}, Anibal and {Walsh}, Catherine and {Wilner}, David J. and {Zhang}, Ke},
        title = "{Molecules with ALMA at Planet-forming Scales (MAPS): A Circumplanetary Disk Candidate in Molecular-line Emission in the AS 209 Disk}",
      journal = {\apjl},
         year = 2022,
        month = aug,
       volume = {934},
       number = {2},
          eid = {L20},
        pages = {L20},
          doi = {10.3847/2041-8213/ac7fa3},
archivePrefix = {arXiv},
       eprint = {2207.05923},
 primaryClass = {astro-ph.EP},
       adsurl = {https://ui.adsabs.harvard.edu/abs/2022ApJ...934L..20B}
}

@INPROCEEDINGS{bae+2023,
       author = {{Bae}, J. and {Isella}, A. and {Zhu}, Z. and {Martin}, R. and {Okuzumi}, S. and {Suriano}, S.},
        title = "{Structured Distributions of Gas and Solids in Protoplanetary Disks}",
    booktitle = {Astronomical Society of the Pacific Conference Series},
         year = 2023,
       editor = {{Inutsuka}, S. and {Aikawa}, Y. and {Muto}, T. and {Tomida}, K. and {Tamura}, M.},
       series = {Astronomical Society of the Pacific Conference Series},
       volume = {534},
        month = jul,
        pages = {423},
       adsurl = {https://ui.adsabs.harvard.edu/abs/2023ASPC..534..423B}
}

@ARTICLE{bae+2025,
       author = {{Bae}, Jaehan and {Flock}, Mario and {Izquierdo}, Andr{\'e}s and {Kanagawa}, Kazuhiro and {Ono}, Tomohiro and {Pinte}, Christophe and {Price}, Daniel J. and {Rosotti}, Giovanni P. and {Wafflard-Fernandez}, Gaylor and {Lesur}, Geoffroy and {Masset}, ‪Fr{\'e}d{\'e}ric and {Andrews}, Sean M. and {Barraza-Alfaro}, Marcelo and {Benisty}, Myriam and {Cataldi}, Gianni and {Cuello}, Nicol{\'a}s and {Curone}, Pietro and {Czekala}, Ian and {Facchini}, Stefano and {Fasano}, Daniele and {Galloway-Sprietsma}, Maria and {Hall}, Cassandra and {Hammond}, Iain and {Huang}, Jane and {Lodato}, Giuseppe and {Longarini}, Cristiano and {Stadler}, Jochen and {Teague}, Richard and {Wilner}, David J. and {Winter}, Andrew J. and {W{\"o}lfer}, Lisa and {Yoshida}, Tomohiro C.},
        title = "{exoALMA. VII. Benchmarking Hydrodynamics and Radiative Transfer Codes}",
      journal = {\apjl},
         year = 2025,
        month = may,
       volume = {984},
       number = {1},
          eid = {L12},
        pages = {L12},
          doi = {10.3847/2041-8213/adc436},
archivePrefix = {arXiv},
       eprint = {2504.18643},
 primaryClass = {astro-ph.EP},
       adsurl = {https://ui.adsabs.harvard.edu/abs/2025ApJ...984L..12B}
}

@ARTICLE{barraza+2021,
       author = {{Barraza-Alfaro}, Marcelo and {Flock}, Mario and {Marino}, Sebastian and {P{\'e}rez}, Sebasti{\'a}n},
        title = "{Observability of the vertical shear instability in protoplanetary disk CO kinematics}",
      journal = {\aap},
         year = 2021,
        month = sep,
       volume = {653},
          eid = {A113},
        pages = {A113},
          doi = {10.1051/0004-6361/202140535},
archivePrefix = {arXiv},
       eprint = {2106.01159},
 primaryClass = {astro-ph.EP},
       adsurl = {https://ui.adsabs.harvard.edu/abs/2021A&A...653A.113B}
}

@ARTICLE{barraza+2025,
       author = {{Barraza-Alfaro}, Marcelo and {Flock}, Mario and {B{\'e}thune}, William and {Teague}, Richard and {Bae}, Jaehan and {Benisty}, Myriam and {Cataldi}, Gianni and {Curone}, Pietro and {Czekala}, Ian and {Facchini}, Stefano and {Fasano}, Daniele and {Fukagawa}, Misato and {Galloway-Sprietsma}, Maria and {Garg}, Himanshi and {Hall}, Cassandra and {Huang}, Jane and {Ilee}, John D. and {Izquierdo}, Andr{\'e}s F. and {Kanagawa}, Kazuhiro and {Koch}, Eric W. and {Lesur}, Geoffroy and {Longarini}, Cristiano and {Loomis}, Ryan A. and {Orihara}, Ryuta and {Pinte}, Christophe and {Price}, Daniel J. and {Rosotti}, Giovanni and {Stadler}, Jochen and {Wafflard-Fernandez}, Gaylor and {Winter}, Andrew J. and {W{\"o}lfer}, Lisa and {Yen}, Hsi-Wei and {Yoshida}, Tomohiro C. and {Zawadzki}, Brianna},
        title = "{exoALMA. XVI. Predicting Signatures of Large-scale Turbulence in Protoplanetary Disks}",
      journal = {\apjl},
         year = 2025,
        month = may,
       volume = {984},
       number = {1},
          eid = {L21},
        pages = {L21},
          doi = {10.3847/2041-8213/adc42d},
archivePrefix = {arXiv},
       eprint = {2504.19853},
 primaryClass = {astro-ph.EP},
       adsurl = {https://ui.adsabs.harvard.edu/abs/2025ApJ...984L..21B}
}

@ARTICLE{baruteau+2019,
       author = {{Baruteau}, Cl{\'e}ment and {Barraza}, Marcelo and {P{\'e}rez}, Sebasti{\'a}n and {Casassus}, Simon and {Dong}, Ruobing and {Lyra}, Wladimir and {Marino}, Sebasti{\'a}n and {Christiaens}, Valentin and {Zhu}, Zhaohuan and {Carmona}, Andr{\'e}s and {Debras}, Florian and {Alarcon}, Felipe},
        title = "{Dust traps in the protoplanetary disc MWC 758: two vortices produced by two giant planets?}",
      journal = {\mnras},
         year = 2019,
        month = jun,
       volume = {486},
       number = {1},
        pages = {304-319},
          doi = {10.1093/mnras/stz802},
archivePrefix = {arXiv},
       eprint = {1903.06537},
 primaryClass = {astro-ph.EP},
       adsurl = {https://ui.adsabs.harvard.edu/abs/2019MNRAS.486..304B}
}

@ARTICLE{benisty+2015,
       author = {{Benisty}, M. and {Juhasz}, A. and {Boccaletti}, A. and {Avenhaus}, H. and {Milli}, J. and {Thalmann}, C. and {Dominik}, C. and {Pinilla}, P. and {Buenzli}, E. and {Pohl}, A. and {Beuzit}, J. -L. and {Birnstiel}, T. and {de Boer}, J. and {Bonnefoy}, M. and {Chauvin}, G. and {Christiaens}, V. and {Garufi}, A. and {Grady}, C. and {Henning}, T. and {Huelamo}, N. and {Isella}, A. and {Langlois}, M. and {M{\'e}nard}, F. and {Mouillet}, D. and {Olofsson}, J. and {Pantin}, E. and {Pinte}, C. and {Pueyo}, L.},
        title = "{Asymmetric features in the protoplanetary disk MWC 758}",
      journal = {\aap},
         year = 2015,
        month = jun,
       volume = {578},
          eid = {L6},
        pages = {L6},
          doi = {10.1051/0004-6361/201526011},
archivePrefix = {arXiv},
       eprint = {1505.05325},
 primaryClass = {astro-ph.SR},
       adsurl = {https://ui.adsabs.harvard.edu/abs/2015A&A...578L...6B}
}

@ARTICLE{benitez+2016,
       author = {{Ben{\'\i}tez-Llambay}, Pablo and {Masset}, Fr{\'e}d{\'e}ric S.},
        title = "{FARGO3D: A New GPU-oriented MHD Code}",
      journal = {\apjs},
         year = 2016,
        month = mar,
       volume = {223},
       number = {1},
          eid = {11},
        pages = {11},
          doi = {10.3847/0067-0049/223/1/11},
archivePrefix = {arXiv},
       eprint = {1602.02359},
 primaryClass = {astro-ph.IM},
       adsurl = {https://ui.adsabs.harvard.edu/abs/2016ApJS..223...11B}
}

@ARTICLE{bethune+2022,
       author = {{B{\'e}thune}, William and {Latter}, Henrik},
        title = "{Gravitoturbulent dynamo in global simulations of gaseous disks}",
      journal = {\aap},
         year = 2022,
        month = jul,
       volume = {663},
          eid = {A138},
        pages = {A138},
          doi = {10.1051/0004-6361/202243219},
archivePrefix = {arXiv},
       eprint = {2206.03917},
 primaryClass = {astro-ph.SR},
       adsurl = {https://ui.adsabs.harvard.edu/abs/2022A&A...663A.138B}
}

@incollection{blum+1967,
  citeseer = {http://citeseer.nj.nec.com/context/77000/0},
  author = {Harry Blum},
  booktitle = {Models for the Perception of Speech and Visual Form},
  editor = {Weiant Wathen-Dunn},
  optstatus = {html doi abstract},
  title = {{A} {T}ransformation for {E}xtracting {N}ew {D}escriptors of
           {S}hape},
  address = {Cambridge},
  publisher = {MIT Press},
  year = {1967},
  pages = {362--380},
}

@ARTICLE{boehler+2018,
       author = {{Boehler}, Y. and {Ricci}, L. and {Weaver}, E. and {Isella}, A. and {Benisty}, M. and {Carpenter}, J. and {Grady}, C. and {Shen}, Bo-Ting and {Tang}, Ya-Wen and {Perez}, L.},
        title = "{The Complex Morphology of the Young Disk MWC 758: Spirals and Dust Clumps around a Large Cavity}",
      journal = {\apj},
         year = 2018,
        month = feb,
       volume = {853},
       number = {2},
          eid = {162},
        pages = {162},
          doi = {10.3847/1538-4357/aaa19c},
archivePrefix = {arXiv},
       eprint = {1712.08845},
 primaryClass = {astro-ph.EP},
       adsurl = {https://ui.adsabs.harvard.edu/abs/2018ApJ...853..162B}
}

@ARTICLE{calcino+2020,
       author = {{Calcino}, Josh and {Christiaens}, Valentin and {Price}, Daniel J. and {Pinte}, Christophe and {Davis}, Tamara M. and {van der Marel}, Nienke and {Cuello}, Nicol{\'a}s},
        title = "{Are the spiral arms in the MWC 758 protoplanetary disc driven by a companion inside the cavity?}",
      journal = {\mnras},
         year = 2020,
        month = oct,
       volume = {498},
       number = {1},
        pages = {639-650},
          doi = {10.1093/mnras/staa2468},
archivePrefix = {arXiv},
       eprint = {2007.06155},
 primaryClass = {astro-ph.EP},
       adsurl = {https://ui.adsabs.harvard.edu/abs/2020MNRAS.498..639C}
}

@ARTICLE{calcino+2022,
       author = {{Calcino}, Josh and {Hilder}, Thomas and {Price}, Daniel J. and {Pinte}, Christophe and {Bollati}, Francesco and {Lodato}, Giuseppe and {Norfolk}, Brodie J.},
        title = "{Mapping the Planetary Wake in HD 163296 with Kinematics}",
      journal = {\apjl},
         year = 2022,
        month = apr,
       volume = {929},
       number = {2},
          eid = {L25},
        pages = {L25},
          doi = {10.3847/2041-8213/ac64a7},
archivePrefix = {arXiv},
       eprint = {2111.07416},
 primaryClass = {astro-ph.EP},
       adsurl = {https://ui.adsabs.harvard.edu/abs/2022ApJ...929L..25C}
}

@ARTICLE{casassus+2019,
       author = {{Casassus}, Simon and {P{\'e}rez}, Sebasti{\'a}n},
        title = "{Kinematic Detections of Protoplanets: A Doppler Flip in the Disk of HD 100546}",
      journal = {\apjl},
         year = 2019,
        month = oct,
       volume = {883},
       number = {2},
          eid = {L41},
        pages = {L41},
          doi = {10.3847/2041-8213/ab4425},
archivePrefix = {arXiv},
       eprint = {1906.06302},
 primaryClass = {astro-ph.EP},
       adsurl = {https://ui.adsabs.harvard.edu/abs/2019ApJ...883L..41C}
}

@ARTICLE{casassus+2021,
       author = {{Casassus}, Simon and {Christiaens}, Valentin and {C{\'a}rcamo}, Miguel and {P{\'e}rez}, Sebasti{\'a}n and {Weber}, Philipp and {Ercolano}, Barbara and {van der Marel}, Nienke and {Pinte}, Christophe and {Dong}, Ruobing and {Baruteau}, Cl{\'e}ment and {Cieza}, Lucas and {van Dishoeck}, Ewine F. and {Jordan}, Andr{\'e}s and {Price}, Daniel J. and {Absil}, Olivier and {Arce-Tord}, Carla and {Faramaz}, Virginie and {Flores}, Christian and {Reggiani}, Maddalena},
        title = "{A dusty filament and turbulent CO spirals in HD 135344B - SAO 206462}",
      journal = {\mnras},
         year = 2021,
        month = nov,
       volume = {507},
       number = {3},
        pages = {3789-3809},
          doi = {10.1093/mnras/stab2359},
archivePrefix = {arXiv},
       eprint = {2104.08379},
 primaryClass = {astro-ph.EP},
       adsurl = {https://ui.adsabs.harvard.edu/abs/2021MNRAS.507.3789C}
}

@ARTICLE{cazzoletti+2018,
       author = {{Cazzoletti}, P. and {van Dishoeck}, E.~F. and {Pinilla}, P. and {Tazzari}, M. and {Facchini}, S. and {van der Marel}, N. and {Benisty}, M. and {Garufi}, A. and {P{\'e}rez}, L.~M.},
        title = "{Evidence for a massive dust-trapping vortex connected to spirals. Multi-wavelength analysis of the HD 135344B protoplanetary disk}",
      journal = {\aap},
         year = 2018,
        month = nov,
       volume = {619},
          eid = {A161},
        pages = {A161},
          doi = {10.1051/0004-6361/201834006},
archivePrefix = {arXiv},
       eprint = {1809.04160},
 primaryClass = {astro-ph.EP},
       adsurl = {https://ui.adsabs.harvard.edu/abs/2018A&A...619A.161C}
}

@ARTICLE{cugno+2024,
       author = {{Cugno}, Gabriele and {Leisenring}, Jarron and {Wagner}, Kevin R. and {Mullin}, Camryn and {Dong}, Ruobing and {Greene}, Thomas and {Johnstone}, Doug and {Meyer}, Michael R. and {Wolff}, Schuyler G. and {Beichman}, Charles and {Boyer}, Martha and {Horner}, Scott and {Hodapp}, Klaus and {Kelly}, Doug and {McCarthy}, Don and {Roellig}, Thomas and {Rieke}, George and {Rieke}, Marcia and {Stansberry}, John and {Young}, Erick},
        title = "{JWST/NIRCam Imaging of Young Stellar Objects. II. Deep Constraints on Giant Planets and a Planet Candidate Outside of the Spiral Disk Around SAO 206462}",
      journal = {\aj},
         year = 2024,
        month = apr,
       volume = {167},
       number = {4},
          eid = {182},
        pages = {182},
          doi = {10.3847/1538-3881/ad1ffc},
archivePrefix = {arXiv},
       eprint = {2401.02834},
 primaryClass = {astro-ph.EP},
       adsurl = {https://ui.adsabs.harvard.edu/abs/2024AJ....167..182C}
}

@ARTICLE{curone+2025,
       author = {{Curone}, Pietro and {Facchini}, Stefano and {Andrews}, Sean M. and {Testi}, Leonardo and {Benisty}, Myriam and {Czekala}, Ian and {Huang}, Jane and {Ilee}, John D. and {Isella}, Andrea and {Lodato}, Giuseppe and {Loomis}, Ryan A. and {Stadler}, Jochen and {Winter}, Andrew J. and {Bae}, Jaehan and {Barraza-Alfaro}, Marcelo and {Cataldi}, Gianni and {Cuello}, Nicol{\'a}s and {Fasano}, Daniele and {Flock}, Mario and {Fukagawa}, Misato and {Galloway-Sprietsma}, Maria and {Garg}, Himanshi and {Hall}, Cassandra and {Izquierdo}, Andr{\'e}s F. and {Kanagawa}, Kazuhiro and {Lesur}, Geoffroy and {Longarini}, Cristiano and {Menard}, Francois and {Orihara}, Ryuta and {Pinte}, Christophe and {Price}, Daniel J. and {Rosotti}, Giovanni and {Teague}, Richard and {Wafflard-Fernandez}, Gaylor and {Wilner}, David J. and {W{\"o}lfer}, Lisa and {Yen}, Hsi-Wei and {Yoshida}, Tomohiro C. and {Zawadzki}, Brianna},
        title = "{exoALMA. IV. Substructures, Asymmetries, and the Faint Outer Disk in Continuum Emission}",
      journal = {\apjl},
         year = 2025,
        month = may,
       volume = {984},
       number = {1},
          eid = {L9},
        pages = {L9},
          doi = {10.3847/2041-8213/adc438},
archivePrefix = {arXiv},
       eprint = {2504.18725},
 primaryClass = {astro-ph.EP},
       adsurl = {https://ui.adsabs.harvard.edu/abs/2025ApJ...984L...9C}
}

@ARTICLE{dartois+2003,
       author = {{Dartois}, E. and {Dutrey}, A. and {Guilloteau}, S.},
        title = "{Structure of the DM Tau Outer Disk: Probing the vertical kinetic temperature gradient}",
      journal = {\aap},
         year = 2003,
        month = feb,
       volume = {399},
        pages = {773-787},
          doi = {10.1051/0004-6361:20021638},
       adsurl = {https://ui.adsabs.harvard.edu/abs/2003A&A...399..773D}
}

@ARTICLE{dong+2015,
       author = {{Dong}, Ruobing and {Zhu}, Zhaohuan and {Rafikov}, Roman R. and {Stone}, James M.},
        title = "{Observational Signatures of Planets in Protoplanetary Disks: Spiral Arms Observed in Scattered Light Imaging Can be Induced by Planets}",
      journal = {\apjl},
         year = 2015,
        month = aug,
       volume = {809},
       number = {1},
          eid = {L5},
        pages = {L5},
          doi = {10.1088/2041-8205/809/1/L5},
archivePrefix = {arXiv},
       eprint = {1507.03596},
 primaryClass = {astro-ph.EP},
       adsurl = {https://ui.adsabs.harvard.edu/abs/2015ApJ...809L...5D}
}

@ARTICLE{dong+2018,
       author = {{Dong}, Ruobing and {Liu}, Sheng-yuan and {Eisner}, Josh and {Andrews}, Sean and {Fung}, Jeffrey and {Zhu}, Zhaohuan and {Chiang}, Eugene and {Hashimoto}, Jun and {Liu}, Hauyu Baobab and {Casassus}, Simon and {Esposito}, Thomas and {Hasegawa}, Yasuhiro and {Muto}, Takayuki and {Pavlyuchenkov}, Yaroslav and {Wilner}, David and {Akiyama}, Eiji and {Tamura}, Motohide and {Wisniewski}, John},
        title = "{The Eccentric Cavity, Triple Rings, Two-armed Spirals, and Double Clumps of the MWC 758 Disk}",
      journal = {\apj},
         year = 2018,
        month = jun,
       volume = {860},
       number = {2},
          eid = {124},
        pages = {124},
          doi = {10.3847/1538-4357/aac6cb},
archivePrefix = {arXiv},
       eprint = {1805.12141},
 primaryClass = {astro-ph.SR},
       adsurl = {https://ui.adsabs.harvard.edu/abs/2018ApJ...860..124D}
}

@ARTICLE{dong+2019,
       author = {{Dong}, Ruobing and {Liu}, Sheng-Yuan and {Fung}, Jeffrey},
        title = "{Observational Signatures of Planets in Protoplanetary Disks: Planet-induced Line Broadening in Gaps}",
      journal = {\apj},
         year = 2019,
        month = jan,
       volume = {870},
       number = {2},
          eid = {72},
        pages = {72},
          doi = {10.3847/1538-4357/aaf38e},
archivePrefix = {arXiv},
       eprint = {1811.09629},
 primaryClass = {astro-ph.EP},
       adsurl = {https://ui.adsabs.harvard.edu/abs/2019ApJ...870...72D}
}

@MISC{dullemond+2012,
       author = {{Dullemond}, C.~P. and {Juhasz}, A. and {Pohl}, A. and {Sereshti}, F. and {Shetty}, R. and {Peters}, T. and {Commercon}, B. and {Flock}, M.},
        title = "{RADMC-3D: A multi-purpose radiative transfer tool}",
 howpublished = {Astrophysics Source Code Library, record ascl:1202.015},
         year = 2012,
        month = feb,
          eid = {ascl:1202.015},
        pages = {ascl:1202.015},
archivePrefix = {ascl},
       eprint = {1202.015},
       adsurl = {https://ui.adsabs.harvard.edu/abs/2012ascl.soft02015D}
}

@ARTICLE{flock+2015,
       author = {{Flock}, M. and {Ruge}, J.~P. and {Dzyurkevich}, N. and {Henning}, Th. and {Klahr}, H. and {Wolf}, S.},
        title = "{Gaps, rings, and non-axisymmetric structures in protoplanetary disks. From simulations to ALMA observations}",
      journal = {\aap},
         year = 2015,
        month = feb,
       volume = {574},
          eid = {A68},
        pages = {A68},
          doi = {10.1051/0004-6361/201424693},
archivePrefix = {arXiv},
       eprint = {1411.2736},
 primaryClass = {astro-ph.EP},
       adsurl = {https://ui.adsabs.harvard.edu/abs/2015A&A...574A..68F}
}

@ARTICLE{flock+2020,
       author = {{Flock}, Mario and {Turner}, Neal J. and {Nelson}, Richard P. and {Lyra}, Wladimir and {Manger}, Natascha and {Klahr}, Hubert},
        title = "{Gas and Dust Dynamics in Starlight-heated Protoplanetary Disks}",
      journal = {\apj},
         year = 2020,
        month = jul,
       volume = {897},
       number = {2},
          eid = {155},
        pages = {155},
          doi = {10.3847/1538-4357/ab9641},
archivePrefix = {arXiv},
       eprint = {2005.11974},
 primaryClass = {astro-ph.EP},
       adsurl = {https://ui.adsabs.harvard.edu/abs/2020ApJ...897..155F}
}

@ARTICLE{foreman+2013,
       author = {{Foreman-Mackey}, Daniel and {Hogg}, David W. and {Lang}, Dustin and {Goodman}, Jonathan},
        title = "{emcee: The MCMC Hammer}",
      journal = {\pasp},
         year = 2013,
        month = mar,
       volume = {125},
       number = {925},
        pages = {306},
          doi = {10.1086/670067},
archivePrefix = {arXiv},
       eprint = {1202.3665},
 primaryClass = {astro-ph.IM},
       adsurl = {https://ui.adsabs.harvard.edu/abs/2013PASP..125..306F}
}

@ARTICLE{gaia+2023,
       author = {{Gaia Collaboration} and {Vallenari}, A. and {Brown}, A.~G.~A. and {Prusti}, T. and {de Bruijne}, J.~H.~J. and {Arenou}, F. and {Babusiaux}, C. and {Biermann}, M. and {Creevey}, O.~L. and {Ducourant}, C. and {Evans}, D.~W. and {Eyer}, L. and {Guerra}, R. and {Hutton}, A. and {Jordi}, C. and {Klioner}, S.~A. and {Lammers}, U.~L. and {Lindegren}, L. and {Luri}, X. and {Mignard}, F. and {Panem}, C. and {Pourbaix}, D. and {Randich}, S. and {Sartoretti}, P. and {Soubiran}, C. and {Tanga}, P. and {Walton}, N.~A. and {Bailer-Jones}, C.~A.~L. and {Bastian}, U. and {Drimmel}, R. and {Jansen}, F. and {Katz}, D. and {Lattanzi}, M.~G. and {van Leeuwen}, F. and {Bakker}, J. and {Cacciari}, C. and {Casta{\~n}eda}, J. and {De Angeli}, F. and {Fabricius}, C. and {Fouesneau}, M. and {Fr{\'e}mat}, Y. and {Galluccio}, L. and {Guerrier}, A. and {Heiter}, U. and {Masana}, E. and {Messineo}, R. and {Mowlavi}, N. and {Nicolas}, C. and {Nienartowicz}, K. and {Pailler}, F. and {Panuzzo}, P. and {Riclet}, F. and {Roux}, W. and {Seabroke}, G.~M. and {Sordo}, R. and {Th{\'e}venin}, F. and {Gracia-Abril}, G. and {Portell}, J. and {Teyssier}, D. and {Altmann}, M. and {Andrae}, R. and {Audard}, M. and {Bellas-Velidis}, I. and {Benson}, K. and {Berthier}, J. and {Blomme}, R. and {Burgess}, P.~W. and {Busonero}, D. and {Busso}, G. and {C{\'a}novas}, H. and {Carry}, B. and {Cellino}, A. and {Cheek}, N. and {Clementini}, G. and {Damerdji}, Y. and {Davidson}, M. and {de Teodoro}, P. and {Nu{\~n}ez Campos}, M. and {Delchambre}, L. and {Dell'Oro}, A. and {Esquej}, P. and {Fern{\'a}ndez-Hern{\'a}ndez}, J. and {Fraile}, E. and {Garabato}, D. and {Garc{\'\i}a-Lario}, P. and {Gosset}, E. and {Haigron}, R. and {Halbwachs}, J.-L. and {Hambly}, N.~C. and {Harrison}, D.~L. and {Hern{\'a}ndez}, J. and {Hestroffer}, D. and {Hodgkin}, S.~T. and {Holl}, B. and {Jan{\ss}en}, K. and {Jevardat de Fombelle}, G. and {Jordan}, S. and {Krone-Martins}, A. and {Lanzafame}, A.~C. and {L{\"o}ffler}, W. and {Marchal}, O. and {Marrese}, P.~M. and {Moitinho}, A. and {Muinonen}, K. and {Osborne}, P. and {Pancino}, E. and {Pauwels}, T. and {Recio-Blanco}, A. and {Reyl{\'e}}, C. and {Riello}, M. and {Rimoldini}, L. and {Roegiers}, T. and {Rybizki}, J. and {Sarro}, L.~M. and {Siopis}, C. and {Smith}, M. and {Sozzetti}, A. and {Utrilla}, E. and {van Leeuwen}, M. and {Abbas}, U. and {{\'A}brah{\'a}m}, P. and {Abreu Aramburu}, A. and {Aerts}, C. and {Aguado}, J.~J. and {Ajaj}, M. and {Aldea-Montero}, F. and {Altavilla}, G. and {{\'A}lvarez}, M.~A. and {Alves}, J. and {Anders}, F. and {Anderson}, R.~I. and {Anglada Varela}, E. and {Antoja}, T. and {Baines}, D. and {Baker}, S.~G. and {Balaguer-N{\'u}{\~n}ez}, L. and {Balbinot}, E. and {Balog}, Z. and {Barache}, C. and {Barbato}, D. and {Barros}, M. and {Barstow}, M.~A. and {Bartolom{\'e}}, S. and {Bassilana}, J.-L. and {Bauchet}, N. and {Becciani}, U. and {Bellazzini}, M. and {Berihuete}, A. and {Bernet}, M. and {Bertone}, S. and {Bianchi}, L. and {Binnenfeld}, A. and {Blanco-Cuaresma}, S. and {Blazere}, A. and {Boch}, T. and {Bombrun}, A. and {Bossini}, D. and {Bouquillon}, S. and {Bragaglia}, A. and {Bramante}, L. and {Breedt}, E. and {Bressan}, A. and {Brouillet}, N. and {Brugaletta}, E. and {Bucciarelli}, B. and {Burlacu}, A. and {Butkevich}, A.~G. and {Buzzi}, R. and {Caffau}, E. and {Cancelliere}, R. and {Cantat-Gaudin}, T. and {Carballo}, R. and {Carlucci}, T. and {Carnerero}, M.~I. and {Carrasco}, J.~M. and {Casamiquela}, L. and {Castellani}, M. and {Castro-Ginard}, A. and {Chaoul}, L. and {Charlot}, P. and {Chemin}, L. and {Chiaramida}, V. and {Chiavassa}, A. and {Chornay}, N. and {Comoretto}, G. and {Contursi}, G. and {Cooper}, W.~J. and {Cornez}, T. and {Cowell}, S. and {Crifo}, F. and {Cropper}, M. and {Crosta}, M. and {Crowley}, C. and {Dafonte}, C. and {Dapergolas}, A. and {David}, M. and {David}, P. and {de Laverny}, P. and {De Luise}, F. and {De March}, R.},
        title = "{Gaia Data Release 3. Summary of the content and survey properties}",
      journal = {\aap},
         year = 2023,
        month = jun,
       volume = {674},
          eid = {A1},
        pages = {A1},
          doi = {10.1051/0004-6361/202243940},
archivePrefix = {arXiv},
       eprint = {2208.00211},
 primaryClass = {astro-ph.GA},
       adsurl = {https://ui.adsabs.harvard.edu/abs/2023A&A...674A...1G}
}

@ARTICLE{galloway+2025,
       author = {{Galloway-Sprietsma}, Maria and {Bae}, Jaehan and {Izquierdo}, Andr{\'e}s F. and {Stadler}, Jochen and {Longarini}, Cristiano and {Teague}, Richard and {Andrews}, Sean M. and {Winter}, Andrew J. and {Benisty}, Myriam and {Facchini}, Stefano and {Rosotti}, Giovanni and {Zawadzki}, Brianna and {Pinte}, Christophe and {Fasano}, Daniele and {Barraza-Alfaro}, Marcelo and {Cataldi}, Gianni and {Cuello}, Nicol{\'a}s and {Curone}, Pietro and {Czekala}, Ian and {Flock}, Mario and {Fukagawa}, Misato and {Gardner}, Charles H. and {Garg}, Himanshi and {Hall}, Cassandra and {Huang}, Jane and {Ilee}, John D. and {Kanagawa}, Kazuhiro and {Lesur}, Geoffroy and {Lodato}, Giuseppe and {Loomis}, Ryan A. and {Menard}, Francois and {Orihara}, Ryuta and {Price}, Daniel J. and {Wafflard-Fernandez}, Gaylor and {Wilner}, David J. and {W{\"o}lfer}, Lisa and {Yen}, Hsi-Wei and {Yoshida}, Tomohiro C.},
        title = "{exoALMA. V. Gaseous Emission Surfaces and Temperature Structures}",
      journal = {\apjl},
         year = 2025,
        month = may,
       volume = {984},
       number = {1},
          eid = {L10},
        pages = {L10},
          doi = {10.3847/2041-8213/adc437},
archivePrefix = {arXiv},
       eprint = {2504.19902},
 primaryClass = {astro-ph.EP},
       adsurl = {https://ui.adsabs.harvard.edu/abs/2025ApJ...984L..10G}
}

@ARTICLE{garufi+2013,
       author = {{Garufi}, A. and {Quanz}, S.~P. and {Avenhaus}, H. and {Buenzli}, E. and {Dominik}, C. and {Meru}, F. and {Meyer}, M.~R. and {Pinilla}, P. and {Schmid}, H.~M. and {Wolf}, S.},
        title = "{Small vs. large dust grains in transitional disks: do different cavity sizes indicate a planet?. SAO 206462 (HD 135344B) in polarized light with VLT/NACO}",
      journal = {\aap},
         year = 2013,
        month = dec,
       volume = {560},
          eid = {A105},
        pages = {A105},
          doi = {10.1051/0004-6361/201322429},
archivePrefix = {arXiv},
       eprint = {1311.4195},
 primaryClass = {astro-ph.EP},
       adsurl = {https://ui.adsabs.harvard.edu/abs/2013A&A...560A.105G}
}

@ARTICLE{grady+2013,
       author = {{Grady}, C.~A. and {Muto}, T. and {Hashimoto}, J. and {Fukagawa}, M. and {Currie}, T. and {Biller}, B. and {Thalmann}, C. and {Sitko}, M.~L. and {Russell}, R. and {Wisniewski}, J. and {Dong}, R. and {Kwon}, J. and {Sai}, S. and {Hornbeck}, J. and {Schneider}, G. and {Hines}, D. and {Moro Mart{\'\i}n}, A. and {Feldt}, M. and {Henning}, Th. and {Pott}, J. -U. and {Bonnefoy}, M. and {Bouwman}, J. and {Lacour}, S. and {Mueller}, A. and {Juh{\'a}sz}, A. and {Crida}, A. and {Chauvin}, G. and {Andrews}, S. and {Wilner}, D. and {Kraus}, A. and {Dahm}, S. and {Robitaille}, T. and {Jang-Condell}, H. and {Abe}, L. and {Akiyama}, E. and {Brandner}, W. and {Brandt}, T. and {Carson}, J. and {Egner}, S. and {Follette}, K.~B. and {Goto}, M. and {Guyon}, O. and {Hayano}, Y. and {Hayashi}, M. and {Hayashi}, S. and {Hodapp}, K. and {Ishii}, M. and {Iye}, M. and {Janson}, M. and {Kandori}, R. and {Knapp}, G. and {Kudo}, T. and {Kusakabe}, N. and {Kuzuhara}, M. and {Mayama}, S. and {McElwain}, M. and {Matsuo}, T. and {Miyama}, S. and {Morino}, J. -I. and {Nishimura}, T. and {Pyo}, T. -S. and {Serabyn}, G. and {Suto}, H. and {Suzuki}, R. and {Takami}, M. and {Takato}, N. and {Terada}, H. and {Tomono}, D. and {Turner}, E. and {Watanabe}, M. and {Yamada}, T. and {Takami}, H. and {Usuda}, T. and {Tamura}, M.},
        title = "{Spiral Arms in the Asymmetrically Illuminated Disk of MWC 758 and Constraints on Giant Planets}",
      journal = {\apj},
         year = 2013,
        month = jan,
       volume = {762},
       number = {1},
          eid = {48},
        pages = {48},
          doi = {10.1088/0004-637X/762/1/48},
archivePrefix = {arXiv},
       eprint = {1212.1466},
 primaryClass = {astro-ph.SR},
       adsurl = {https://ui.adsabs.harvard.edu/abs/2013ApJ...762...48G}
}

@ARTICLE{hacar+2016,
       author = {{Hacar}, A. and {Alves}, J. and {Burkert}, A. and {Goldsmith}, P.},
        title = "{Opacity broadening and interpretation of suprathermal CO linewidths: Macroscopic turbulence and tangled molecular clouds}",
      journal = {\aap},
         year = 2016,
        month = jun,
       volume = {591},
          eid = {A104},
        pages = {A104},
          doi = {10.1051/0004-6361/201527319},
archivePrefix = {arXiv},
       eprint = {1603.08521},
 primaryClass = {astro-ph.GA},
       adsurl = {https://ui.adsabs.harvard.edu/abs/2016A&A...591A.104H}
}

@ARTICLE{hall+2020,
       author = {{Hall}, C. and {Dong}, R. and {Teague}, R. and {Terry}, J. and {Pinte}, C. and {Paneque-Carre{\~n}o}, T. and {Veronesi}, B. and {Alexander}, R.~D. and {Lodato}, G.},
        title = "{Predicting the Kinematic Evidence of Gravitational Instability}",
      journal = {\apj},
         year = 2020,
        month = dec,
       volume = {904},
       number = {2},
          eid = {148},
        pages = {148},
          doi = {10.3847/1538-4357/abac17},
archivePrefix = {arXiv},
       eprint = {2007.15686},
 primaryClass = {astro-ph.SR},
       adsurl = {https://ui.adsabs.harvard.edu/abs/2020ApJ...904..148H}
}

@ARTICLE{izquierdo+2021,
       author = {{Izquierdo}, A.~F. and {Testi}, L. and {Facchini}, S. and {Rosotti}, G.~P. and {van Dishoeck}, E.~F.},
        title = "{The Disc Miner. I. A statistical framework to detect and quantify kinematical perturbations driven by young planets in discs}",
      journal = {\aap},
         year = 2021,
        month = jun,
       volume = {650},
          eid = {A179},
        pages = {A179},
          doi = {10.1051/0004-6361/202140779},
archivePrefix = {arXiv},
       eprint = {2104.09596},
 primaryClass = {astro-ph.EP},
       adsurl = {https://ui.adsabs.harvard.edu/abs/2021A&A...650A.179I}
}

@ARTICLE{izquierdo+2022,
       author = {{Izquierdo}, Andr{\'e}s F. and {Facchini}, Stefano and {Rosotti}, Giovanni P. and {van Dishoeck}, Ewine F. and {Testi}, Leonardo},
        title = "{A New Planet Candidate Detected in a Dust Gap of the Disk around HD 163296 through Localized Kinematic Signatures: An Observational Validation of the DISCMINER}",
      journal = {\apj},
         year = 2022,
        month = mar,
       volume = {928},
       number = {1},
          eid = {2},
        pages = {2},
          doi = {10.3847/1538-4357/ac474d},
archivePrefix = {arXiv},
       eprint = {2111.06367},
 primaryClass = {astro-ph.EP},
       adsurl = {https://ui.adsabs.harvard.edu/abs/2022ApJ...928....2I}
}

@ARTICLE{izquierdo+2023,
       author = {{Izquierdo}, A.~F. and {Testi}, L. and {Facchini}, S. and {Rosotti}, G.~P. and {van Dishoeck}, E.~F. and {W{\"o}lfer}, L. and {Paneque-Carre{\~n}o}, T.},
        title = "{The Disc Miner. II. Revealing gas substructures and kinematic signatures from planet-disc interaction through line profile analysis}",
      journal = {\aap},
         year = 2023,
        month = jun,
       volume = {674},
          eid = {A113},
        pages = {A113},
          doi = {10.1051/0004-6361/202245425},
archivePrefix = {arXiv},
       eprint = {2304.03607},
 primaryClass = {astro-ph.EP},
       adsurl = {https://ui.adsabs.harvard.edu/abs/2023A&A...674A.113I}
}

@ARTICLE{izquierdo+2025,
       author = {{Izquierdo}, Andr{\'e}s F. and {Stadler}, Jochen and {Galloway-Sprietsma}, Maria and {Benisty}, Myriam and {Pinte}, Christophe and {Bae}, Jaehan and {Teague}, Richard and {Facchini}, Stefano and {W{\"o}lfer}, Lisa and {Longarini}, Cristiano and {Curone}, Pietro and {Andrews}, Sean M. and {Barraza-Alfaro}, Marcelo and {Cataldi}, Gianni and {Cuello}, Nicol{\'a}s and {Czekala}, Ian and {Fasano}, Daniele and {Flock}, Mario and {Fukagawa}, Misato and {Garg}, Himanshi and {Hall}, Cassandra and {Hammond}, Iain and {Hilder}, Thomas and {Huang}, Jane and {Ilee}, John D. and {Isella}, Andrea and {Kanagawa}, Kazuhiro and {Lesur}, Geoffroy and {Lodato}, Giuseppe and {Loomis}, Ryan A. and {Orihara}, Ryuta and {Price}, Daniel J. and {Rosotti}, Giovanni and {Testi}, Leonardo and {Yen}, Hsi-Wei and {Wafflard-Fernandez}, Gaylor and {Wilner}, David J. and {Winter}, Andrew J. and {Yoshida}, Tomohiro C. and {Zawadzki}, Brianna},
        title = "{exoALMA. III. Line-intensity Modeling and System Property Extraction from Protoplanetary Disks}",
      journal = {\apjl},
         year = 2025,
        month = may,
       volume = {984},
       number = {1},
          eid = {L8},
        pages = {L8},
          doi = {10.3847/2041-8213/adc439},
archivePrefix = {arXiv},
       eprint = {2504.19986},
 primaryClass = {astro-ph.EP},
       adsurl = {https://ui.adsabs.harvard.edu/abs/2025ApJ...984L...8I}
}

@ARTICLE{izquierdo+2026,
       author = {{Izquierdo}, Andr{\'e}s F. and {Bae}, Jaehan and {Galloway-Sprietsma}, Maria and {van Dishoeck}, Ewine F. and {Facchini}, Stefano and {Rosotti}, Giovanni and {Stadler}, Jochen and {Benisty}, Myriam and {Testi}, Leonardo},
        title = "{Circumplanetary Disk Candidate in the Disk of HD 163296 Traced by Localized Emission from Simple Organics}",
      journal = {\apjl},
         year = 2026,
        month = jan,
       volume = {997},
       number = {1},
          eid = {L2},
        pages = {L2},
          doi = {10.3847/2041-8213/ae2f59},
archivePrefix = {arXiv},
       eprint = {2601.10631},
 primaryClass = {astro-ph.EP},
       adsurl = {https://ui.adsabs.harvard.edu/abs/2026ApJ...997L...2I}
}

@ARTICLE{koch+2015,
   author = {{Koch}, E.~W. and {Rosolowsky}, E.~W.},
    title = "{Filament identification through mathematical morphology}",
  journal = {\mnras},
archivePrefix = "arXiv",
   eprint = {1507.02289},
     year = 2015,
    month = oct,
   volume = 452,
    pages = {3435-3450},
      doi = {10.1093/mnras/stv1521},
   adsurl = {http://adsabs.harvard.edu/abs/2015MNRAS.452.3435K}
}

@ARTICLE{law+2023,
       author = {{Law}, Charles J. and {Booth}, Alice S. and {{\"O}berg}, Karin I.},
        title = "{SO and SiS Emission Tracing an Embedded Planet and Compact $^{12}$CO and $^{13}$CO Counterparts in the HD 169142 Disk}",
      journal = {\apjl},
         year = 2023,
        month = jul,
       volume = {952},
       number = {1},
          eid = {L19},
        pages = {L19},
          doi = {10.3847/2041-8213/acdfd0},
archivePrefix = {arXiv},
       eprint = {2306.13710},
 primaryClass = {astro-ph.EP},
       adsurl = {https://ui.adsabs.harvard.edu/abs/2023ApJ...952L..19L}
}

@INPROCEEDINGS{lesur+2023,
       author = {{Lesur}, G. and {Flock}, M. and {Ercolano}, B. and {Lin}, M. and {Yang}, C. and {Barranco}, J.~A. and {Benitez-Llambay}, P. and {Goodman}, J. and {Johansen}, A. and {Klahr}, H. and {Laibe}, G. and {Lyra}, W. and {Marcus}, P.~S. and {Nelson}, R.~P. and {Squire}, J. and {Simon}, J.~B. and {Turner}, N.~J. and {Umurhan}, O.~M. and {Youdin}, A.~N.},
        title = "{Hydro-, Magnetohydro-, and Dust-Gas Dynamics of Protoplanetary Disks}",
    booktitle = {Astronomical Society of the Pacific Conference Series},
         year = 2023,
       editor = {{Inutsuka}, S. and {Aikawa}, Y. and {Muto}, T. and {Tomida}, K. and {Tamura}, M.},
       series = {Astronomical Society of the Pacific Conference Series},
       volume = {534},
        month = jul,
        pages = {465},
       adsurl = {https://ui.adsabs.harvard.edu/abs/2023ASPC..534..465L}
}

@article{leys+2013,
title = {Detecting outliers: Do not use standard deviation around the mean, use absolute deviation around the median},
journal = {Journal of Experimental Social Psychology},
volume = {49},
number = {4},
pages = {764-766},
year = {2013},
issn = {0022-1031},
doi = {https://doi.org/10.1016/j.jesp.2013.03.013},
url = {https://www.sciencedirect.com/science/article/pii/S0022103113000668},
author = {Christophe Leys and Christophe Ley and Olivier Klein and Philippe Bernard and Laurent Licata}
}

@ARTICLE{longarini+2025,
       author = {{Longarini}, Cristiano and {Lodato}, Giuseppe and {Rosotti}, Giovanni and {Andrews}, Sean and {Winter}, Andrew and {Stadler}, Jochen and {Izquierdo}, Andr{\'e}s and {Galloway-Sprietsma}, Maria and {Facchini}, Stefano and {Curone}, Pietro and {Benisty}, Myriam and {Teague}, Richard and {Bae}, Jaehan and {Barraza-Alfaro}, Marcelo and {Cataldi}, Gianni and {Czekala}, Ian and {Cuello}, Nicol{\'a}s and {Fasano}, Daniele and {Flock}, Mario and {Fukagawa}, Misato and {Garg}, Himanshi and {Hall}, Cassandra and {Hammond}, Iain and {Hardiman}, Caitlyn and {Hilder}, Thomas and {Huang}, Jane and {Ilee}, John D. and {Isella}, Andrea and {Kanagawa}, Kazuhiro and {Lesur}, Geoffroy and {Loomis}, Ryan A. and {M{\'e}nard}, Francois and {Orihara}, Ryuta and {Pinte}, Christophe and {Price}, Daniel and {Testi}, Leonardo and {Fernandez}, Gaylor Wafflard- and {W{\"o}lfer}, Lisa and {Yen}, Hsi-Wei and {Yoshida}, Tomohiro C. and {Zawadzki}, Brianna},
        title = "{exoALMA. XII. Weighing and Sizing exoALMA Disks with Rotation Curve Modelling}",
      journal = {\apjl},
         year = 2025,
        month = may,
       volume = {984},
       number = {1},
          eid = {L17},
        pages = {L17},
          doi = {10.3847/2041-8213/adc431},
archivePrefix = {arXiv},
       eprint = {2504.18726},
 primaryClass = {astro-ph.EP},
       adsurl = {https://ui.adsabs.harvard.edu/abs/2025ApJ...984L..17L}
}

@ARTICLE{loomis+2025,
       author = {{Loomis}, Ryan A. and {Facchini}, Stefano and {Benisty}, Myriam and {Curone}, Pietro and {Ilee}, John D. and {Cataldi}, Gianni and {Yen}, Hsi-Wei and {Teague}, Richard and {Pinte}, Christophe and {Huang}, Jane and {Garg}, Himanshi and {Orihara}, Ryuta and {Czekala}, Ian and {Zawadzki}, Brianna and {Andrews}, Sean M. and {Wilner}, David J. and {Bae}, Jaehan and {Barraza-Alfaro}, Marcelo and {Fasano}, Daniele and {Flock}, Mario and {Fukagawa}, Misato and {Galloway-Sprietsma}, Maria and {Izquierdo}, Andr{\'e}s F. and {Kanagawa}, Kazuhiro and {Lesur}, Geoffroy and {Longarini}, Cristiano and {Menard}, Francois and {Price}, Daniel J. and {Rosotti}, Giovanni and {Stadler}, Jochen and {Wafflard-Fernandez}, Gaylor and {W{\"o}lfer}, Lisa and {Yoshida}, Tomohiro C.},
        title = "{exoALMA. II. Data Calibration and Imaging Pipeline}",
      journal = {\apjl},
         year = 2025,
        month = may,
       volume = {984},
       number = {1},
          eid = {L7},
        pages = {L7},
          doi = {10.3847/2041-8213/adc43a},
archivePrefix = {arXiv},
       eprint = {2504.19870},
 primaryClass = {astro-ph.EP},
       adsurl = {https://ui.adsabs.harvard.edu/abs/2025ApJ...984L...7L}
}

@ARTICLE{maio+2025,
       author = {{Maio}, F. and {Fedele}, D. and {Roccatagliata}, V. and {Facchini}, S. and {Lodato}, G. and {Desidera}, S. and {Garufi}, A. and {Mesa}, D. and {Ruzza}, A. and {Toci}, C. and {Testi}, L. and {Zurlo}, A. and {Rosotti}, G.},
        title = "{Unveiling a protoplanet candidate embedded in the HD 135344B disk with VLT/ERIS}",
      journal = {\aap},
         year = 2025,
        month = jul,
       volume = {699},
          eid = {L10},
        pages = {L10},
          doi = {10.1051/0004-6361/202554472},
       adsurl = {https://ui.adsabs.harvard.edu/abs/2025A&A...699L..10M}
}

@ARTICLE{mao+2023,
       author = {{Mao}, Shunyuan and {Dong}, Ruobing and {Lu}, Lu and {Yi}, Kwang Moo and {Wang}, Sifan and {Perdikaris}, Paris},
        title = "{PPDONet: Deep Operator Networks for Fast Prediction of Steady-state Solutions in Disk-Planet Systems}",
      journal = {\apjl},
         year = 2023,
        month = jun,
       volume = {950},
       number = {2},
          eid = {L12},
        pages = {L12},
          doi = {10.3847/2041-8213/acd77f},
archivePrefix = {arXiv},
       eprint = {2305.11111},
 primaryClass = {astro-ph.EP},
       adsurl = {https://ui.adsabs.harvard.edu/abs/2023ApJ...950L..12M}
}

@ARTICLE{mignone+2007,
       author = {{Mignone}, A. and {Bodo}, G. and {Massaglia}, S. and {Matsakos}, T. and {Tesileanu}, O. and {Zanni}, C. and {Ferrari}, A.},
        title = "{PLUTO: A Numerical Code for Computational Astrophysics}",
      journal = {\apjs},
         year = 2007,
        month = may,
       volume = {170},
       number = {1},
        pages = {228-242},
          doi = {10.1086/513316},
archivePrefix = {arXiv},
       eprint = {astro-ph/0701854},
 primaryClass = {astro-ph},
       adsurl = {https://ui.adsabs.harvard.edu/abs/2007ApJS..170..228M}
}

@INPROCEEDINGS{miotello+2023,
       author = {{Miotello}, A. and {Kamp}, I. and {Birnstiel}, T. and {Cleeves}, L.~C. and {Kataoka}, A.},
        title = "{Setting the Stage for Planet Formation: Measurements and Implications of the Fundamental Disk Properties}",
    booktitle = {Astronomical Society of the Pacific Conference Series},
         year = 2023,
       editor = {{Inutsuka}, S. and {Aikawa}, Y. and {Muto}, T. and {Tomida}, K. and {Tamura}, M.},
       series = {Astronomical Society of the Pacific Conference Series},
       volume = {534},
        month = jul,
        pages = {501},
          doi = {10.48550/arXiv.2203.09818},
archivePrefix = {arXiv},
       eprint = {2203.09818},
 primaryClass = {astro-ph.EP},
       adsurl = {https://ui.adsabs.harvard.edu/abs/2023ASPC..534..501M}
}

@ARTICLE{muto+2012,
       author = {{Muto}, T. and {Grady}, C.~A. and {Hashimoto}, J. and {Fukagawa}, M. and {Hornbeck}, J.~B. and {Sitko}, M. and {Russell}, R. and {Werren}, C. and {Cur{\'e}}, M. and {Currie}, T. and {Ohashi}, N. and {Okamoto}, Y. and {Momose}, M. and {Honda}, M. and {Inutsuka}, S. and {Takeuchi}, T. and {Dong}, R. and {Abe}, L. and {Brandner}, W. and {Brandt}, T. and {Carson}, J. and {Egner}, S. and {Feldt}, M. and {Fukue}, T. and {Goto}, M. and {Guyon}, O. and {Hayano}, Y. and {Hayashi}, M. and {Hayashi}, S. and {Henning}, T. and {Hodapp}, K.~W. and {Ishii}, M. and {Iye}, M. and {Janson}, M. and {Kandori}, R. and {Knapp}, G.~R. and {Kudo}, T. and {Kusakabe}, N. and {Kuzuhara}, M. and {Matsuo}, T. and {Mayama}, S. and {McElwain}, M.~W. and {Miyama}, S. and {Morino}, J. -I. and {Moro-Martin}, A. and {Nishimura}, T. and {Pyo}, T. -S. and {Serabyn}, E. and {Suto}, H. and {Suzuki}, R. and {Takami}, M. and {Takato}, N. and {Terada}, H. and {Thalmann}, C. and {Tomono}, D. and {Turner}, E.~L. and {Watanabe}, M. and {Wisniewski}, J.~P. and {Yamada}, T. and {Takami}, H. and {Usuda}, T. and {Tamura}, M.},
        title = "{Discovery of Small-scale Spiral Structures in the Disk of SAO 206462 (HD 135344B): Implications for the Physical State of the Disk from Spiral Density Wave Theory}",
      journal = {\apjl},
         year = 2012,
        month = apr,
       volume = {748},
       number = {2},
          eid = {L22},
        pages = {L22},
          doi = {10.1088/2041-8205/748/2/L22},
archivePrefix = {arXiv},
       eprint = {1202.6139},
 primaryClass = {astro-ph.EP},
       adsurl = {https://ui.adsabs.harvard.edu/abs/2012ApJ...748L..22M}
}

@ARTICLE{oberg+2021,
       author = {{{\"O}berg}, Karin I. and {Guzm{\'a}n}, Viviana V. and {Walsh}, Catherine and {Aikawa}, Yuri and {Bergin}, Edwin A. and {Law}, Charles J. and {Loomis}, Ryan A. and {Alarc{\'o}n}, Felipe and {Andrews}, Sean M. and {Bae}, Jaehan and {Bergner}, Jennifer B. and {Boehler}, Yann and {Booth}, Alice S. and {Bosman}, Arthur D. and {Calahan}, Jenny K. and {Cataldi}, Gianni and {Cleeves}, L. Ilsedore and {Czekala}, Ian and {Furuya}, Kenji and {Huang}, Jane and {Ilee}, John D. and {Kurtovic}, Nicolas T. and {Le Gal}, Romane and {Liu}, Yao and {Long}, Feng and {M{\'e}nard}, Fran{\c{c}}ois and {Nomura}, Hideko and {P{\'e}rez}, Laura M. and {Qi}, Chunhua and {Schwarz}, Kamber R. and {Sierra}, Anibal and {Teague}, Richard and {Tsukagoshi}, Takashi and {Yamato}, Yoshihide and {van't Hoff}, Merel L.~R. and {Waggoner}, Abygail R. and {Wilner}, David J. and {Zhang}, Ke},
        title = "{Molecules with ALMA at Planet-forming Scales (MAPS). I. Program Overview and Highlights}",
      journal = {\apjs},
         year = 2021,
        month = nov,
       volume = {257},
       number = {1},
          eid = {1},
        pages = {1},
          doi = {10.3847/1538-4365/ac1432},
archivePrefix = {arXiv},
       eprint = {2109.06268},
 primaryClass = {astro-ph.EP},
       adsurl = {https://ui.adsabs.harvard.edu/abs/2021ApJS..257....1O}
}

@ARTICLE{oberg+2023,
       author = {{{\"O}berg}, Karin I. and {Facchini}, Stefano and {Anderson}, Dana E.},
        title = "{Protoplanetary Disk Chemistry}",
      journal = {\araa},
         year = 2023,
        month = aug,
       volume = {61},
        pages = {287-328},
          doi = {10.1146/annurev-astro-022823-040820},
archivePrefix = {arXiv},
       eprint = {2309.05685},
 primaryClass = {astro-ph.EP},
       adsurl = {https://ui.adsabs.harvard.edu/abs/2023ARA&A..61..287O}
}

@ARTICLE{perez+2018,
       author = {{P{\'e}rez}, Sebasti{\'a}n and {Casassus}, S. and {Ben{\'\i}tez-Llambay}, P.},
        title = "{Observability of planet-disc interactions in CO kinematics}",
      journal = {\mnras},
         year = 2018,
        month = oct,
       volume = {480},
       number = {1},
        pages = {L12-L17},
          doi = {10.1093/mnrasl/sly109},
archivePrefix = {arXiv},
       eprint = {1806.05125},
 primaryClass = {astro-ph.EP},
       adsurl = {https://ui.adsabs.harvard.edu/abs/2018MNRAS.480L..12P}
}

@ARTICLE{pinte+2018_kink,
       author = {{Pinte}, C. and {Price}, D.~J. and {M{\'e}nard}, F. and {Duch{\^e}ne}, G. and {Dent}, W.~R.~F. and {Hill}, T. and {de Gregorio-Monsalvo}, I. and {Hales}, A. and {Mentiplay}, D.},
        title = "{Kinematic Evidence for an Embedded Protoplanet in a Circumstellar Disk}",
      journal = {\apjl},
         year = 2018,
        month = jun,
       volume = {860},
       number = {1},
          eid = {L13},
        pages = {L13},
          doi = {10.3847/2041-8213/aac6dc},
archivePrefix = {arXiv},
       eprint = {1805.10293},
 primaryClass = {astro-ph.SR},
       adsurl = {https://ui.adsabs.harvard.edu/abs/2018ApJ...860L..13P}
}

@ARTICLE{pinte+2019,
       author = {{Pinte}, C. and {van der Plas}, G. and {M{\'e}nard}, F. and
         {Price}, D.~J. and {Christiaens}, V. and {Hill}, T. and
         {Mentiplay}, D. and {Ginski}, C. and {Choquet}, E. and {Boehler}, Y. and
         {Duch{\^e}ne}, G. and {Perez}, S. and {Casassus}, S.},
        title = "{Kinematic detection of a planet carving a gap in a protoplanetary disk}",
      journal = {Nature Astronomy},
         year = 2019,
        month = aug,
       volume = {3},
        pages = {1109-1114},
          doi = {10.1038/s41550-019-0852-6},
archivePrefix = {arXiv},
       eprint = {1907.02538},
 primaryClass = {astro-ph.SR},
       adsurl = {https://ui.adsabs.harvard.edu/abs/2019NatAs...3.1109P}
}

@INPROCEEDINGS{pinte+2023,
       author = {{Pinte}, C. and {Teague}, R. and {Flaherty}, K. and {Hall}, C. and {Facchini}, S. and {Casassus}, S.},
        title = "{Kinematic Structures in Planet-Forming Disks}",
    booktitle = {Astronomical Society of the Pacific Conference Series},
         year = 2023,
       editor = {{Inutsuka}, S. and {Aikawa}, Y. and {Muto}, T. and {Tomida}, K. and {Tamura}, M.},
       series = {Astronomical Society of the Pacific Conference Series},
       volume = {534},
        month = jul,
        pages = {645},
       adsurl = {https://ui.adsabs.harvard.edu/abs/2023ASPC..534..645P}
}

@ARTICLE{pinte+2025,
       author = {{Pinte}, Christophe and {Ilee}, John D. and {Huang}, Jane and {Benisty}, Myriam and {Facchini}, Stefano and {Fukagawa}, Misato and {Teague}, Richard and {Bae}, Jaehan and {Barraza-Alfaro}, Marcelo and {Cataldi}, Gianni and {Cuello}, Nicol{\'a}s and {Curone}, Pietro and {Czekala}, Ian and {Fasano}, Daniele and {Flock}, Mario and {Galloway-Sprietsma}, Maria and {Garg}, Himanshi and {Hall}, Cassandra and {Hammond}, Iain and {Hardiman}, Caitlyn and {Hilder}, Thomas and {Izquierdo}, Andr{\'e}s F. and {Kanagawa}, Kazuhiro and {Lesur}, Geoffroy and {Lodato}, Giuseppe and {Longarini}, Cristiano and {Loomis}, Ryan A. and {Masset}, Fr{\'e}d{\'e}ric and {Menard}, Francois and {Orihara}, Ryuta and {Price}, Daniel J. and {Rosotti}, Giovanni and {Stadler}, Jochen and {Yen}, Hsi-Wei and {Wafflard-Fernandez}, Gaylor and {Wilner}, David J. and {Winter}, Andrew J. and {W{\"o}lfer}, Lisa and {Yoshida}, Tomohiro C. and {Zawadzki}, Brianna},
        title = "{exoALMA. X. Channel Maps Reveal Complex $^{12}$CO Abundance Distributions and a Variety of Kinematic Structures with Evidence for Embedded Planets}",
      journal = {\apjl},
         year = 2025,
        month = may,
       volume = {984},
       number = {1},
          eid = {L15},
        pages = {L15},
          doi = {10.3847/2041-8213/adc433},
archivePrefix = {arXiv},
       eprint = {2504.18717},
 primaryClass = {astro-ph.EP},
       adsurl = {https://ui.adsabs.harvard.edu/abs/2025ApJ...984L..15P}
}

@ARTICLE{rabago+2021,
       author = {{Rabago}, Ian and {Zhu}, Zhaohuan},
        title = "{Constraining protoplanetary disc accretion and young planets using ALMA kinematic observations}",
      journal = {\mnras},
         year = 2021,
        month = apr,
       volume = {502},
       number = {4},
        pages = {5325-5339},
          doi = {10.1093/mnras/stab447},
archivePrefix = {arXiv},
       eprint = {2102.03007},
 primaryClass = {astro-ph.EP},
       adsurl = {https://ui.adsabs.harvard.edu/abs/2021MNRAS.502.5325R}
}

@ARTICLE{ragusa+2024,
       author = {{Ragusa}, Enrico and {Lynch}, Elliot and {Laibe}, Guillaume and {Longarini}, Cristiano and {Ceppi}, Simone},
        title = "{Probing the eccentricity in protostellar discs: Modelling kinematics and morphologies}",
      journal = {\aap},
         year = 2024,
        month = jun,
       volume = {686},
          eid = {A264},
        pages = {A264},
          doi = {10.1051/0004-6361/202449583},
archivePrefix = {arXiv},
       eprint = {2404.02958},
 primaryClass = {astro-ph.EP},
       adsurl = {https://ui.adsabs.harvard.edu/abs/2024A&A...686A.264R}
}

@ARTICLE{reggiani+2018,
       author = {{Reggiani}, M. and {Christiaens}, V. and {Absil}, O. and {Mawet}, D. and {Huby}, E. and {Choquet}, E. and {Gomez Gonzalez}, C.~A. and {Ruane}, G. and {Femenia}, B. and {Serabyn}, E. and {Matthews}, K. and {Barraza}, M. and {Carlomagno}, B. and {Defr{\`e}re}, D. and {Delacroix}, C. and {Habraken}, S. and {Jolivet}, A. and {Karlsson}, M. and {Orban de Xivry}, G. and {Piron}, P. and {Surdej}, J. and {Vargas Catalan}, E. and {Wertz}, O.},
        title = "{Discovery of a point-like source and a third spiral arm in the transition disk around the Herbig Ae star MWC 758}",
      journal = {\aap},
         year = 2018,
        month = mar,
       volume = {611},
          eid = {A74},
        pages = {A74},
          doi = {10.1051/0004-6361/201732016},
archivePrefix = {arXiv},
       eprint = {1710.11393},
 primaryClass = {astro-ph.EP},
       adsurl = {https://ui.adsabs.harvard.edu/abs/2018A&A...611A..74R}
}

@ARTICLE{ren+2020,
       author = {{Ren}, Bin and {Dong}, Ruobing and {van Holstein}, Rob G. and {Ruffio}, Jean-Baptiste and {Calvin}, Benjamin A. and {Girard}, Julien H. and {Benisty}, Myriam and {Boccaletti}, Anthony and {Esposito}, Thomas M. and {Choquet}, {\'E}lodie and {Mawet}, Dimitri and {Pueyo}, Laurent and {Stolker}, Tomas and {Chiang}, Eugenede and {Boer}, Jozua and {Debes}, John H. and {Garufi}, Antonio and {Grady}, Carol A. and {Hines}, Dean C. and {Maire}, Anne-Lise and {M{\'e}nard}, Fran{\c{c}}ois and {Millar-Blanchaer}, Maxwell A. and {Perrin}, Marshall D. and {Poteet}, Charles A. and {Schneider}, Glenn},
        title = "{Dynamical Evidence of a Spiral Arm-driving Planet in the MWC 758 Protoplanetary Disk}",
      journal = {\apjl},
         year = 2020,
        month = aug,
       volume = {898},
       number = {2},
          eid = {L38},
        pages = {L38},
          doi = {10.3847/2041-8213/aba43e},
archivePrefix = {arXiv},
       eprint = {2007.04980},
 primaryClass = {astro-ph.EP},
       adsurl = {https://ui.adsabs.harvard.edu/abs/2020ApJ...898L..38R}
}

@ARTICLE{ren+2023,
       author = {{Ren}, Bin B. and {Benisty}, Myriam and {Ginski}, Christian and {Tazaki}, Ryo and {Wallack}, Nicole L. and {Milli}, Julien and {Garufi}, Antonio and {Bae}, Jaehan and {Facchini}, Stefano and {M{\'e}nard}, Fran{\c{c}}ois and {Pinilla}, Paola and {Swastik}, C. and {Teague}, Richard and {Wahhaj}, Zahed},
        title = "{Protoplanetary disks in K$_{s}$-band total intensity and polarized light}",
      journal = {\aap},
         year = 2023,
        month = dec,
       volume = {680},
          eid = {A114},
        pages = {A114},
          doi = {10.1051/0004-6361/202347353},
archivePrefix = {arXiv},
       eprint = {2310.08589},
 primaryClass = {astro-ph.EP},
       adsurl = {https://ui.adsabs.harvard.edu/abs/2023A&A...680A.114R}
}

@ARTICLE{rosotti+2020,
       author = {{Rosotti}, Giovanni P. and {Teague}, Richard and {Dullemond}, Cornelis and {Booth}, Richard A. and {Clarke}, Cathie J.},
        title = "{The efficiency of dust trapping in ringed protoplanetary discs}",
      journal = {\mnras},
         year = 2020,
        month = jun,
       volume = {495},
       number = {1},
        pages = {173-181},
          doi = {10.1093/mnras/staa1170},
archivePrefix = {arXiv},
       eprint = {2004.11394},
 primaryClass = {astro-ph.EP},
       adsurl = {https://ui.adsabs.harvard.edu/abs/2020MNRAS.495..173R}
}

@ARTICLE{speedie+2024,
       author = {{Speedie}, Jessica and {Dong}, Ruobing and {Hall}, Cassandra and {Longarini}, Cristiano and {Veronesi}, Benedetta and {Paneque-Carre{\~n}o}, Teresa and {Lodato}, Giuseppe and {Tang}, Ya-Wen and {Teague}, Richard and {Hashimoto}, Jun},
        title = "{Gravitational instability in a planet-forming disk}",
      journal = {\nat},
         year = 2024,
        month = sep,
       volume = {633},
       number = {8028},
        pages = {58-62},
          doi = {10.1038/s41586-024-07877-0},
archivePrefix = {arXiv},
       eprint = {2409.02196},
 primaryClass = {astro-ph.EP},
       adsurl = {https://ui.adsabs.harvard.edu/abs/2024Natur.633...58S}
}

@ARTICLE{stadler+2023,
       author = {{Stadler}, J. and {Benisty}, M. and {Izquierdo}, A. and {Facchini}, S. and {Teague}, R. and {Kurtovic}, N. and {Pinilla}, P. and {Bae}, J. and {Ansdell}, M. and {Loomis}, R. and {Mayama}, S. and {Perez}, L.~M. and {Testi}, L.},
        title = "{A kinematically detected planet candidate in a transition disk}",
      journal = {\aap},
         year = 2023,
        month = feb,
       volume = {670},
          eid = {L1},
        pages = {L1},
          doi = {10.1051/0004-6361/202245381},
archivePrefix = {arXiv},
       eprint = {2301.01684},
 primaryClass = {astro-ph.EP},
       adsurl = {https://ui.adsabs.harvard.edu/abs/2023A&A...670L...1S}
}

@ARTICLE{stadler+2025,
       author = {{Stadler}, Jochen and {Benisty}, Myriam and {Winter}, Andrew J. and {Izquierdo}, Andr{\'e}s F. and {Longarini}, Cristiano and {Galloway-Sprietsma}, Maria and {Curone}, Pietro and {Andrews}, Sean M. and {Bae}, Jaehan and {Facchini}, Stefano and {Rosotti}, Giovanni and {Teague}, Richard and {Barraza-Alfaro}, Marcelo and {Cataldi}, Gianni and {Cuello}, Nicol{\'a}s and {Czekala}, Ian and {Fasano}, Daniele and {Flock}, Mario and {Fukagawa}, Misato and {Garg}, Himanshi and {Hall}, Cassandra and {Hammond}, Iain and {Hilder}, Thomas and {Huang}, Jane and {Ilee}, John D. and {Kanagawa}, Kazuhiro and {Lesur}, Geoffroy and {Lodato}, Giuseppe and {Loomis}, Ryan A. and {Menard}, Francois and {Orihara}, Ryuta and {Pinte}, Christophe and {Price}, Daniel J. and {Yen}, Hsi-Wei and {Wafflard-Fernandez}, Gaylor and {Wilner}, David J. and {W{\"o}lfer}, Lisa and {Yoshida}, Tomohiro C. and {Zawadzki}, Brianna},
        title = "{exoALMA. VI. Rotating under Pressure: Rotation Curves, Azimuthal Velocity Substructures, and Gas Pressure Variations}",
      journal = {\apjl},
         year = 2025,
        month = may,
       volume = {984},
       number = {1},
          eid = {L11},
        pages = {L11},
          doi = {10.3847/2041-8213/adb152},
archivePrefix = {arXiv},
       eprint = {2504.20036},
 primaryClass = {astro-ph.EP},
       adsurl = {https://ui.adsabs.harvard.edu/abs/2025ApJ...984L..11S}
}

@ARTICLE{stolker+2016,
       author = {{Stolker}, T. and {Dominik}, C. and {Avenhaus}, H. and {Min}, M. and {de Boer}, J. and {Ginski}, C. and {Schmid}, H.~M. and {Juhasz}, A. and {Bazzon}, A. and {Waters}, L.~B.~F.~M. and {Garufi}, A. and {Augereau}, J. -C. and {Benisty}, M. and {Boccaletti}, A. and {Henning}, Th. and {Langlois}, M. and {Maire}, A. -L. and {M{\'e}nard}, F. and {Meyer}, M.~R. and {Pinte}, C. and {Quanz}, S.~P. and {Thalmann}, C. and {Beuzit}, J. -L. and {Carbillet}, M. and {Costille}, A. and {Dohlen}, K. and {Feldt}, M. and {Gisler}, D. and {Mouillet}, D. and {Pavlov}, A. and {Perret}, D. and {Petit}, C. and {Pragt}, J. and {Rochat}, S. and {Roelfsema}, R. and {Salasnich}, B. and {Soenke}, C. and {Wildi}, F.},
        title = "{Shadows cast on the transition disk of HD 135344B. Multiwavelength VLT/SPHERE polarimetric differential imaging}",
      journal = {\aap},
         year = 2016,
        month = nov,
       volume = {595},
          eid = {A113},
        pages = {A113},
          doi = {10.1051/0004-6361/201528039},
archivePrefix = {arXiv},
       eprint = {1603.00481},
 primaryClass = {astro-ph.EP},
       adsurl = {https://ui.adsabs.harvard.edu/abs/2016A&A...595A.113S}
}

@ARTICLE{teague+2018_bettermoments,
       author = {{Teague}, Richard and {Foreman-Mackey}, Daniel},
        title = "{A Robust Method to Measure Centroids of Spectral Lines}",
      journal = {Research Notes of the American Astronomical Society},
         year = 2018,
        month = sep,
       volume = {2},
       number = {3},
          eid = {173},
        pages = {173},
          doi = {10.3847/2515-5172/aae265},
archivePrefix = {arXiv},
       eprint = {1809.10295},
 primaryClass = {astro-ph.IM},
       adsurl = {https://ui.adsabs.harvard.edu/abs/2018RNAAS...2..173T}
}

@ARTICLE{teague+2018a,
       author = {{Teague}, Richard and {Bae}, Jaehan and {Bergin}, Edwin A. and
         {Birnstiel}, Tilman and {Foreman-Mackey}, Daniel},
        title = "{A Kinematical Detection of Two Embedded Jupiter-mass Planets in HD 163296}",
      journal = {\apjl},
         year = 2018,
        month = jun,
       volume = {860},
       number = {1},
          eid = {L12},
        pages = {L12},
          doi = {10.3847/2041-8213/aac6d7},
archivePrefix = {arXiv},
       eprint = {1805.10290},
 primaryClass = {astro-ph.EP},
       adsurl = {https://ui.adsabs.harvard.edu/abs/2018ApJ...860L..12T}
}

@ARTICLE{teague+2019nat,
       author = {{Teague}, Richard and {Bae}, Jaehan and {Bergin}, Edwin A.},
        title = "{Meridional flows in the disk around a young star}",
      journal = {\nat},
         year = 2019,
        month = oct,
       volume = {574},
       number = {7778},
        pages = {378-381},
          doi = {10.1038/s41586-019-1642-0},
archivePrefix = {arXiv},
       eprint = {1910.06980},
 primaryClass = {astro-ph.EP},
       adsurl = {https://ui.adsabs.harvard.edu/abs/2019Natur.574..378T}
}

@ARTICLE{teague+2021,
       author = {{Teague}, Richard and {Bae}, Jaehan and {Aikawa}, Yuri and {Andrews}, Sean M. and {Bergin}, Edwin A. and {Bergner}, Jennifer B. and {Boehler}, Yann and {Booth}, Alice S. and {Bosman}, Arthur D. and {Cataldi}, Gianni and {Czekala}, Ian and {Guzm{\'a}n}, Viviana V. and {Huang}, Jane and {Ilee}, John D. and {Law}, Charles J. and {Le Gal}, Romane and {Long}, Feng and {Loomis}, Ryan A. and {M{\'e}nard}, Fran{\c{c}}ois and {{\"O}berg}, Karin I. and {P{\'e}rez}, Laura M. and {Schwarz}, Kamber R. and {Sierra}, Anibal and {Walsh}, Catherine and {Wilner}, David J. and {Yamato}, Yoshihide and {Zhang}, Ke},
        title = "{Molecules with ALMA at Planet-forming Scales (MAPS). XVIII. Kinematic Substructures in the Disks of HD 163296 and MWC 480}",
      journal = {\apjs},
         year = 2021,
        month = nov,
       volume = {257},
       number = {1},
          eid = {18},
        pages = {18},
          doi = {10.3847/1538-4365/ac1438},
archivePrefix = {arXiv},
       eprint = {2109.06218},
 primaryClass = {astro-ph.EP},
       adsurl = {https://ui.adsabs.harvard.edu/abs/2021ApJS..257...18T}
}

@ARTICLE{teague+2025,
       author = {{Teague}, Richard and {Benisty}, Myriam and {Facchini}, Stefano and {Fukagawa}, Misato and {Pinte}, Christophe and {Andrews}, Sean M. and {Bae}, Jaehan and {Barraza-Alfaro}, Marcelo and {Cataldi}, Gianni and {Cuello}, Nicol{\'a}s and {Curone}, Pietro and {Czekala}, Ian and {Fasano}, Daniele and {Flock}, Mario and {Galloway-Sprietsma}, Maria and {Garg}, Himanshi and {Hall}, Cassandra and {Hammond}, Iain and {Hilder}, Thomas and {Huang}, Jane and {Ilee}, John D. and {Izquierdo}, Andr{\'e}s F. and {Kanagawa}, Kazuhiro and {Lesur}, Geoffroy and {Lodato}, Giuseppe and {Longarini}, Cristiano and {Loomis}, Ryan A. and {Masset}, Fr{\'e}d{\'e}ric and {Menard}, Francois and {Orihara}, Ryuta and {Price}, Daniel J. and {Rosotti}, Giovanni and {Stadler}, Jochen and {Testi}, Leonardo and {Yen}, Hsi-Wei and {Wafflard-Fernandez}, Gaylor and {Wilner}, David J. and {Winter}, Andrew J. and {W{\"o}lfer}, Lisa and {Yoshida}, Tomohiro C. and {Zawadzki}, Brianna},
        title = "{exoALMA. I. Science Goals, Project Design, and Data Products}",
      journal = {\apjl},
         year = 2025,
        month = may,
       volume = {984},
       number = {1},
          eid = {L6},
        pages = {L6},
          doi = {10.3847/2041-8213/adc43b},
archivePrefix = {arXiv},
       eprint = {2504.18688},
 primaryClass = {astro-ph.EP},
       adsurl = {https://ui.adsabs.harvard.edu/abs/2025ApJ...984L...6T}
}

@ARTICLE{vandermarel+2016,
       author = {{van der Marel}, N. and {Cazzoletti}, P. and {Pinilla}, P. and {Garufi}, A.},
        title = "{Vortices and Spirals in the HD135344B Transition Disk}",
      journal = {\apj},
         year = 2016,
        month = dec,
       volume = {832},
       number = {2},
          eid = {178},
        pages = {178},
          doi = {10.3847/0004-637X/832/2/178},
archivePrefix = {arXiv},
       eprint = {1607.05775},
 primaryClass = {astro-ph.EP},
       adsurl = {https://ui.adsabs.harvard.edu/abs/2016ApJ...832..178V}
}

@ARTICLE{wagner+2023,
       author = {{Wagner}, Kevin and {Stone}, Jordan and {Skemer}, Andrew and {Ertel}, Steve and {Dong}, Ruobing and {Apai}, D{\'a}niel and {Spalding}, Eckhart and {Leisenring}, Jarron and {Sitko}, Michael and {Kratter}, Kaitlin and {Barman}, Travis and {Marley}, Mark and {Miles}, Brittany and {Boccaletti}, Anthony and {Assani}, Korash and {Bayyari}, Ammar and {Uyama}, Taichi and {Woodward}, Charles E. and {Hinz}, Phil and {Briesemeister}, Zackery and {Lawson}, Kellen and {M{\'e}nard}, Fran{\c{c}}ois and {Pantin}, Eric and {Russell}, Ray W. and {Skrutskie}, Michael and {Wisniewski}, John},
        title = "{Direct images and spectroscopy of a giant protoplanet driving spiral arms in MWC 758}",
      journal = {Nature Astronomy},
         year = 2023,
        month = jul,
          doi = {10.1038/s41550-023-02028-3},
archivePrefix = {arXiv},
       eprint = {2307.04021},
 primaryClass = {astro-ph.EP},
       adsurl = {https://ui.adsabs.harvard.edu/abs/2023NatAs.tmp..146W}
}

@ARTICLE{winter+2025,
       author = {{Winter}, Andrew J. and {Benisty}, Myriam and {Izquierdo}, Andr{\'e}s F. and {Lodato}, Giuseppe and {Teague}, Richard and {Kimmig}, Carolin N. and {Andrews}, Sean M. and {Bae}, Jaehan and {Barraza-Alfaro}, Marcelo and {Cuello}, Nicol{\'a}s and {Curone}, Pietro and {Czekala}, Ian and {Facchini}, Stefano and {Fasano}, Daniele and {Hall}, Cassandra and {Hardiman}, Caitlyn and {Hilder}, Thomas and {Ilee}, John D. and {Fukagawa}, Misato and {Longarini}, Cristiano and {M{\'e}nard}, Fran{\c{c}}ois and {Orihara}, Ryuta and {Pinte}, Christophe and {Price}, Daniel J. and {Rosotti}, Giovanni and {Stadler}, Jochen and {Wilner}, David J. and {W{\"o}lfer}, Lisa and {Yen}, Hsi-Wei and {Yoshida}, Tomohiro C. and {Zawadzki}, Brianna},
        title = "{exoALMA. XVIII. Interpreting Large-scale Kinematic Structures as Moderate Warping}",
      journal = {\apjl},
         year = 2025,
        month = sep,
       volume = {990},
       number = {1},
          eid = {L10},
        pages = {L10},
          doi = {10.3847/2041-8213/adf113},
archivePrefix = {arXiv},
       eprint = {2507.11669},
 primaryClass = {astro-ph.EP},
       adsurl = {https://ui.adsabs.harvard.edu/abs/2025ApJ...990L..10W}
}

@ARTICLE{woelfer+2023,
       author = {{W{\"o}lfer}, L. and {Facchini}, S. and {van der Marel}, N. and {van Dishoeck}, E.~F. and {Benisty}, M. and {Bohn}, A.~J. and {Francis}, L. and {Izquierdo}, A.~F. and {Teague}, R.~D.},
        title = "{Kinematics and brightness temperatures of transition discs. A survey of gas substructures as seen with ALMA}",
      journal = {\aap},
         year = 2023,
        month = feb,
       volume = {670},
          eid = {A154},
        pages = {A154},
          doi = {10.1051/0004-6361/202243601},
archivePrefix = {arXiv},
       eprint = {2208.09494},
 primaryClass = {astro-ph.EP},
       adsurl = {https://ui.adsabs.harvard.edu/abs/2023A&A...670A.154W}
}

@ARTICLE{woelfer+2025,
       author = {{W{\"o}lfer}, Lisa and {Barraza-Alfaro}, Marcelo and {Teague}, Richard and {Curone}, Pietro and {Benisty}, Myriam and {Fukagawa}, Misato and {Bae}, Jaehan and {Cataldi}, Gianni and {Czekala}, Ian and {Facchini}, Stefano and {Fasano}, Daniele and {Flock}, Mario and {Galloway-Sprietsma}, Maria and {Garg}, Himanshi and {Hall}, Cassandra and {Huang}, Jane and {Ilee}, John D. and {Izquierdo}, Andr{\'e}s F. and {Kanagawa}, Kazuhiro and {Lesur}, Geoffroy and {Longarini}, Cristiano and {Loomis}, Ryan A. and {Menard}, Francois and {Nath}, Anika and {Orihara}, Ryuta and {Pinte}, Christophe and {Price}, Daniel J. and {Rosotti}, Giovanni and {Stadler}, Jochen and {Wafflard-Fernandez}, Gaylor and {Winter}, Andrew J. and {Yen}, Hsi-Wei and {Yoshida}, Tomohiro C. and {Zawadzki}, Brianna},
        title = "{exoALMA. XVII. Characterizing the Gas Dynamics around Dust Asymmetries}",
      journal = {\apjl},
         year = 2025,
        month = may,
       volume = {984},
       number = {1},
          eid = {L22},
        pages = {L22},
          doi = {10.3847/2041-8213/adc42c},
archivePrefix = {arXiv},
       eprint = {2504.20023},
 primaryClass = {astro-ph.EP},
       adsurl = {https://ui.adsabs.harvard.edu/abs/2025ApJ...984L..22W}
}

@ARTICLE{xie+2024,
       author = {{Xie}, Chen and {Xie}, Chengyan and {Ren}, Bin B. and {Benisty}, Myriam and {Ginski}, Christian and {Fang}, Taotao and {Casassus}, Simon and {Bae}, Jaehan and {Facchini}, Stefano and {M{\'e}nard}, Fran{\c{c}}ois and {van Holstein}, Rob G.},
        title = "{Disk Evolution Study Through Imaging of Nearby Young Stars (DESTINYS): Dynamical Evidence of a Spiral-Arm-Driving and Gap-Opening Protoplanet from SAO 206462 Spiral Motion}",
      journal = {Universe},
         year = 2024,
        month = dec,
       volume = {10},
       number = {12},
          eid = {465},
        pages = {465},
          doi = {10.3390/universe10120465},
archivePrefix = {arXiv},
       eprint = {2412.14402},
 primaryClass = {astro-ph.EP},
       adsurl = {https://ui.adsabs.harvard.edu/abs/2024Univ...10..465X}
}

@ARTICLE{casa,
       author = {{CASA Team} and {Bean}, Ben and {Bhatnagar}, Sanjay and {Castro}, Sandra and {Donovan Meyer}, Jennifer and {Emonts}, Bjorn and {Garcia}, Enrique and {Garwood}, Robert and {Golap}, Kumar and {Gonzalez Villalba}, Justo and {Harris}, Pamela and {Hayashi}, Yohei and {Hoskins}, Josh and {Hsieh}, Mingyu and {Jagannathan}, Preshanth and {Kawasaki}, Wataru and {Keimpema}, Aard and {Kettenis}, Mark and {Lopez}, Jorge and {Marvil}, Joshua and {Masters}, Joseph and {McNichols}, Andrew and {Mehringer}, David and {Miel}, Renaud and {Moellenbrock}, George and {Montesino}, Federico and {Nakazato}, Takeshi and {Ott}, Juergen and {Petry}, Dirk and {Pokorny}, Martin and {Raba}, Ryan and {Rau}, Urvashi and {Schiebel}, Darrell and {Schweighart}, Neal and {Sekhar}, Srikrishna and {Shimada}, Kazuhiko and {Small}, Des and {Steeb}, Jan-Willem and {Sugimoto}, Kanako and {Suoranta}, Ville and {Tsutsumi}, Takahiro and {van Bemmel}, Ilse M. and {Verkouter}, Marjolein and {Wells}, Akeem and {Xiong}, Wei and {Szomoru}, Arpad and {Griffith}, Morgan and {Glendenning}, Brian and {Kern}, Jeff},
        title = "{CASA, the Common Astronomy Software Applications for Radio Astronomy}",
      journal = {\pasp},
         year = 2022,
        month = nov,
       volume = {134},
       number = {1041},
          eid = {114501},
        pages = {114501},
          doi = {10.1088/1538-3873/ac9642},
archivePrefix = {arXiv},
       eprint = {2210.02276},
 primaryClass = {astro-ph.IM},
       adsurl = {https://ui.adsabs.harvard.edu/abs/2022PASP..134k4501C}
}

@ARTICLE{Hunter_mpl,
  author={Hunter, John D.},
  journal={Computing in Science   Engineering}, 
  title={Matplotlib: A 2D Graphics Environment}, 
  year={2007},
  volume={9},
  number={3},
  pages={90-95},
  doi={10.1109/MCSE.2007.55}}

@Article{harris_np,
 title         = {Array programming with {NumPy}},
 author        = {Charles R. Harris and K. Jarrod Millman and St{\'{e}}fan J.
                 van der Walt and Ralf Gommers and Pauli Virtanen and David
                 Cournapeau and Eric Wieser and Julian Taylor and Sebastian
                 Berg and Nathaniel J. Smith and Robert Kern and Matti Picus
                 and Stephan Hoyer and Marten H. van Kerkwijk and Matthew
                 Brett and Allan Haldane and Jaime Fern{\'{a}}ndez del
                 R{\'{i}}o and Mark Wiebe and Pearu Peterson and Pierre
                 G{\'{e}}rard-Marchant and Kevin Sheppard and Tyler Reddy and
                 Warren Weckesser and Hameer Abbasi and Christoph Gohlke and
                 Travis E. Oliphant},
 year          = {2020},
 month         = sep,
 journal       = {Nature},
 volume        = {585},
 number        = {7825},
 pages         = {357--362},
 doi           = {10.1038/s41586-020-2649-2},
 publisher     = {Springer Science and Business Media {LLC}},
 url           = {https://doi.org/10.1038/s41586-020-2649-2}
}

@MISC{Virtanen_scipy,
       author = {{Virtanen}, Pauli and {Gommers}, Ralf and {Burovski}, Evgeni and {Oliphant}, Travis E. and {Weckesser}, Warren and {Cournapeau}, David and {Alexbrc} and {Peterson}, Pearu and {Reddy}, Tyler and {Wilson}, Josh and {Haberland}, Matt and {Mayorov}, Nikolay and {Endolith} and {Nelson}, Andrew and {Van Der Walt}, Stefan and {Laxalde}, Denis and {Brett}, Matthew and {Polat}, Ilhan and {Larson}, Eric and {Millman}, Jarrod and {Lars} and {Van Mulbregt}, Paul and {Eric-Jones} and {Carey}, CJ and {Moore}, Eric and {Kern}, Robert and {Leslie}, Tim and {Perktold}, Josef and {Striega}, Kai and {Feng}, Yu},
        title = "{scipy/scipy: SciPy 1.5.3}",
         year = 2020,
        month = oct,
          eid = {10.5281/zenodo.4100507},
          doi = {10.5281/zenodo.4100507},
      version = {v1.5.3},
    publisher = {Zenodo},
       adsurl = {https://ui.adsabs.harvard.edu/abs/2020zndo...4100507V}
}

@ARTICLE{Astropy_2022,
       author = {{Astropy Collaboration} and {Price-Whelan}, Adrian M. and {Lim}, Pey Lian and {Earl}, Nicholas and {Starkman}, Nathaniel and {Bradley}, Larry and {Shupe}, David L. and {Patil}, Aarya A. and {Corrales}, Lia and {Brasseur}, C.~E. and {N{\"o}the}, Maximilian and {Donath}, Axel and {Tollerud}, Erik and {Morris}, Brett M. and {Ginsburg}, Adam and {Vaher}, Eero and {Weaver}, Benjamin A. and {Tocknell}, James and {Jamieson}, William and {van Kerkwijk}, Marten H. and {Robitaille}, Thomas P. and {Merry}, Bruce and {Bachetti}, Matteo and {G{\"u}nther}, H. Moritz and {Aldcroft}, Thomas L. and {Alvarado-Montes}, Jaime A. and {Archibald}, Anne M. and {B{\'o}di}, Attila and {Bapat}, Shreyas and {Barentsen}, Geert and {Baz{\'a}n}, Juanjo and {Biswas}, Manish and {Boquien}, M{\'e}d{\'e}ric and {Burke}, D.~J. and {Cara}, Daria and {Cara}, Mihai and {Conroy}, Kyle E. and {Conseil}, Simon and {Craig}, Matthew W. and {Cross}, Robert M. and {Cruz}, Kelle L. and {D'Eugenio}, Francesco and {Dencheva}, Nadia and {Devillepoix}, Hadrien A.~R. and {Dietrich}, J{\"o}rg P. and {Eigenbrot}, Arthur Davis and {Erben}, Thomas and {Ferreira}, Leonardo and {Foreman-Mackey}, Daniel and {Fox}, Ryan and {Freij}, Nabil and {Garg}, Suyog and {Geda}, Robel and {Glattly}, Lauren and {Gondhalekar}, Yash and {Gordon}, Karl D. and {Grant}, David and {Greenfield}, Perry and {Groener}, Austen M. and {Guest}, Steve and {Gurovich}, Sebastian and {Handberg}, Rasmus and {Hart}, Akeem and {Hatfield-Dodds}, Zac and {Homeier}, Derek and {Hosseinzadeh}, Griffin and {Jenness}, Tim and {Jones}, Craig K. and {Joseph}, Prajwel and {Kalmbach}, J. Bryce and {Karamehmetoglu}, Emir and {Ka{\l}uszy{\'n}ski}, Miko{\l}aj and {Kelley}, Michael S.~P. and {Kern}, Nicholas and {Kerzendorf}, Wolfgang E. and {Koch}, Eric W. and {Kulumani}, Shankar and {Lee}, Antony and {Ly}, Chun and {Ma}, Zhiyuan and {MacBride}, Conor and {Maljaars}, Jakob M. and {Muna}, Demitri and {Murphy}, N.~A. and {Norman}, Henrik and {O'Steen}, Richard and {Oman}, Kyle A. and {Pacifici}, Camilla and {Pascual}, Sergio and {Pascual-Granado}, J. and {Patil}, Rohit R. and {Perren}, Gabriel I. and {Pickering}, Timothy E. and {Rastogi}, Tanuj and {Roulston}, Benjamin R. and {Ryan}, Daniel F. and {Rykoff}, Eli S. and {Sabater}, Jose and {Sakurikar}, Parikshit and {Salgado}, Jes{\'u}s and {Sanghi}, Aniket and {Saunders}, Nicholas and {Savchenko}, Volodymyr and {Schwardt}, Ludwig and {Seifert-Eckert}, Michael and {Shih}, Albert Y. and {Jain}, Anany Shrey and {Shukla}, Gyanendra and {Sick}, Jonathan and {Simpson}, Chris and {Singanamalla}, Sudheesh and {Singer}, Leo P. and {Singhal}, Jaladh and {Sinha}, Manodeep and {Sip{\H{o}}cz}, Brigitta M. and {Spitler}, Lee R. and {Stansby}, David and {Streicher}, Ole and {{\v{S}}umak}, Jani and {Swinbank}, John D. and {Taranu}, Dan S. and {Tewary}, Nikita and {Tremblay}, Grant R. and {de Val-Borro}, Miguel and {Van Kooten}, Samuel J. and {Vasovi{\'c}}, Zlatan and {Verma}, Shresth and {de Miranda Cardoso}, Jos{\'e} Vin{\'\i}cius and {Williams}, Peter K.~G. and {Wilson}, Tom J. and {Winkel}, Benjamin and {Wood-Vasey}, W.~M. and {Xue}, Rui and {Yoachim}, Peter and {Zhang}, Chen and {Zonca}, Andrea and {Astropy Project Contributors}},
        title = "{The Astropy Project: Sustaining and Growing a Community-oriented Open-source Project and the Latest Major Release (v5.0) of the Core Package}",
      journal = {\apj},
         year = 2022,
        month = aug,
       volume = {935},
       number = {2},
          eid = {167},
        pages = {167},
          doi = {10.3847/1538-4357/ac7c74},
archivePrefix = {arXiv},
       eprint = {2206.14220},
 primaryClass = {astro-ph.IM},
       adsurl = {https://ui.adsabs.harvard.edu/abs/2022ApJ...935..167A}
}

@ARTICLE{cmasher+2020,
       author = {{van der Velden}, Ellert},
        title = "{CMasher: Scientific colormaps for making accessible, informative and 'cmashing' plots}",
      journal = {The Journal of Open Source Software},
         year = 2020,
        month = feb,
       volume = {5},
       number = {46},
          eid = {2004},
        pages = {2004},
          doi = {10.21105/joss.02004},
archivePrefix = {arXiv},
       eprint = {2003.01069},
 primaryClass = {eess.IV},
       adsurl = {https://ui.adsabs.harvard.edu/abs/2020JOSS....5.2004V}
}

@article{scikit-image,
 title = {scikit-image: image processing in {P}ython},
 author = {van der Walt, {S}t\'efan and {S}ch\"onberger, {J}ohannes {L}. and
           {Nunez-Iglesias}, {J}uan and {B}oulogne, {F}ran\c{c}ois and {W}arner,
           {J}oshua {D}. and {Y}ager, {N}eil and {G}ouillart, {E}mmanuelle and
           {Y}u, {T}ony and the scikit-image contributors},
 year = {2014},
 month = {6},
 volume = {2},
 pages = {e453},
 journal = {PeerJ},
 issn = {2167-8359},
 url = {https://doi.org/10.7717/peerj.453},
 doi = {10.7717/peerj.453}
}

@software{radio-beam,
       author = {{Koch}, Eric and {Ginsburg}, Adam and {AKL} and {Rosolowsky}, Erik and {Sip{\H{o}}cz}, Brigitta and {Robitaille}, Thomas and {O'Brien}, Andrew and {Adamginsburg} and {Privon}, George C. and {De Val-Borro}, Miguel},
        title = "{radio-astro-tools/radio-beam: v0.3.3}",
         year = 2021,
        month = mar,
          eid = {10.5281/zenodo.4623788},
          doi = {10.5281/zenodo.4623788},
      version = {v0.3.3},
    publisher = {Zenodo},
       adsurl = {https://ui.adsabs.harvard.edu/abs/2021zndo...4623788K}
}

@software{spectral-cube,
       author = {{Ginsburg}, Adam and {Koch}, Eric and {Robitaille}, Thomas and {Beaumont}, Chris and {Adamginsburg} and {Sip{\H{o}}cz}, Brigitta and {ZuHone}, John and {Patra}, Sushobhana and {Jones}, Craig and {Lim}, P.~L. and {Stern}, Kris and {Rosolowsky}, Erik and {Earl}, Nicholas and {De Val-Borro}, Miguel and {Jrobbfed} and {Shuokong} and {Kepley}, Amanda and {Sokolov}, Vlas and {Badger}, The Gitter and {Maret}, S{\'e}bastien and {Garrido}, Juli{\'a}n and {Booker}, Joseph and {Tollerud}, Erik},
        title = "{radio-astro-tools/spectral-cube: Release v0.4.5}",
         year = 2019,
        month = nov,
          eid = {10.5281/zenodo.3558614},
          doi = {10.5281/zenodo.3558614},
      version = {v0.4.5},
    publisher = {Zenodo},
       adsurl = {https://ui.adsabs.harvard.edu/abs/2019zndo...3558614G}
}
\bibliographystyle{aasjournal}

\appendix
\label{sec:supporting_figures}

\begin{figure*}
   \centering
    \includegraphics[width=1\textwidth]{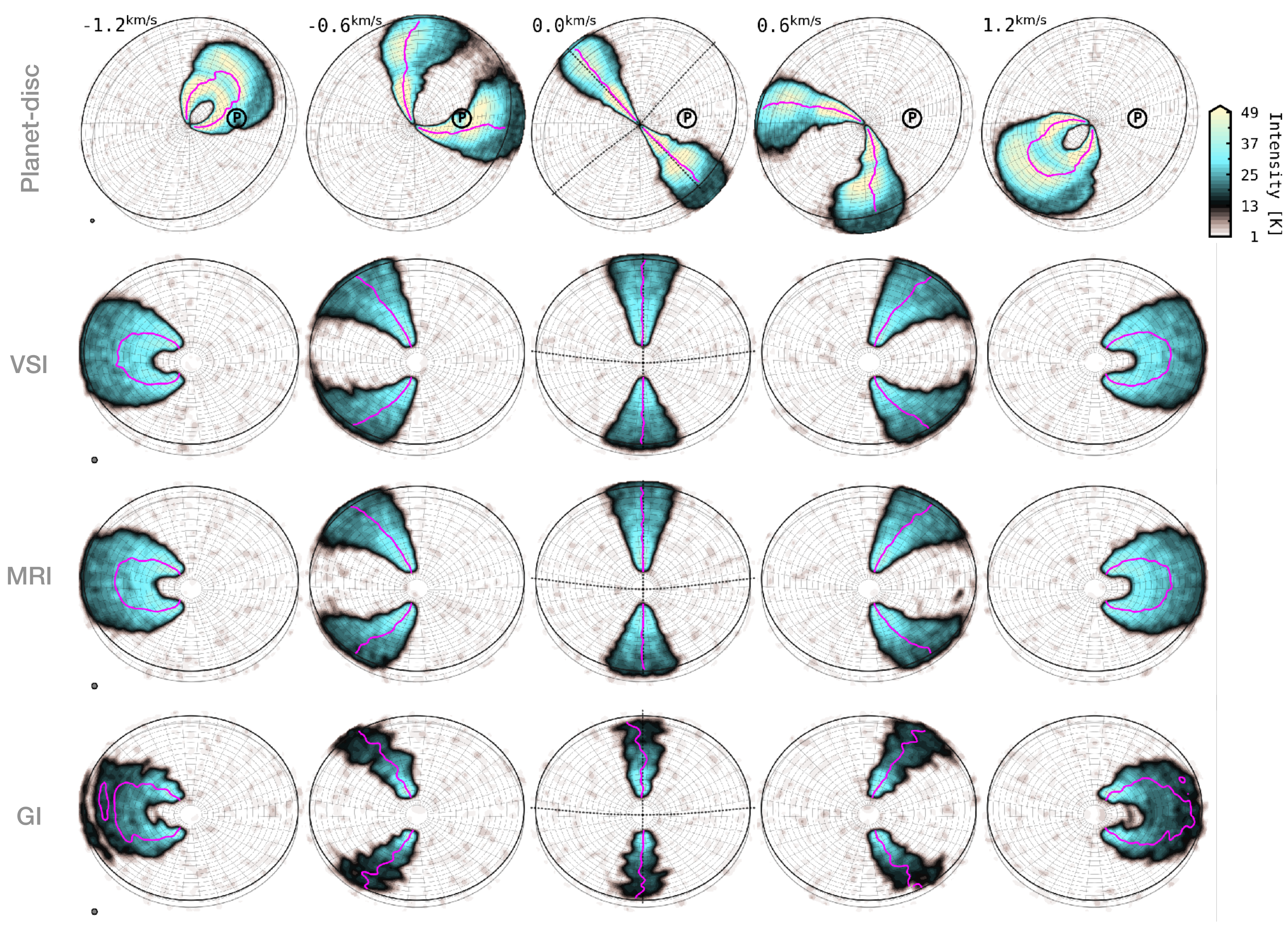}
      \caption{Selected \twCOfull{} channel maps for the physical mechanisms studied in Sect.~\ref{sec:hydro}, shown for a disc inclination of $i = -30^\circ$. The channels in the top row correspond to synthetic observations of the planet-disc interaction model with a planet-to-star mass ratio of $q = 2 \times 10^{-3}$ and a planet azimuth of $\phi_p = -45^\circ$ in the disc reference frame; {the projected planet location is labeled ``P''. The beam size is shown by the filled circle in the lower left of the first-column panels.} The magenta line denotes the isovelocity contour at the velocity of each channel, {measured from the corresponding centroid velocity map.}
              }
         \label{fig:channels_mechanisms}
   \end{figure*}
\begin{figure*}
   \centering    \includegraphics[width=0.6\textwidth]{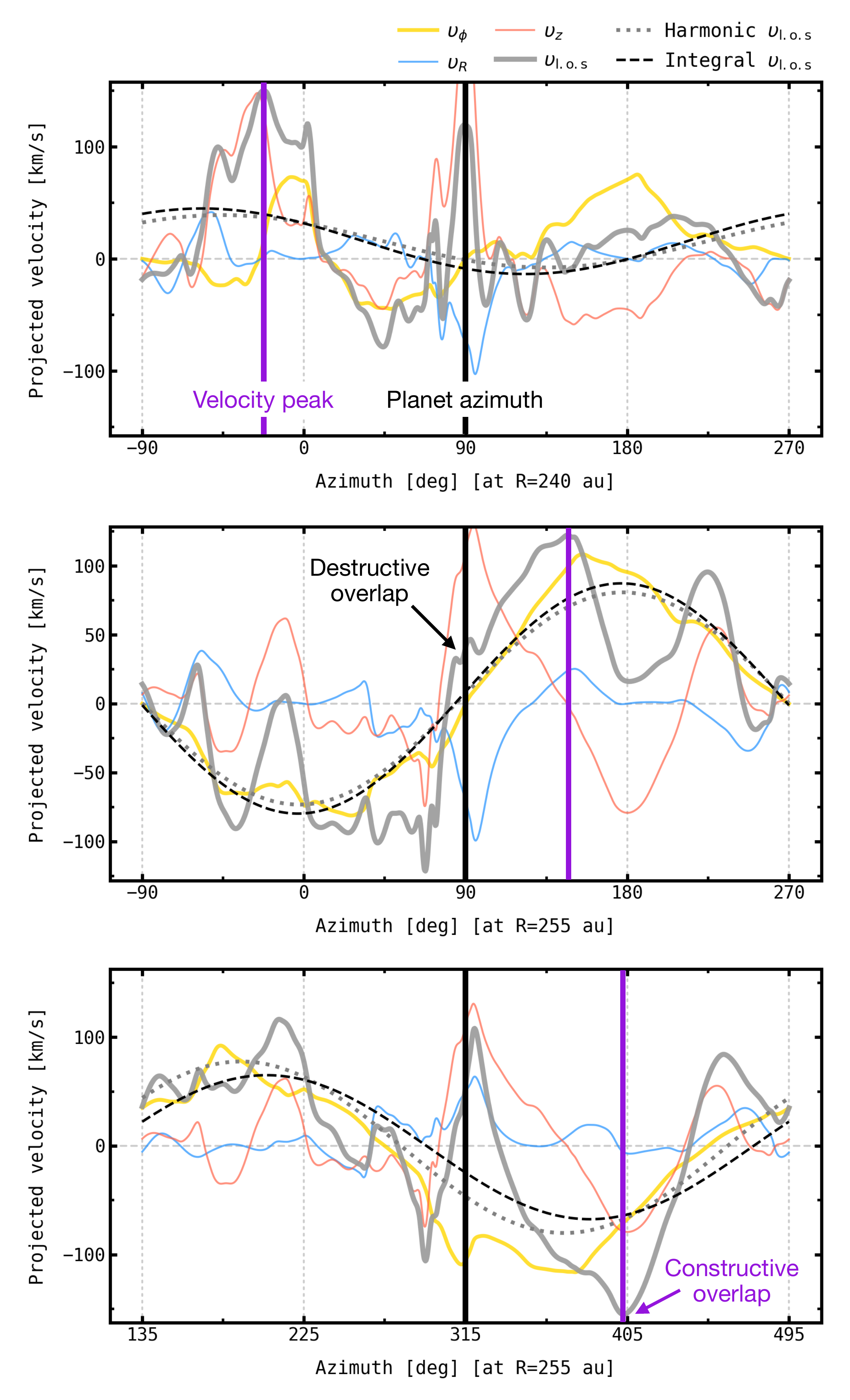}
      \caption{
      Illustrating how the combined contributions of velocity components from a planet-disc interaction simulation ($q=2\times10^{-3}$, before postprocessing), after subtraction of Keplerian motion, can naturally produce peak line-of-sight velocity residuals at azimuthal locations far from the embedded planet, either through constructive overlap of the projected velocities away from the planet or destructive addition near it. The violet and black vertical lines denote the azimuth of the projected peak velocity and the true planet azimuth, respectively. The top two panels show azimuthal profiles of the projected velocities at radii of 240\,au (top) and 255\,au (middle) for a planet located at an azimuth of $\phi_p = 90^\circ$ and orbital radius $R_p = 240$\,au. The bottom row is analogous to the middle but for a planet at $\phi_p = 315^\circ$. All projected velocities were computed assuming a disc inclination of $-30^\circ$ and an emission surface at $z/r = 0.2$. \textit{Harmonic} $v_{\rm los}$ refers to the reconstructed velocity profile obtained by forward fitting the velocity components in the $v_{\rm los}$ equation (i.e. Eq. 3 of \citealt{izquierdo+2025}), assuming they are constant in azimuth, while \textit{Integral} $v_{\rm los}$ follows the analytic solution introduced in \citet{izquierdo+2023}.
              } 
       \label{fig:vlos_profiles_planet}
   \end{figure*}
\begin{figure*}
   \centering
    \includegraphics[width=0.89\textwidth]{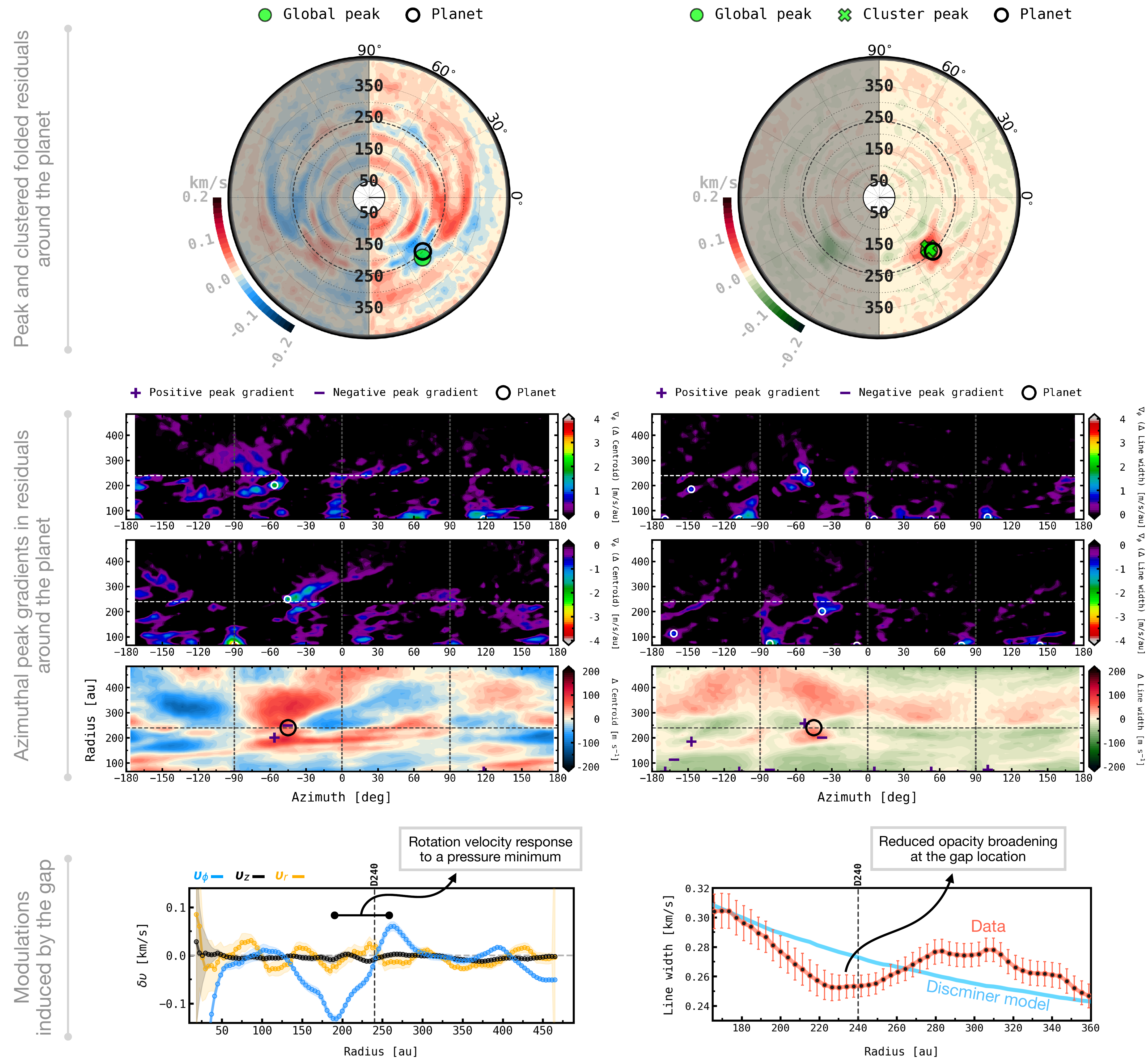}
      \caption{
      Summary of planet-driven signatures identified in the velocity and line-width residuals traced by \twCO{} emission from a planet-disc interaction simulation with a planet-to-star mass ratio of $q\!\sim\!2\times10^{-3}$, {assuming a disc inclination of $i=-30^\circ$ and postprocessed with a fiducial beam size of $0\farcs{15}$.} The planet is embedded at an orbital radius of 240\,au and an azimuth of $-45^\circ$. {See Section \ref{sec:hydro} for details of the simulation and postprocessing setups.} 
              }
    \label{fig:hydro_signatures_12co}
   \end{figure*}
\begin{figure*}
   \centering
    \includegraphics[width=0.8\textwidth]{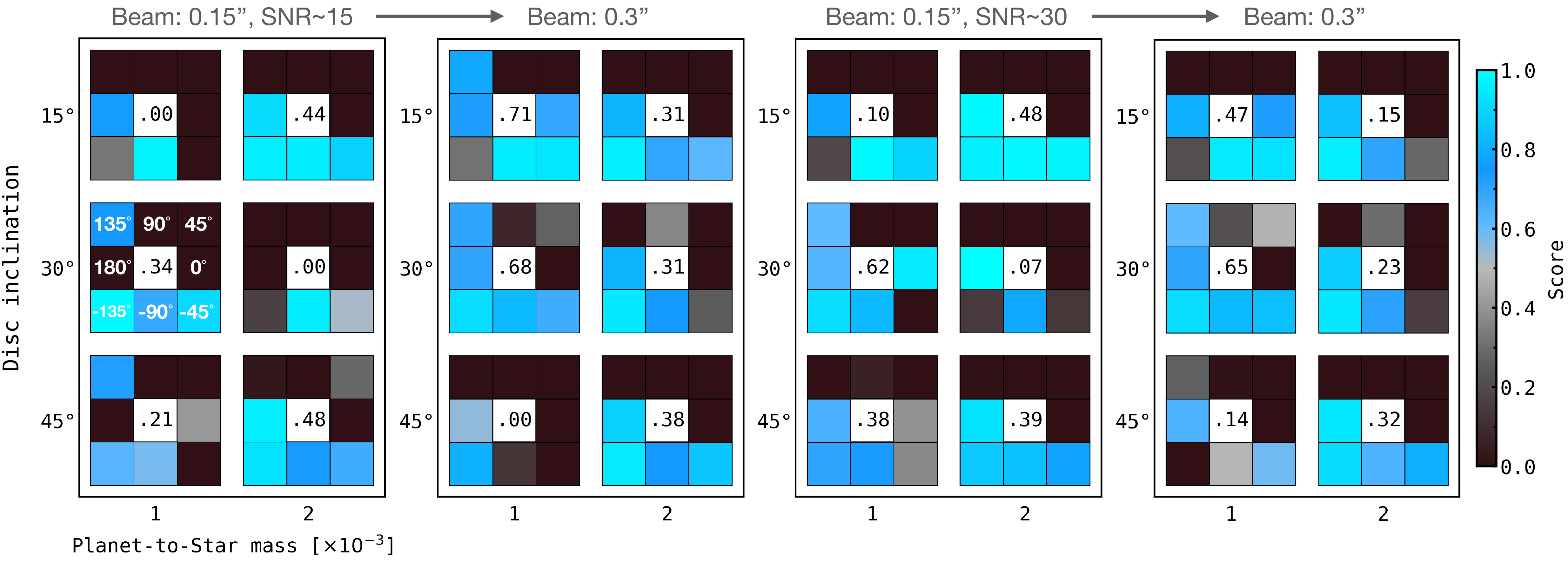}
      \caption{As Fig. \ref{fig:scores_linewidth} but for detections based on localized velocity residuals.
              }
         \label{fig:scores_velocity}
   \end{figure*}

Here we provide complementary figures that support the results presented in the main text. Figure \ref{fig:channels_mechanisms} illustrates selected \twCO{} channel maps from the synthetic observations of the planet-disc interaction and disc-instability simulations introduced in Sect. \ref{sec:hydro}. Figure \ref{fig:vlos_profiles_planet} presents azimuthal profiles of projected velocity components from planet-disc interaction simulations, demonstrating how the constructive or destructive overlap of velocities can induce strong signals in this observable away from the planet location. Figure \ref{fig:hydro_signatures_12co} summarizes the key line diagnostics used to distinguish planet-driven signatures in velocity or line-width residual maps. Figure \ref{fig:scores_velocity} presents the detection score (defined in Sect. \ref{sec:hydro_linewidths}) for planet-driven velocity peaks, complementing the scores based on line-width enhancements (Fig. \ref{fig:scores_linewidth}). 

\begin{figure*}
   \centering
    \includegraphics[width=1.0\textwidth]{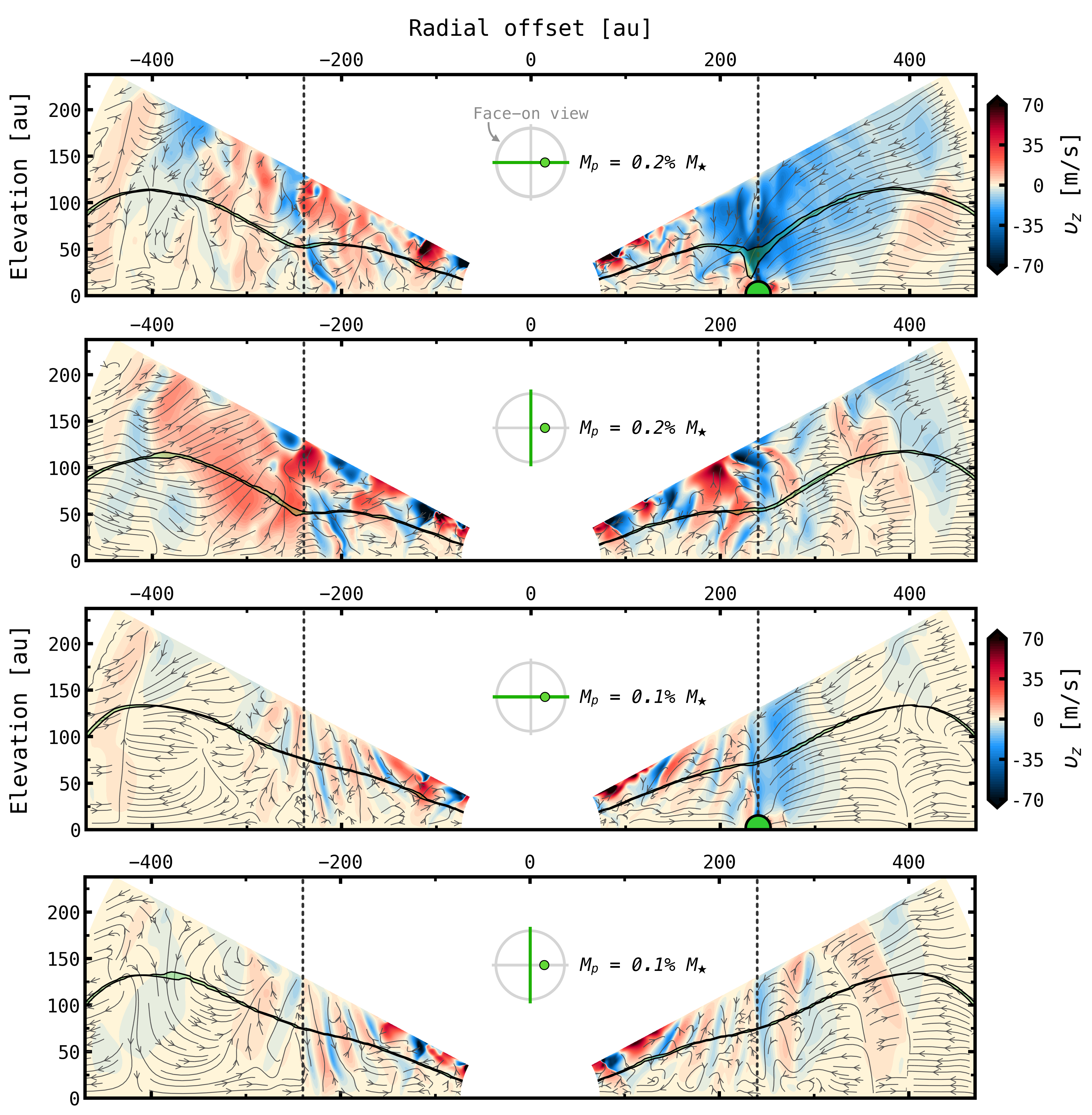}
      \caption{
      Edge-on slices of the vertical velocity component from the planet-disc interaction simulations presented in this work, overlaid with streamlines that incorporate the radial velocity field. The first and third panels from top to bottom show slices taken along the planet azimuth, while the second and fourth panels depict slices perpendicular to it. Optical depth $\tau = 1$ surfaces representative of the \twCO{} emitting region are overlaid as solid black curves, with the region between the minimum and maximum $\tau = 1$ elevations within a $\pm 10^{\circ}$ wedge shaded in green. The strong localized downward flows around the planets are the primary drivers of the line-width enhancements and asymmetries identified in Sect. \ref{sec:hydro}.
              }
    \label{fig:edgeon_velocities_tau}
   \end{figure*}

\begin{figure*}
   \centering
    \includegraphics[width=1.0\textwidth]{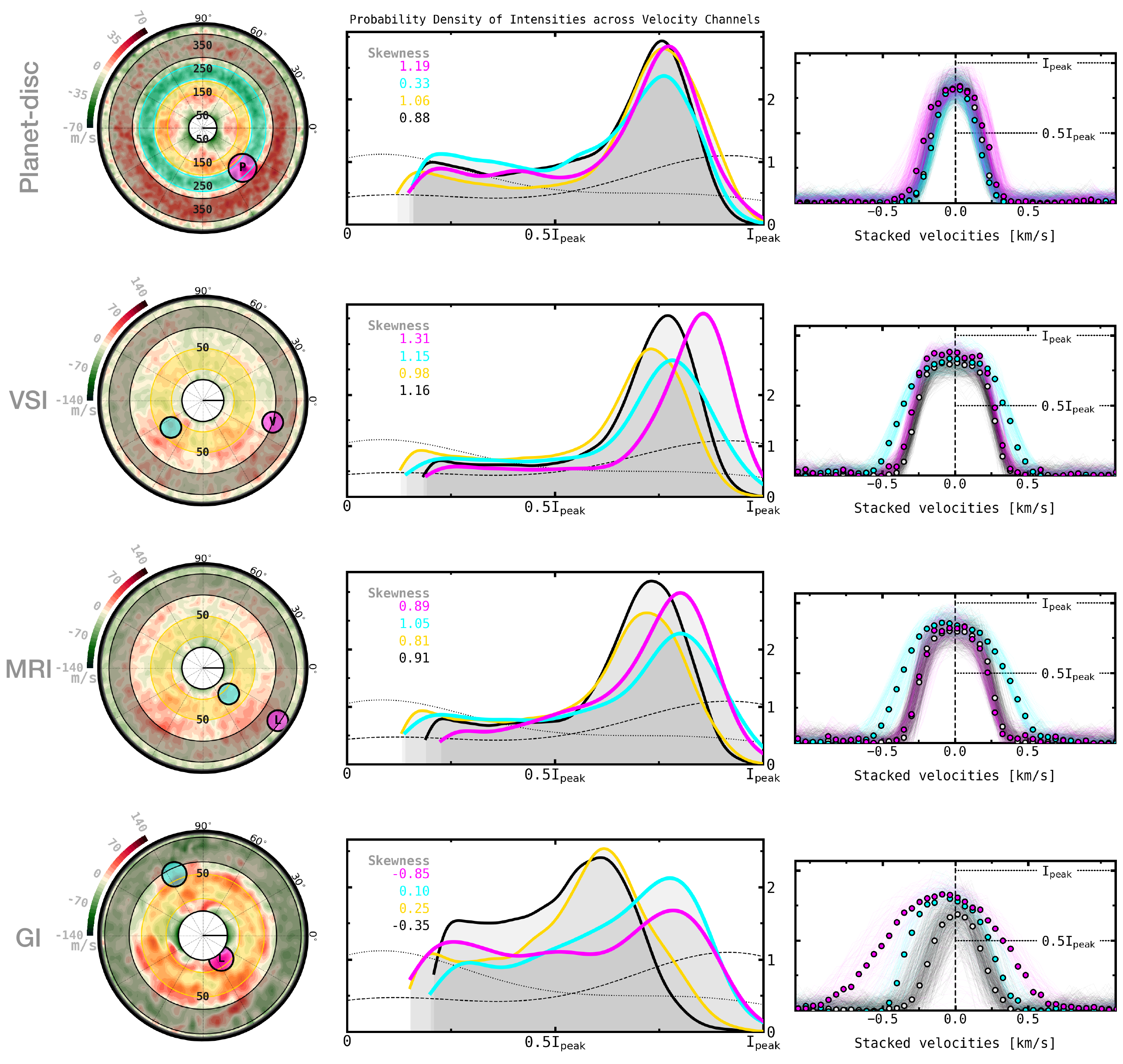}
      \caption{As Fig. \ref{fig:intensdistrib} but for synthetic cubes of \thCOfull{} line emission, assuming a \twCO{}/\thCO{} abundance ratio of 77.
              }
         \label{fig:intensdistrib_13co}
   \end{figure*}

Figure \ref{fig:edgeon_velocities_tau} illustrates edge-on slices of the vertical velocity component from planet-disc interaction simulations with planet masses of 0.1 and 0.2\% of the stellar mass. This figure demonstrates that converging downward flows are localized to the planet's vicinity and are responsible for the line asymmetries discussed in Sect. \ref{sec:line_asymmetries}. Figure \ref{fig:intensdistrib_13co} shows the absence of line asymmetries in optically thin emission, where planet-driven flows influence the line profile equally from both the front and back sides of the disc. This contrasts with the optically thick case, where a redshifted excess is induced in the lines around the planet due to the dominant front-side contribution (Fig. \ref{fig:intensdistrib}). Line broadening associated with localized features in the disc-instability simulations remains symmetric in both optically thin and thick emission. 

\begin{figure*}
   \centering
    \includegraphics[width=0.7\textwidth]{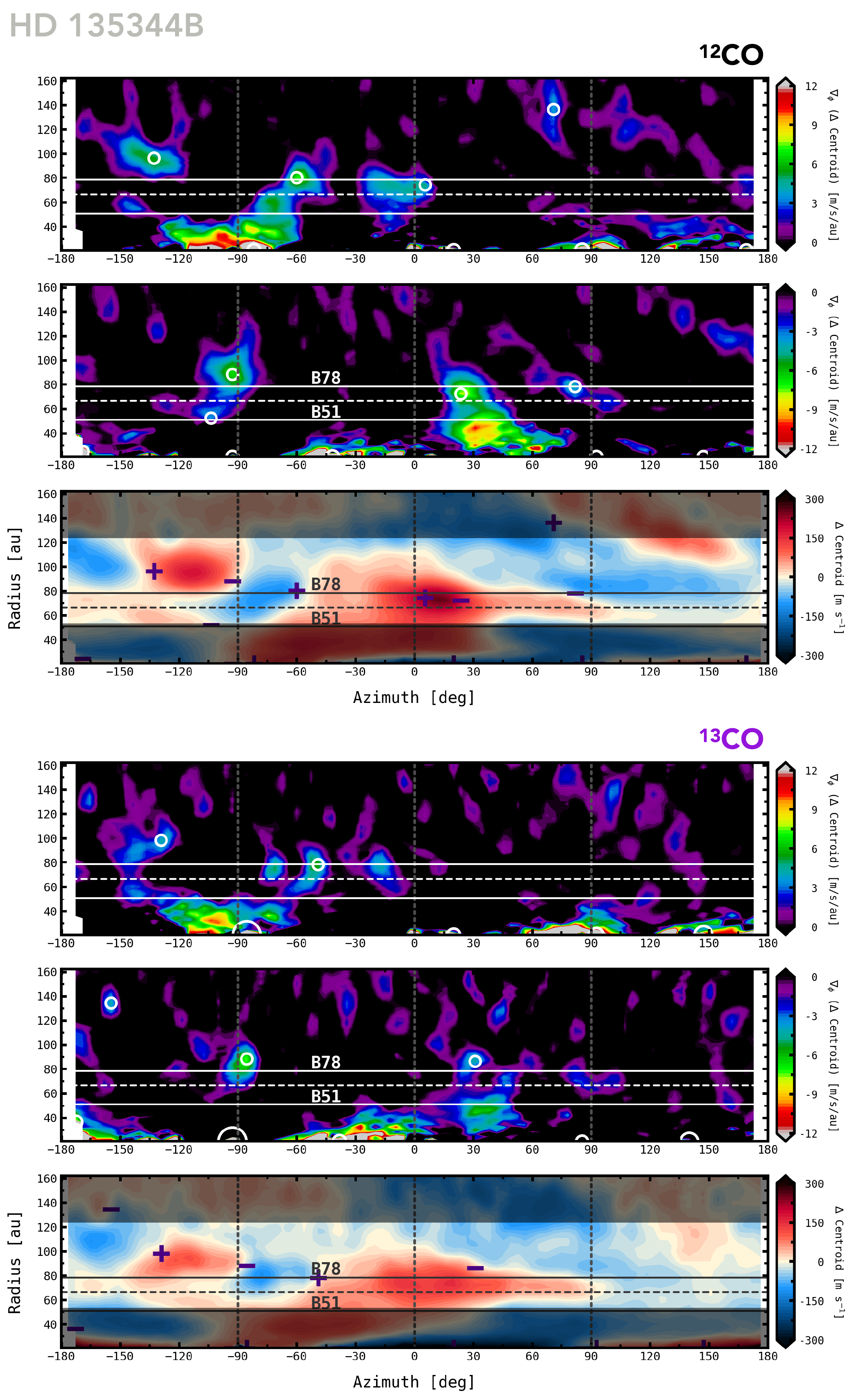}
      \caption{Polar maps illustrating the azimuthal gradients of the velocity residuals in the disc of \hdone{}, derived from \twCO{} and \thCO{} observations. The top and middle sub-panels show the positive and negative gradient components, respectively, while the bottom panels display the corresponding velocity residuals, enabling the identification of Doppler-flip signatures in the kinematics. White circles mark the locations of the steepest gradients exceeding a threshold of $2.5$\,m\,s$^{-1}$\,au$^{-1}$, corresponding to velocity shifts greater than half the channel spacing per beam size (i.e. $50$\,m\,s$^{-1}$\,beam$^{-1}$). Pluses and minuses overlaid on the velocity residuals indicate the sign of the peak azimuthal gradients and highlight the central locations of the Doppler flips. Solid and dashed lines denote the radial separations of millimeter dust rings and gaps, respectively.
              }
         \label{fig:peak_gradients_hd135344}
   \end{figure*}

\begin{figure*}
   \centering
    \includegraphics[width=0.85\textwidth]{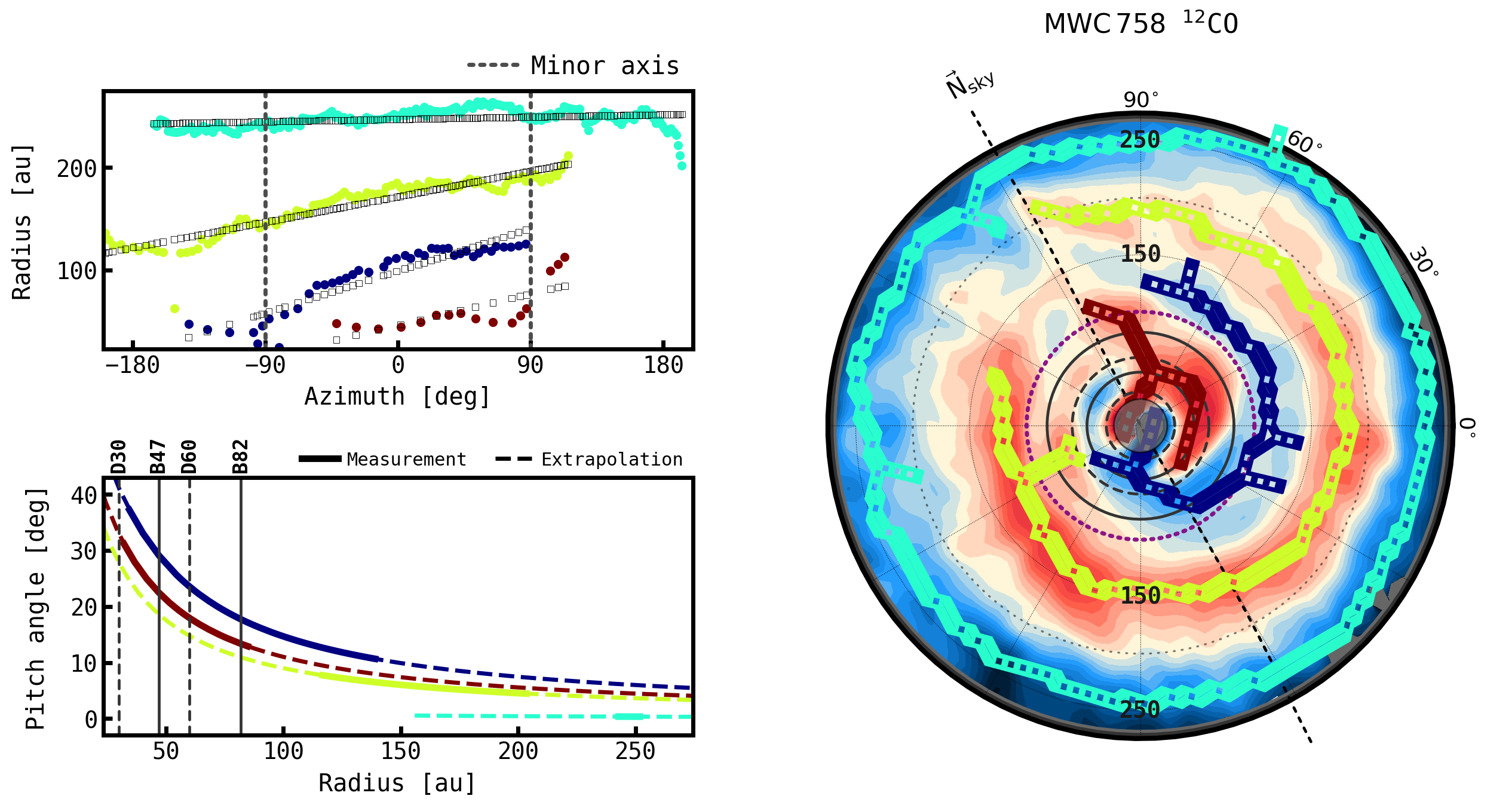} \\ \vspace{0.5cm}
    \includegraphics[width=0.65\textwidth]{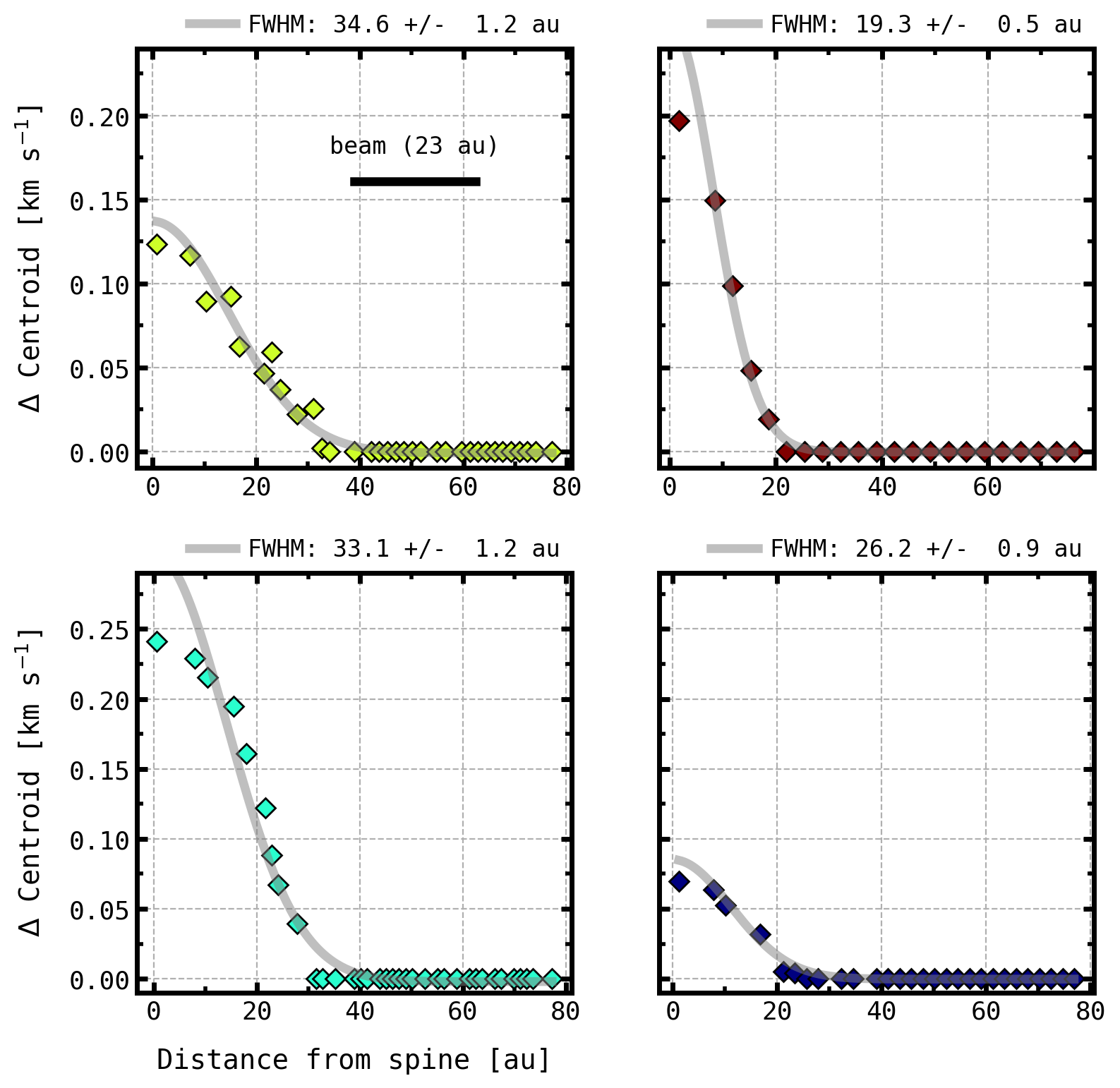}
      \caption{Extraction and characterization of coherent large-scale velocity substructures traced by \twCOfull{} in the disc of \mwcsev{} using \filfinder{}. The top panels illustrate the spatial configuration of the detected filaments, together with linear fits of the form $r = a + b\phi$ used to estimate the pitch angles of the observed substructures. The main non-axisymmetric signature, traced in yellow across all panels, is well described by a tightly wound spiral with pitch angles $<10^\circ$ and dominated by downward vertical motions. The bottom panels display the average widths of the substructures, measured perpendicular to the medial axes of the filaments. Only the longest and outermost signatures (shown in yellow and cyan) are resolved by more than one resolution element.
      }
         \label{fig:filaments_mwc758_12co}
   \end{figure*}

Figure \ref{fig:peak_gradients_hd135344} shows the locations of peak velocity gradients in the azimuthal direction for the disc of \hdone{}, which help identify Doppler flips potentially associated with embedded planets (see Sect. \ref{sec:dopper_flip}). Finally, Figure \ref{fig:filaments_mwc758_12co} summarizes the extraction of large-scale kinematic structures in the disc of \mwcsev{}, together with their pitch-angle and width profiles.

\end{document}